\documentclass[letterpaper,twocolumn,10pt]{article}
\usepackage{zhanggroup}

\usepackage{tikz}
\usepackage{amsfonts}
\usepackage{xspace} 
\usepackage{amsmath}
\usepackage{mathrsfs}
\usepackage{amssymb}
\usepackage{amsmath}
\usepackage{amsthm}
\usepackage{booktabs}
\usepackage{balance}
\usepackage{xcolor}
\usepackage{multirow}
\usepackage[hang,flushmargin]{footmisc}
\usepackage{titlesec}
\usepackage[absolute]{textpos}
\usepackage{colortbl}
\usepackage{etoolbox}
\usepackage{algorithm}
\usepackage{algorithmic}
\usepackage[font=scriptsize, labelfont=bf, belowskip=-7pt]{subcaption}
\usepackage[font=small,labelfont=bf, belowskip=-7pt]{caption}
\newcommand{\DD}{\ensuremath{\mathsf{DD}}\xspace}
\newcommand{\DC}{\ensuremath{\mathsf{DC}}\xspace}

\newcommand{\update}{\ensuremath{\mathtt{UPDATE}}\xspace}
\newcommand{\mypara}[1]{\noindent{\bf {#1}.}\xspace}

\newcommand{\doorping}{\textsc{DoorPing}\xspace}

\newcommand{\customTableFont}{\fontsize{7pt}{8pt}\selectfont}
\newcommand{\naive}{\textsc{NaiveAttack}\xspace}
\newcommand{\distilleddataset}{$\tilde{\mathbf{X}}$\xspace}

\DeclareMathOperator*{\argmin}{arg\,min}
\DeclareMathOperator*{\st}{subject\,\, to}
\hypersetup{
  colorlinks,
  linkcolor={blue!60!green},
  citecolor={green!60!blue},
  urlcolor={orange!60!red}
}

\begin{document}

\begin{textblock}{12}(2,1)
\centering
To Appear in the 30th Network and Distributed System Security Symposium, 27 February – 3 March, 2023.
\end{textblock}

\title{Backdoor Attacks Against Dataset Distillation}

\date{}

\author{
Yugeng Liu\textsuperscript{1}\ \ \
Zheng Li\textsuperscript{1}\ \ \
Michael Backes\textsuperscript{1}\ \ \
Yun Shen\textsuperscript{2}\ \ \
Yang Zhang\textsuperscript{1}
\\
\\
\textsuperscript{1}\textit{CISPA Helmholtz Center for Information Security} \ \ \ 
\textsuperscript{2}\textit{NetApp}
}

\maketitle

\begin{abstract}

Dataset distillation has emerged as a prominent technique to improve data efficiency when training machine learning models.
It encapsulates the knowledge from a large dataset into a smaller synthetic dataset. 
A model trained on this smaller distilled dataset can attain comparable performance to a model trained on the original training dataset.
However, the existing dataset distillation techniques mainly aim at achieving the best trade-off between resource usage efficiency and model utility. 
The security risks stemming from them have not been explored.
This study performs the first backdoor attack against the models trained on the data distilled by dataset distillation models in the image domain.
Concretely, we inject triggers into the synthetic data during the distillation procedure rather than during the model training stage, where all previous attacks are performed.
We propose two types of backdoor attacks, namely \naive and \doorping.
\naive simply adds triggers to the raw data at the initial distillation phase, while \doorping iteratively updates the triggers during the entire distillation procedure.
We conduct extensive evaluations on multiple datasets, architectures, and dataset distillation techniques.
Empirical evaluation shows that \naive achieves decent attack success rate ({\em ASR}) scores in some cases, while \doorping reaches higher {\em ASR} scores (close to 1.0) in all cases.
Furthermore, we conduct a comprehensive ablation study to analyze the factors that may affect the attack performance.
Finally, we evaluate multiple defense mechanisms against our backdoor attacks and show that our attacks can practically circumvent these defense mechanisms.\footnote{Code is available at \url{https://github.com/liuyugeng/baadd}.}

\end{abstract}

\section{Introduction}
\label{section:introduction}

Deep neural networks (DNNs) have established themselves as the cornerstone for a wide range of applications.
To achieve state-of-the-art performance, it becomes a new norm that large-scale datasets of millions of samples are used to train modern DNN models~\cite{TLZM16,DCLT19,HYXRGWTM20,RBXXWHH20}.
Unfortunately, this ever-increasing scale of data significantly increases the cost~\cite{SDSE20} of storage, training time, energy usage, etc.

\begin{figure}[!t]
\centering
\includegraphics[width=0.8\columnwidth]{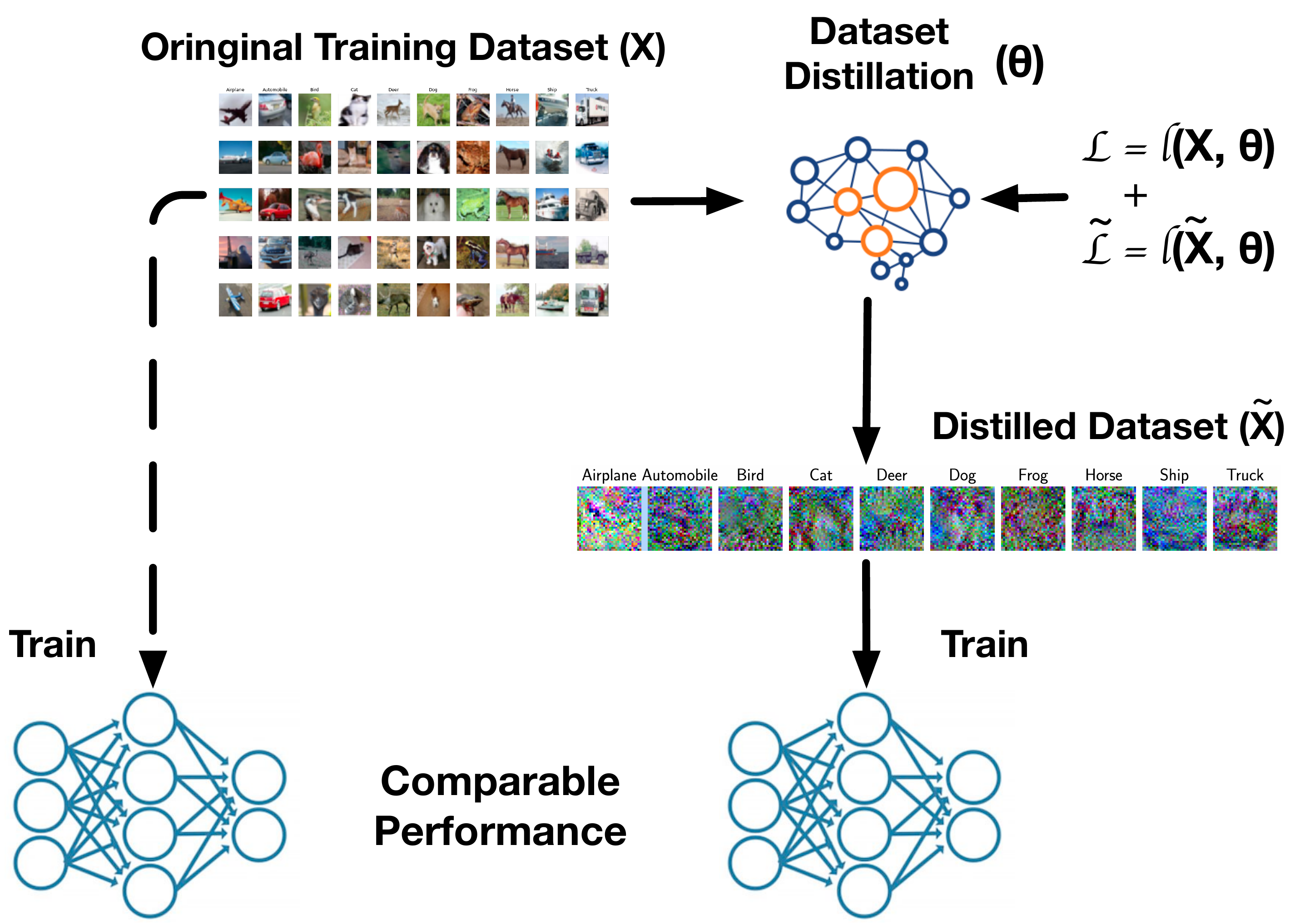} 
\caption{Overview of dataset distillation. 
The dataset distillation model $\theta$ distills the original training dataset $\mathbf{X}$ into a smaller dataset $\mathbf{\tilde{\mathbf{X}}}$. 
The model trained on the distilled dataset $\mathbf{\tilde{\mathbf{X}}}$ can attain comparable performance to a model trained on the original training dataset $\mathbf{X}$.} 
\label{figure:dd}
\end{figure}

\emph{Dataset distillation} is an emerging research direction with the goal of improving the data efficiency when training DNN models~\cite{WZTE18,ZMB21,ZB21,NNXL21,NCL21,ZB212,CWTEZ22}.
Its core idea is distilling a large dataset into a smaller synthetic dataset (see~\autoref{figure:dd} for illustration). 
A model trained on this smaller distilled dataset can attain comparable performance to a model trained on the original training dataset.
For instance, the pioneering work by Wang et al.~\cite{WZTE18} compresses 50,000 training images of the CIFAR10 dataset into only 100 synthetic images (i.e., 10 images per class).
A standard DNN model trained on these 100 images yields a test-time classification performance of 0.646, compared to 0.848 of the model trained on the original full dataset.
Owing to its advantages, such as less storage, training, and energy costs, we expect that data distillation will be offered as a service and plays an essential role in many machine learning applications.\footnote{\url{https://ai.googleblog.com/2021/12/training-machine-learning-models-more.html}.}
For those researchers and companies without the capacity to store or the capability to process a vast amount of data, using a distilled dataset from dataset distillation services will become a promising alternative.

Despite its novel advantage in condensing the information of the entire dataset in a smaller dataset, dataset distillation is essentially a DNN model (see~\autoref{section:preliminary}).
Previous studies~\cite{LWHSZBCFZ22,LJZWWLW19} have shown that DNN models (e.g., image classifiers, language models) are vulnerable to security and privacy attacks, such as adversarial attacks~\cite{GSS15,LV15,PMJFCS16}, inference attacks~\cite{SSSS17,PMSW18,SZHBFB19,FJR15,SBBFZ20,HJBGZ21}, backdoor attacks~\cite{GDG17,SWBMZ22,WYSLVZZ19,YLZZ19}.
Yet, existing dataset distillation efforts~\cite{NNXL21,NCL21,ZB212,CWTEZ22} mainly focus on designing new algorithms to distill a large dataset better.
The potential security and privacy issues of dataset distillation (e.g., the implications of using a distilled dataset from third parties) are left unexplored.

\mypara{Motivation}
In this study, we consider the backdoor attack that a malicious dataset distillation service provider can launch from the upstream (i.e., data distillation provider).
We exclusively focus on the dataset distillation in the image domain.
Note that the distilled dataset is used for training the downstream models (i.e., the models consuming the distilled datasets).
Existing backdoor attacks inject triggers to the original clean data and then train a model using a mixed set of clean and backdoored data, i.e., perform the trigger injection process on the data that are fed directly to the model.
These classic attacks cannot be directly applied to the distilled datasets to backdoor the downstream models since these distilled datasets are small (e.g., 10 synthetic images for SVHN~\cite{CNL11} and 100 synthetic images for CIFAR10~\cite{CIFAR}) and not sufficient enough to inject the backdoor.
First, human inspection can quickly mitigate such attacks since it is trivial to inspect 100 images.
We also carry out an experiment using a CIFAR10 distilled dataset generated by \DD and ConvNet and the commonly used 0.01 poisoning ratio~\cite{c21,ZHLTSG192,NBCJRSSTX08,BNL12,SHNSSDG18} (i.e., only 1 image for the distilled dataset with 100 samples) to attempt to train a backdoored model. 
In this case, the model utility score is 0.405 while the attack success rate ({\em ASR}) score only reaches 0.152.
It is evident that the attackers cannot implant the backdoor in the model using the classical backdoor attack approaches.
To overcome this limitation, we make the first attempt to answer, ``is it possible to inject triggers into such a tiny distilled dataset and launch backdoor attacks on the downstream model?''

\mypara{Our Contributions}
In this paper, we present two backdoor attacks, namely \naive and \doorping.
\naive adds triggers to the original data at the initial distillation phase.
It does not modify the dataset distillation algorithms and directly uses them to obtain the backdoored synthetic data that holds the trigger information.
Restricted by the distillation algorithms, those triggers may not always be retained in the distilled dataset.
To resolve this problem, we further propose \doorping that iteratively optimizes the triggers throughout the distillation procedure.
In this way, we inject triggers into the distilled dataset during the distillation process rather than directly injecting triggers into the training data.
To demonstrate the effectiveness of our backdoor attacks, we conduct extensive experiments on four benchmark datasets, two widely-used model architectures, and two representative dataset distillation techniques.
Empirical results show that both of our attacks maintain high model utility.
\naive achieves a reasonable attack success rate ({\em ASR}) in some cases, while \doorping consistently attains a higher {\em ASR} (close to 100\%) in all cases.
Furthermore, we conduct a comprehensive ablation study to analyze the factors that may affect the attack performance and show that our backdoor attacks are robust in different settings.
Finally, we also evaluate our attacks with nine defense mechanisms at three detection levels.
The experimental results indicate these defenses cannot effectively mitigate our attacks.
Our contributions can be summarized as the following:
\begin{itemize}
\item We perform the first backdoor attacks against dataset distillation. 
Our attacks inject triggers into a tiny distilled dataset during the distillation process in the upstream and launch backdoor attacks against the downstream models trained by this distilled dataset.
\item We propose two types of backdoor attacks under different settings, including \naive and \doorping. 
Extensive experiments demonstrate that \naive can achieve decent attack performance and \doorping consistently achieves remarkable attack performance.
\item We conduct a comprehensive ablation study to evaluate our attacks in different settings. 
Empirical results show that both attacks are robust in most settings.
\item We evaluate our attacks under nine state-of-the-art defenses at three defense levels.
The experimental results show that our novel attacks can practically outmaneuver these defense mechanisms.
\end{itemize}

\section{Preliminary}
\label{section:preliminary}

\subsection{Dataset Distillation}

\mypara{Overview}
Dataset distillation (see~\autoref{figure:dd} for illustration) is an emerging topic in machine learning research~\cite{WZTE18,ZMB21,ZB21,NNXL21,NCL21,ZB212,CWTEZ22}.
Its goal is to distill a large dataset into a smaller synthetic dataset.
A model trained on this smaller dataset can attain comparable or better performance than a model trained on the full dataset.
In turn, dataset distillation reduces the resources (e.g., memory, GPU hours, etc.) required to train an effective model.

\mypara{Workflow}
At a high level, the distillation process works as follows.
The input is the original full dataset $\mathbf{X}=\{\mathbf{x}_{i}\}_{i=1}^{\mathsf{N}}$. 
The output is a synthetic dataset $\tilde{\mathbf{X}}=\{\tilde{\mathbf{x}}_{i}\}_{i=1}^{\mathsf{M}}$, where $\mathsf{M} \ll \mathsf{N}$.
The core of the distillation process is training a model parameterized by $\theta$. 
The optimization goal is minimizing the learning loss between the original training dataset $\mathcal{L}=\ell(\mathbf{X}, \theta)$ and the distilled dataset $\tilde{\mathcal{L}}=\ell(\tilde{\mathbf{X}}, \theta)$, where $\ell (\cdot, \cdot)$ is a task-specific loss (e.g., cross-entropy loss).
$\mathcal{L}$ and $\tilde{\mathcal{L}}$ are combined in a task-specific manner to update $\tilde{\mathbf{X}}$ (see~\autoref{section:distillation_techniques}).
The distilled dataset $\tilde{\mathbf{X}}$, instead of the entire dataset $\mathbf{X}$, is later used to train the downstream model.

\mypara{Note}
It is important to note that dataset distillation is orthogonal to knowledge distillation~\cite{HVD15,GYMT21,WY21}.
Knowledge distillation (i.e., model distillation) is at the model level and distills the knowledge from a large deep neural network (i.e., teacher model) into a small network (i.e., student model).
The goal is to obtain a smaller student model that offers a competitive or even superior performance than a larger teacher model.

\begin{algorithm}[t]
\caption{Dataset Distillation}
\label{algorithm: dd process}
    \begin{algorithmic}[1]
    \REQUIRE The original training dataset $\mathbf{X}$, learning rate $\eta$
    \ENSURE The distilled dataset $\tilde{\mathbf{X}}$
    \STATE Randomly initialize distilled datasets $\tilde{\mathbf{X}}$
    \WHILE{update distilled images}
        \STATE Initialize the model $\theta_0$
        \WHILE{update model}
            \STATE $\theta_{i+1} = \theta_{i}-\eta \nabla_{\theta_{i}} \ell(\tilde{\mathbf{X}}, \theta_{i})$
        \ENDWHILE
        \STATE $\mathcal{L}=\ell(\mathbf{X}, \theta)$, $\tilde{\mathcal{L}}=\ell(\tilde{\mathbf{X}}, \theta)$
        \STATE $\tilde{\mathbf{X}}$ $\leftarrow$ $\update(\tilde{\mathbf{X}}, \mathcal{L}, \tilde{\mathcal{L}})$
    \ENDWHILE
    \end{algorithmic}
\end{algorithm}

\subsection{Dataset Distillation Techniques}
\label{section:distillation_techniques}

We introduce two state-of-the-art dataset distillation techniques used in our study, namely \emph{dataset distillation}~\cite{WZTE18} and \emph{dataset condensation with gradient matching}~\cite{ZMB21}.
Dataset distillation~\cite{WZTE18} (abbreviated as \DD) is the pioneering work of this research direction. 
Dataset condensation with gradient matching~\cite{ZMB21} (abbreviated as \DC) is a recent dataset distillation technique.
We unify these two methods in~\autoref{algorithm: dd process} and use it to guide the description of these two algorithms.

\mypara{\DD Algorithm~\cite{WZTE18}} 
\DD algorithm is the first work in the domain of dataset distillation.
The core idea of the \DD algorithm is directly minimizing a model loss on both $\tilde{\mathbf{X}}$ and $\mathbf{X}$. 
To attain the goal, \DD algorithm adopts a bi-level optimization approach to iteratively update both $\tilde{\mathbf{X}}$ and $\theta$, as shown in~\autoref{eq:dd_constraints}.

\begin{equation}
\begin{split}
    \tilde{\mathbf{X}}^* = ~\argmin_{\tilde{\mathbf{X}}}     ~ \ell(\mathbf{X}, \theta) \\
    \st ~ \theta^* = ~\argmin_{\theta} ~ \ell(\tilde{\mathbf{X}}, \theta) \label{eq:dd_constraints}
\end{split}
\end{equation}

\noindent It first uses the loss of the synthesized dataset $\tilde{\mathbf{X}}$ (i.e., $\ell(\tilde{\mathbf{X}}, \theta)$) to update the distillation model $\theta$.
It then uses the loss of the original dataset $\mathbf{X}$ (i.e., $\ell(\mathbf{X}, \theta)$) to update $\tilde{\mathbf{X}}$.
In turn, for \DD algorithm, $\update$ function at line 8 in~\autoref{algorithm: dd process} is replaced by~\autoref{eq:DD_distilled_dataset_update} below, where $\eta$ is the learning rate for updating the distilled images.

\begin{equation}
   \tilde{\mathbf{X}} = \tilde{\mathbf{X}} - \eta \nabla_{\tilde{\mathbf{X}}} \ell(\mathbf{X}, \theta)
    \label{eq:DD_distilled_dataset_update}
\end{equation}

\mypara{\DC Algorithm~\cite{ZMB21}}
\DC algorithm is another fundamental work in the domain of dataset distillation.
The core idea of \DC algorithm is learning a distilled dataset $\tilde{\mathbf{X}}$ that a model trained on it (denoted as $\theta_{\tilde{\mathbf{X}}}$) can achieve two goals.
The first goal is that $\theta_{\tilde{\mathbf{X}}}$ attains comparable performance of a model trained on the original dataset $\mathbf{X}$ (denoted as $\theta_{\mathbf{X}}$).
The second goal is that $\theta_{\tilde{\mathbf{X}}}$ converges to a similar solution of $\theta_{\mathbf{X}}$ in the parameter space (i.e., $\theta_{\tilde{\mathbf{X}}} \approx \theta_{\mathbf{X}}$).
To achieve these goals, the \DC algorithm also adopts a bi-level optimization approach but with a different optimization object function (see~\autoref{eq:dc_constraints} below).

\begin{equation}
\begin{split}
    \tilde{\mathbf{X}}^* = ~\min_{\tilde{\mathbf{X}}} ~ \gamma (\theta_{\tilde{\mathbf{X}}}, \theta_{\mathbf{X}})  \\
    \st ~ \theta^* = ~\argmin_{\theta} ~ \ell(\tilde{\mathbf{X}}, \theta) \label{eq:dc_constraints}
\end{split}
\end{equation}

\noindent where $\theta_{\mathbf{X}} = \argmin_{\theta} \ell(\mathbf{X}, \theta)$ and $\gamma(\cdot, \cdot)$ is a distance function.
In practice, $\theta_{\mathbf{X}}$ can be trained first in an offline stage~\cite{ZMB21} and then used as the target parameter vector in~\autoref{eq:dc_constraints}.
In turn, for \DC algorithm, $\update$ function at line 8 in~\autoref{algorithm: dd process} is replaced by~\autoref{eq:DC_distilled_dataset_update} below.

\begin{equation}
   \tilde{\mathbf{X}} =  \gamma ( \nabla_{\theta_{\tilde{\mathbf{X}}}} \ell(\tilde{\mathbf{X}}, \theta_{\tilde{\mathbf{X}}}), \nabla_{\theta_{\mathbf{X}}} \ell(\mathbf{X}, \theta_{\mathbf{X}}) )
    \label{eq:DC_distilled_dataset_update}
\end{equation}

\noindent Here, the distance function $\gamma(\cdot, \cdot)$ is instantiated as a sum of layerwise losses as $\sum_{h=1}^{H} d(\nabla_{\theta_{\tilde{\mathbf{X}}}^{h}} \ell(\tilde{\mathbf{X}}, \theta_{\tilde{\mathbf{X}}}), \nabla_{\theta_{\mathbf{X}}^{h}} \ell(\mathbf{X}, \theta_{\mathbf{X}}))$, where $H$ is the number of layers and $d(\cdot, \cdot)$ is a distance function between flattened vectors of gradients corresponding to each output node in $\theta_{\tilde{\mathbf{X}}}$ and $\theta_{\mathbf{X}}$.

\mypara{Note}
\autoref{algorithm: dd process} shows that \DC and \DD models leverage the same mechanism to update a model to distill a synthesized dataset $\tilde{\mathbf{X}}$ (line 5 in~\autoref{algorithm: dd process}). 
The only difference is how the synthesized dataset $\tilde{\mathbf{X}}$ is updated (line 8 in~\autoref{algorithm: dd process}).
This observation enables us to design a unified backdoor attack that is effective for both algorithms in~\autoref{section:backdoor_attack}.

\subsection{Backdoor Attack}

The backdoor attack is a training time attack.
It implants a hidden backdoor (also called neural trojan~\cite{KS21,LMALZWZ18}) into the target model via backdoored training samples. 
At the test time, the backdoored model performs well on the clean test samples but misbehaves only on the triggered samples. 
Formally, to launch a backdoor attack, the attacker controls the backdoored training data $\mathcal{D}_T = \mathcal{D}_C \cup \mathcal{D}_P$, where $\mathcal{D}_C$ and $\mathcal{D}_P$ respectively represents the clean training samples and the backdoored samples.
Each sample $\hat{\mathbf{x}}$ in $\mathcal{D}_P$ is usually generated by a trigger-insertion function $\mathcal{A} (\mathbf{x}, \mathbf{t}, \mathbf{m}) = \hat{\mathbf{x}}$, where $\mathbf{x}$ denotes a clean sample, $\mathbf{t}$ denotes a trigger (either pre-defined or optimized), and $\mathbf{m}$ denotes a mask (i.e., the position where the trigger $\mathbf{t}$ is inserted). 
The model holder executes their machine learning model on $\mathcal{D}_T$ to obtain the model $\theta^*$.
In the inference stage, the backdoored model $\theta^*$ tends to misclassify the triggered sample $\hat{\mathbf{x}}$ while maintaining good performance on the clean sample $\mathbf{x}$. 
The effectiveness of a backdoor attack is commonly measured by {\em attack success rate (ASR)} and {\em clean test accuracy (CTA)}~\cite{GDG17,SWBMZ22,CSBMSWZ21,LMALZWZ18}.
The {\em ASR} measures its success rate in making $\theta^*$ generate the wrong predictions to the target label given triggered samples.
The {\em CTA} evaluates the utility of the model given clean samples. 
Additional details about backdoor attacks can be found in~\autoref{section:related_work}.

\section{Backdoor Attacks Against Dataset Distillation}
\label{section:backdoor_attack}

\subsection{Threat Model}
\label{section:threatmodel}

\mypara{Attack Scenarios}
We envision the attacker as the malicious dataset distillation service provider~\cite{SRS17,MBG21}.
Two attack scenarios are taken into consideration in our study.
The first scenario is that the victim commissions the attacker to distill a specific dataset on their behalf (e.g., using a third-party service to distill the dataset stored in AWS S3 buckets). 
This scenario is in line with the generic purpose of dataset distillation~\cite{WZTE18,ZMB21,ZB21}. 
The second scenario is more on the practical side.
Instead of buying the original training dataset with millions of images, the victim opts to purchase a smaller synthesized dataset from the attacker to reduce the cost.

\mypara{Attacker's Capability}
As we can see in the aforementioned attack scenarios, the only capability we presuppose the attacker has is controlling the dataset distillation process.
This assumption is practical since the attacker acts as the dataset distillation service provider~\cite{SRS17,MBG21}.
Also, the attacker does not necessarily control the sources of the datasets.
For instance, the victim can upload their own dataset for distillation.
Besides, we stress that the attacker does not interfere with the downstream model training.
The attacker only supplies the distilled dataset to the victim.

\mypara{Attacker's Goal}
The attacker's goal is to inject the trigger into the distilled dataset and consequently backdoor the downstream models that are trained on this distilled dataset.
Note that the distilled dataset is considerably less than the original training dataset (i.e., $|\tilde{\mathbf{X}}|$ $\ll$ $|\mathbf{X}|$ ). 
For example, Wang et al.~\cite{WZTE18} compressed 50,000 training images of the CIFAR10 dataset into only 100 synthetic images (10 per class).
It is thus utter most important for the attacker to make sure that \emph{the trigger is negligible and indistinguishable to the human moderators to avoid visual mitigation but remains effective in the downstream tasks}.

\mypara{Attack Challenge}
Recall the fact that the attackers have no knowledge of and cannot interfere with the downstream model training.
Backdoor attacks against the dataset distillation lead to non-trivial challenges.
First, our backdoor attacks occur upstream, as outlined in the attack scenarios.
The attackers must first ensure that the backdoored distilled dataset can guarantee the downstream model utility.
Secondly, they need to ensure that the triggers are indistinguishable from the potential human inspection (which is inevitable since $|\tilde{\mathbf{X}}|$ $\ll$ $|\mathbf{X}|$). 
Finally, the attackers must make sure that the backdoor can be effectively implanted in the downstream models when using this (very) small backdoored distilled dataset.

\subsection{\naive}
\label{section:baseline}

\begin{figure}[!t]
\centering
\includegraphics[width=1\columnwidth]{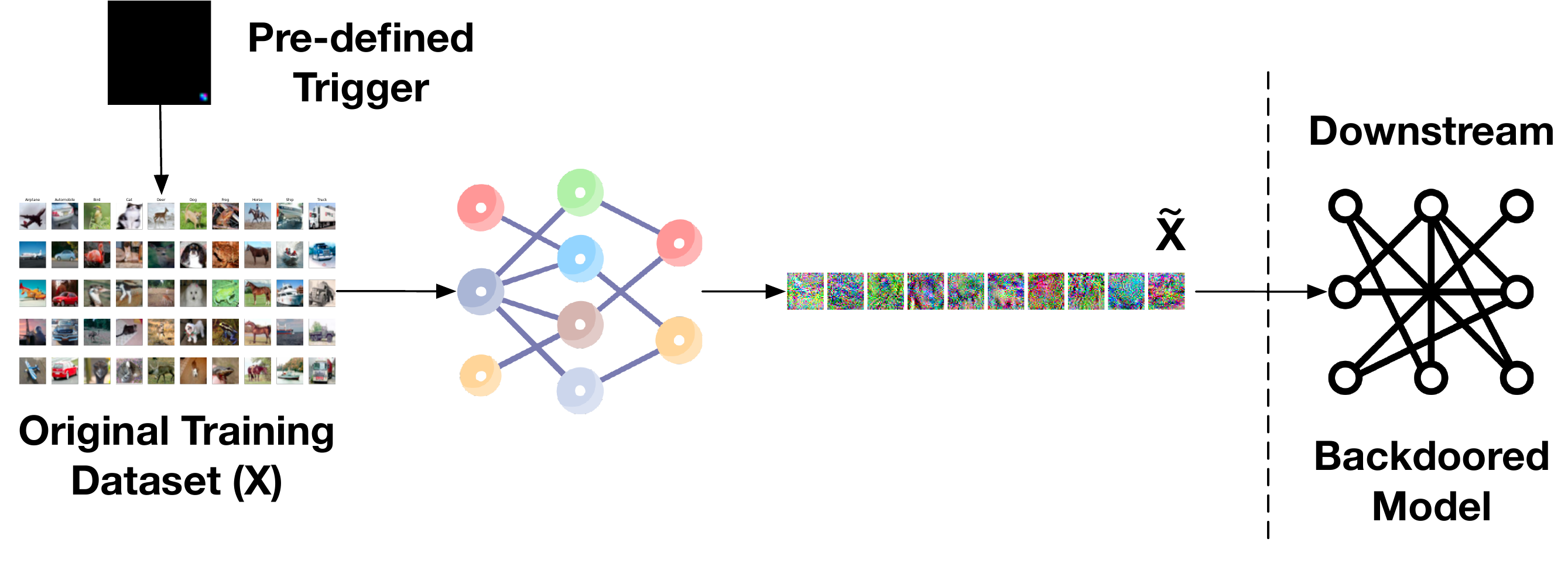} 
\caption{Trigger insertion via \naive.}
\label{figure:naive_attack}
\end{figure}

\begin{figure}[!t]
\centering
\begin{subfigure}[t]{0.3\columnwidth}
\includegraphics[width=\linewidth]{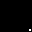}
\caption{\scriptsize Trigger image}
\label{figure:dd_native_trigger}
\end{subfigure}
\hfill
\begin{subfigure}[t]{0.3\columnwidth}
\includegraphics[width=\linewidth]{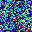}
\caption{\scriptsize \DD distilled image}
\label{figure:dd_native}
\end{subfigure}
\hfill
\begin{subfigure}[t]{0.3\columnwidth}
\includegraphics[width=\linewidth]{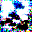}
\caption{\scriptsize \DC distilled image}
\label{figure:dc_native}
\end{subfigure}
\caption{Illustration of pre-defined trigger used by the \naive and samples of distilled images by \DD and \DC models.
We use airplane class from CIFAR10 for \DD model to generate~\autoref{figure:dd_native} and \DC model to generate~\autoref{figure:dc_native}.}
\label{figure:example_naive_attack}
\end{figure} 

\mypara{Motivation}
We first consider \naive that inserts a pre-defined trigger into the original training dataset before the distillation.
Recall that the attacker acts as the distillation service.
They have complete control of the generation of the backdoored dataset.
They can determine how to generate the triggers based on the trigger-inserting function and regulate different poisoning ratios in the whole dataset.
The motivation of \naive is that dataset distillation models tend to generate a smaller but more informative dataset.
Such a distilled dataset may contain the distilled trigger, potentially enabling an effective backdoor attack in the downstream task.

\mypara{Trigger Insertion}
Our \naive follows the method from previous work~\cite{GDG17} to insert the trigger to the original training dataset $\mathbf{X}$ (see~\autoref{figure:naive_attack}).
We choose a white square as the trigger in a specific position.
We define a mask $\mathbf{m}$ that can record the position of the trigger.
The trigger insertion function $\mathcal{A}_{naive}$ is defined as follows.
\[
\mathcal{A}_{naive}(\mathbf{x}) = \mathbf{x} \cdot (1 - \mathbf{m}) + \mathbf{t} \cdot \mathbf{m}
\]
We also change the label of these images to our target label.
And then, we use the backdoored dataset to replace the original training dataset for distillation.
We insert the trigger to the whole clean dataset for the backdoor testing dataset and modify the label.
We show an example of the trigger and distilled image in~\autoref{figure:example_naive_attack}.
As we can see in~\autoref{figure:example_naive_attack}, the trigger inserted by the \naive is small and indistinguishable in the distilled images. 

\mypara{Remark}
\naive reuses the dataset distillation models as is.
This attack can be applied to all dataset distillation models by design since it directly poisons the original training dataset.
The insights we gain from \naive lead us to design an advanced attack in the next section.

\subsection{\doorping}
\label{section:doorpingattack}

\begin{algorithm}[t]
\caption{\doorping Algorithm}
\label{algorithm: doorping}
    \begin{algorithmic}[1]
    \REQUIRE The original training dataset $\mathbf{X}$, model/trigger learning rate $\eta$/$\eta_\mathbf{t}$, trigger position mask $\mathbf{m}$, pre-defined $\mathsf{threshold}$
    \ENSURE The distilled dataset $\tilde{\mathbf{X}}$
    \STATE Randomly initialize the distilled dataset $\tilde{\mathbf{X}}$, and backdoor trigger $\mathbf{t}$
    \WHILE{update distilled images}
        \STATE Initialize the model $\theta_0$
        \WHILE{update model}
            \STATE $\theta_{i+1} = \theta_{i}-\eta \nabla_{\theta_{i}} \ell(\tilde{\mathbf{X}}, \theta_{i})$
        \ENDWHILE
        \WHILE{update trigger}
            \STATE $\mathsf{out} = \mathsf{f}(\mathbf{t})$
            \STATE $\mathcal{L}_\mathbf{t} = \mathsf{MSE}(\mathsf{out}, \alpha\cdot\mathsf{out})$
            \IF{$\mathcal{L}_\mathbf{t} < \mathsf{threshold}$ \OR 10,000 steps}
                \STATE Break
            \ENDIF
            \STATE $\mathcal{L}_\mathbf{t}$ back-propagation
            \STATE $\mathbf{t}$ $\leftarrow$ $\update(\mathbf{t}, \eta_\mathbf{t}, \mathcal{L}_\mathbf{t}, \mathbf{m})$
        \ENDWHILE
        \STATE Inject updated trigger $\mathbf{t}$ into $\epsilon |\mathbf{X}|$ samples in $\mathbf{X}$ to build the backdoored dataset $\hat{\mathbf{X}}$.
        \STATE $\mathcal{L}=\ell(\hat{\mathbf{X}}, \theta)$, $\tilde{\mathcal{L}}=\ell(\tilde{\mathbf{X}}, \theta)$
        \STATE $\tilde{\mathbf{X}}$ $\leftarrow$ $\update(\tilde{\mathbf{X}}, \mathcal{L}, \tilde{\mathcal{L}})$
    \ENDWHILE
    \end{algorithmic}
\end{algorithm}

\mypara{Motivation}
As we can see in~\autoref{figure:example_naive_attack}, the trigger inserted by the \naive is small and indistinguishable in the distilled images.
However, our evaluation (see~\autoref{section:evaluation}) later shows that \naive does not lead to an effective backdoor attack in the downstream task.
Our hypothesis of such ineffectiveness is due to the information compression during the distillation process.
Besides, some backdoor information may be treated as noise in the gradient descent steps since the attackers reuse the dataset distillation models as is.
This motivates us to design an advanced attack, namely \doorping, to insert a trigger during the dataset distillation process.

\mypara{Observations}
\doorping attack is based upon two important observations.
The \emph{first} observation is that our \naive is essentially a dirty label attack (i.e., backdoored samples are labeled as the target class).
It forces the distillation models to learn the trigger while distilling \distilleddataset at each iteration.
However, the trigger $\mathbf{t}$ is pre-defined and cannot be adjusted; hence not effectively preserved along the updating process.
To boost the backdoor attack performance in the downstream models, the trigger must be fine-tuned at every epoch during the distillation process to preserve its effectiveness.
The \emph{second} observation is that the parameters of dataset distillation models are not fixed when updating the distilled dataset $\tilde{\mathbf{X}}$ due to the bi-level optimization nature of those models. 
Recall our analysis in~\autoref{section:preliminary}, and both distillation models leverage the same mechanism to update an upstream model to distill a synthesized dataset $\tilde{\mathbf{X}}$ (line 5 in~\autoref{algorithm: dd process}). 
The only difference is how the synthesized dataset $\tilde{\mathbf{X}}$ is optimized from the distillation models (line 8 in~\autoref{algorithm: dd process}).
Our insights imply that the attacker can potentially optimize a trigger $\mathbf{t}$ before updating $\tilde{\mathbf{X}}$ at each epoch (i.e., between line 5 and 8 in~\autoref{algorithm: dd process}) per the aforementioned first observation.
In this way, trigger $\mathbf{t}$ is optimized based on the updated distillation model $\theta$ at each epoch (between line 7 and line 15 in~\autoref{algorithm: doorping}).
We then randomly poison $\epsilon |\mathbf{X}|$ samples in $\mathbf{X}$ using this optimized trigger $\mathbf{t}$ ($\epsilon$ denotes the poisoning ratio).
Finally, we use the backdoored training dataset to update $\tilde{\mathbf{X}}$ (between line 17 and line 18 in~\autoref{algorithm: doorping}).

\begin{figure}[!t]
\centering
\includegraphics[width=1\columnwidth]{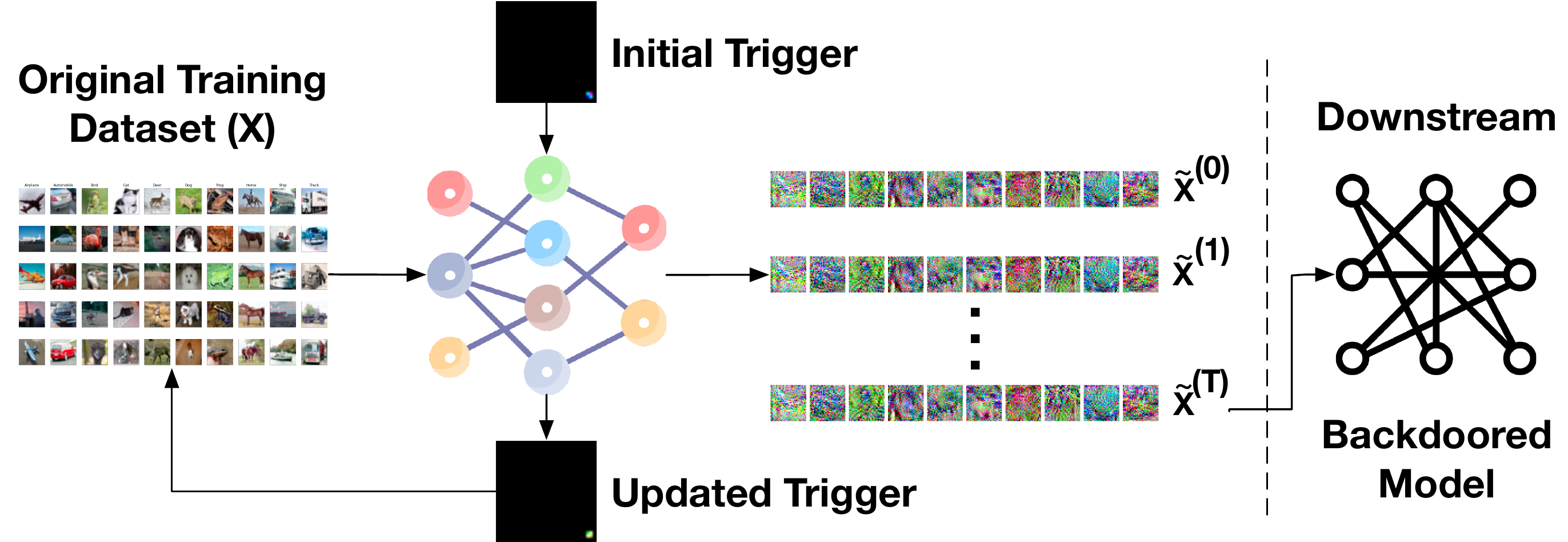}
\caption{Trigger updating on \doorping.}
\label{figure:doorping_attack}
\end{figure}

\begin{figure}[!t]
\centering
\begin{subfigure}[t]{0.3\columnwidth}
\includegraphics[width=\linewidth]{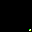}
\caption{\scriptsize \DD trigger image}
\label{figure:dd_trigger}
\end{subfigure}
\hfill
\begin{subfigure}[t]{0.3\columnwidth}
\includegraphics[width=\linewidth]{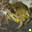}
\caption{\scriptsize \DD training image}
\label{figure:dd_image_with_trigger}
\end{subfigure}
\hfill
\begin{subfigure}[t]{0.3\columnwidth}
\includegraphics[width=\linewidth]{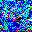} 
\caption{\scriptsize \DD distilled image}
\label{figure:dd_distilled_trigger_image}
\end{subfigure}

\vspace{0.3cm}
\begin{subfigure}[t]{0.3\columnwidth}
\includegraphics[width=\linewidth]{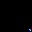}
\caption{\scriptsize \DC trigger image}
\label{figure:dc_trigger}
\end{subfigure}
\hfill
\begin{subfigure}[t]{0.3\columnwidth}
\includegraphics[width=\linewidth]{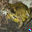}
\caption{\scriptsize \DC training image}
\label{figure:dc_image_with_trigger}
\end{subfigure}
\hfill
\begin{subfigure}[t]{0.3\columnwidth}
\includegraphics[width=\linewidth]{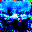} 
\caption{\scriptsize \DC distilled image}
\label{figure:dc_distilled_trigger_image}
\end{subfigure}
\caption{Illustration of the optimized trigger by \doorping attack and samples of distilled images by \DD and \DC models.
We use the \emph{airplane} class from CIFAR10 as our target backdoor class when employing \doorping.}
\label{figure:example_doorping_attack}
\end{figure}

\mypara{Trigger Insertion}
We illustrate the overall workflow of \doorping attack in~\autoref{figure:doorping_attack} and outline \doorping attack in~\autoref{algorithm: doorping}.
Our goal is to optimize the trigger $\mathbf{t}$ so that it can be better preserved during the distillation process.
The rationale is that the better trigger information can be preserved in the distillation dataset, the higher probability a backdoor attack can be successfully launched at the downstream model.
To this end, we first randomly initialize a trigger $\mathbf{t}$ and put it into the model to get the output (line 8,~\autoref{algorithm: doorping}), 
\[
\mathsf{out} = \mathsf{f}(\mathbf{t})
\]
where $\mathsf{f} = \theta_{1:\mathsf{layer}}$, and $\mathsf{layer}$ denotes the second to the last layer of $\theta$ (i.e., the layer before the softmax layer) in our study.
We then re-organize the values of $\mathsf{out}$ in descending order based on the sum of the weights of the associated parameters.
We choose the top-$k$ values from $\mathsf{out}$.
The rationale here is to identify top-$k$ neurons that cause the distillation model to misbehave.
Finally, we calculate the mean squared error (MSE) loss between the output and the output multiplied by a magnification factor $\alpha$, and then the trigger image using~\autoref{eq:trigger_loss} (line 9,~\autoref{algorithm: doorping}).
Note that we use $\alpha$ to magnify the output by these top-$k$ neurons purposely.
In our main experiments, we empirically set $k$ to 1 (see~\autoref{section:top_k_selection}) and $\alpha$ to 10 (see~\autoref{section:alpha}).

\begin{equation}
    \mathcal{L}_\mathbf{t} = \mathsf{MSE}(\mathsf{out}, \alpha\cdot\mathsf{out}) 
    \label{eq:trigger_loss}
\end{equation}

In summary, the above process enables the trigger $\mathbf{t}$ to learn from the top-$k$ neurons that cause the distillation model to misbehave.
Once we obtain this optimized trigger $\mathbf{t}$ (line 14,~\autoref{algorithm: doorping}), we use it to randomly poison $\epsilon |\mathbf{X}|$ samples in $\mathbf{X}$ (line 16,~\autoref{algorithm: doorping}). 
Then we use this backdoored dataset $\hat{\mathbf{X}}$ to update the distilled dataset $\tilde{\mathbf{X}}$ (line 18,~\autoref{algorithm: doorping}).

\mypara{Analysis}
\doorping trigger insertion can be mathematically summarized by~\autoref{eq:dd_doorping} and~\autoref{eq:dd_doorping_update_trigger}. 
Note that~\autoref{eq:dd_doorping} distills a set of prospective distilled data and 
\autoref{eq:dd_doorping_update_trigger} can be treated as \doorping trigger insertion function $\mathcal{A}_{\doorping}$ and insert an optimized trigger into the aforementioned prospective distilled data.

\begin{equation}
\begin{split}
    \tilde{\mathbf{X}}^* = \mathcal{L}_{\theta} (\hat{\mathbf{X}}, \tilde{\mathbf{X}}) \\
    \st ~ \theta^* = ~\argmin_{\theta} ~ \ell(\tilde{\mathbf{X}}, \theta) \label{eq:dd_doorping}
\end{split}
\end{equation}

\noindent where $\mathcal{L}_{\theta}$ denotes the model-specific distillation loss and $\hat{\mathbf{X}}$ is the original training dataset with backdoor samples, which is defined in~\autoref{eq:dd_doorping_update_trigger}.

\begin{equation}
\begin{split}
    \hat{\mathbf{X}} = \epsilon \cdot \mathbf{X} \cdot (1 - \mathbf{m}) + \mathbf{t} \cdot \mathbf{m}  \\
    \st ~ \mathbf{t}^* = \mathbf{t} - \mathbf{m}\cdot\eta_\mathbf{t} \nabla_{\mathbf{t}} \mathcal{L}_\mathbf{t}(\mathbf{t}, \theta) \label{eq:dd_doorping_update_trigger}
\end{split}
\end{equation}

\noindent where $\mathcal{L}_\mathbf{t}(\cdot, \cdot)$ is defined in~\autoref{eq:trigger_loss}.
It is straightforward to observe that the \doorping attack is also universally applicable to the different dataset distillation models.
\autoref{figure:example_doorping_attack} illustrates the different optimized triggers and distilled images for \DD and \DC models.

\mypara{Note}
Our method is different from~\cite{LMALZWZ18}. 
\doorping continuously optimizes the trigger $\mathbf{t}$ in every iteration to ensure the trigger is preserved in the synthetic dataset \distilleddataset.
As we show in the experiments (see~\autoref{section:evaluation}), directly applying the technique from~\cite{LMALZWZ18} (i.e., using a one-time updated trigger) leads to a sub-optimal performance, i.e., the {\em ASR} plunges after several epochs.
\doorping enables us to optimize the trigger $\mathbf{t}$ to maximize its effectiveness in the distilled dataset \distilleddataset.
This is particularly important since \doorping does not interfere with the model-specific dataset update process (line 18 in~\autoref{algorithm: doorping}).
For instance, as we can see in~\autoref{figure:example_doorping_attack}, different distillation models lead to different optimized triggers and considerably different distilled images given the same target class (i.e., airplane). 
It is also important to note that \doorping allows the attacker to keep a trigger trajectory (i.e., a collection of triggers) during the distillation process (line 14,~\autoref{algorithm: doorping}).
This unique capability enables the attackers to outmaneuver input-level defense mechanisms, as we later show in~\autoref{section:input_level_defense}.

\section{Experimental Settings}
\label{section:experiments}

\mypara{Datasets}
We use four widely used benchmark datasets in our study.
\begin{itemize}
\item \textbf{Fashion-MNIST (FMNIST)~\cite{XRV17}} is an image dataset containing 70,000, 28$\times$28, gray-scale images. Each class contains 7,000 images. 
The classes include T-shirt, trouser, pullover, dress, coat, sandal, shirt, sneaker, bag, and ankle boot.
\item \textbf{CIFAR10~\cite{CIFAR}} consists of 60,000, 32$\times$32 color images in 10 classes, with 6,000 images per class. There are 50,000 training images and 10,000 test images.
The classes are airplane, automobile, bird, cat, deer, dog, frog, horse, ship, and truck.
\item \textbf{STL10~\cite{CNL11}} is a 10-class image dataset similar to CIFAR10. 
Each class contains 1,300 images.
The size of each sample is 96$\times$96.
The classes include airplane, bird, car, cat, deer, dog, horse, monkey, ship, and truck.
\item \textbf{SVHN~\cite{NWCBWN11}} is a digit classification benchmark dataset that contains the images of printed digits (from 0 to 9) cropped from pictures of house number plates.
The size of each sample is 32$\times$32.
Among the dataset, 73,257 digits are for training, while 26,032 digits are for testing.
\end{itemize}

\noindent All the samples in the datasets are re-sized to 32$\times$32 pixels.
This is a common practice to ensure that the comparison among different datasets is fair~\cite{LWHSZBCFZ22}.

\mypara{Dataset Distillation Models}
\label{section:targetmodel}
In this paper, we utilize two different model architectures for dataset distillation - AlexNet~\cite{KSH12} and 128-width ConvNet.
These two models have been widely used in the domain of dataset distillation~\cite{WZTE18,ZMB21,ZB21,NNXL21,NCL21,ZB212,CWTEZ22}.
For ConvNet, it contains five different layers.
The first three are the convolutional layers with ReLU activation, and the last two layers are the fully connected layers.
For the distillation process, we first randomly initialize 10 different images for each class, with 100 images in total as our default settings for both \DD and \DC algorithms, which is the same distilled images per class as the original works~\cite{WZTE18,ZMB21}.
Then we use these images to train the models.

\mypara{Hyperparameters of Dataset Distillation}
We reuse the default settings from the respective distillation methods as outlined in~\cite{WZTE18,ZMB21}.
In particular, 400 epochs are used in \DD where Adam is used as the optimizer.
The batch size for the original training dataset is 1,024.
We run \DC for 1,000 epochs and employ stochastic gradient descent (SGD) as the optimizer.
Note that \DC has an additional SGD optimizer for updating the images.

\mypara{Backdoor Attack Settings}
We outline our backdoor attack settings below.

\begin{itemize}
\item \textbf{\naive.}
As we mentioned in~\autoref{section:baseline}, we add backdoor triggers before distillation.
The trigger is a $2 \times 2$ white patch (i.e., 4 pixels in total).
We insert the trigger in the bottom right corner of an image.
\item \textbf{\doorping.}
We first randomly initialize a $2 \times 2$ trigger and insert it in the bottom right corner of an image.
When optimizing the triggers (see~\autoref{algorithm: doorping}), we use Adam as the optimizer and MSE as the loss function.
We train the trigger up to 10,000 epochs if the MSE loss is not less than the threshold.
Empirically, we set this threshold to 0.5 as this value is small enough for the loss function, and the corresponding trigger is also good enough for our attacks.
More concretely, if the MSE loss is less than this threshold, it indicates a less effective trigger can be learned from the selected neurons.
Thus, the algorithm makes an early stop to accelerate the learning process.
Note that the threshold is not related to any datasets or models since it is only used to reduce the trigger optimizing process.
We set the poisoning ratio $\epsilon$ to commonly used value 0.01 by default~\cite{c21,ZHLTSG192,NBCJRSSTX08,BNL12,SHNSSDG18}.
\end{itemize}

\mypara{Evaluation Metrics}
In this paper, we adopt {\em attack success rate (ASR)} and {\em clean test accuracy (CTA)} as our evaluation metrics.
\begin{itemize}
\item The {\em ASR} measures the attack effectiveness of the backdoored model on a triggered testing dataset.
\item The {\em CTA} assesses the utility of the backdoored model on the clean testing dataset.
\end{itemize}
Both {\em ASR} and {\em CTA} scores are normalized between 0.0 and 1.0.
The higher the {\em ASR} score is, the better the backdoor trigger injected.
The closer the {\em CTA} score of the backdoored model to the one of a clean model, i.e., a model trained using clean data only, the better the backdoored model’s utility.

\mypara{Downstream Models}
Note that dataset distillation tailors the distilled dataset for a given architecture.
Due to this limitation of dataset distillation, all of the downstream models should be the same architecture as the dataset distillation models.
In our evaluation, the downstream models are also AlexNet and 128-width ConvNet and share the same architectural design as the dataset distillation models (see~\autoref{section:targetmodel}). 

\mypara{Runtime Configuration}
Unless otherwise mentioned, we consider the following parameter settings for both \naive and \doorping by default: $2 \times 2$ trigger size, 0.01 of the poisoning ratio, and 10 images of each class in distilled dataset.
All the experiments in this paper are repeated 10 times.
For each run, we follow the same experimental setup laid out before. 
We report the mean and standard deviation of each metric to evaluate the attack performance.

\mypara{Remark}
We outline additional hyper-parameters and experimental settings in~\autoref{section:hypara}.

\section{Evaluation}
\label{section:evaluation}

\begin{figure*}[!t]
\centering
\includegraphics[width=1.9\columnwidth]{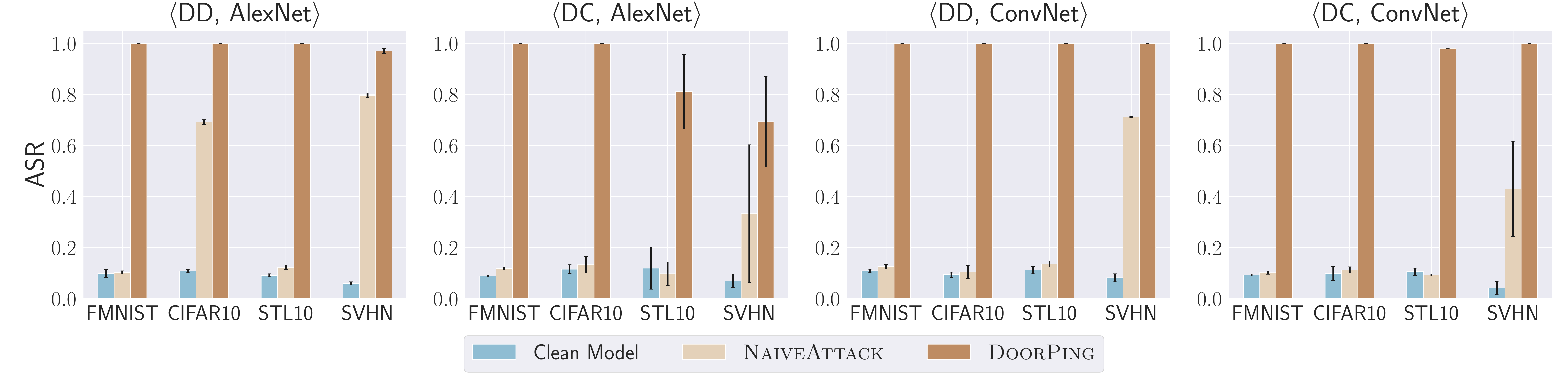}
\caption{{\em ASR} of clean model, \naive and \doorping for different distillation algorithms and different model architectures.}
\label{figure:basic_asr}
\end{figure*}

\begin{figure*}[!t]
\centering
\includegraphics[width=1.9\columnwidth]{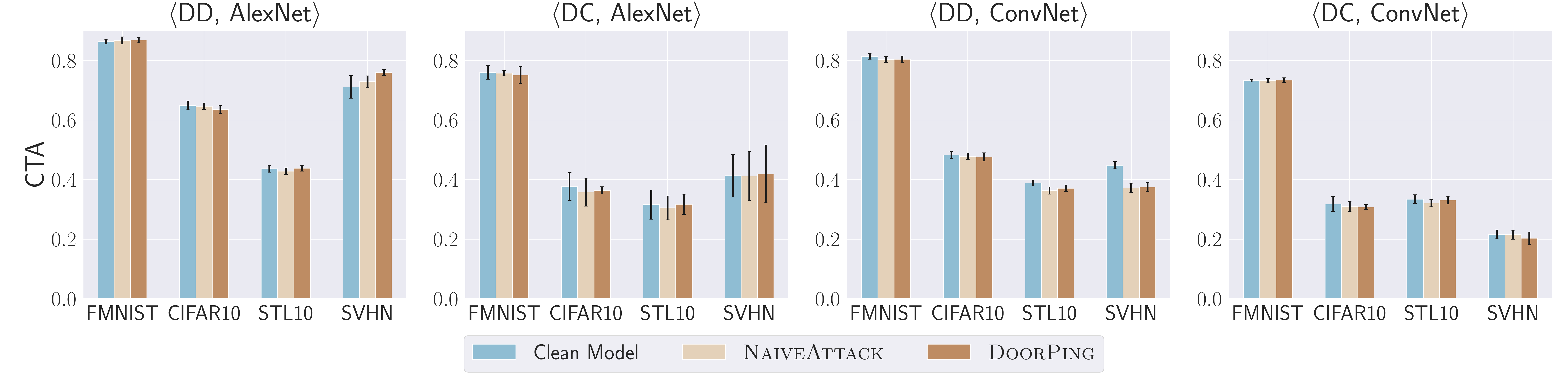}
\caption{{\em CTA} score of clean model, \naive and \doorping for different distillation algorithms and different model architectures.}
\label{figure:basic_cta}
\end{figure*}

In this section, we present the performance of \naive and \doorping against dataset distillation.
We conduct extensive experiments to answer the following research questions (RQs):
\begin{itemize}
\item {\em RQ1:} Do both \naive and \doorping achieve high attack performance?
\item {\em RQ2:} Do both \naive and \doorping preserve the model utility?
\end{itemize}
Concretely, we first evaluate the attack performance ({\em ASR} score) of \naive and \doorping on all tasks, model architectures, and distillation methods.
We then evaluate the utility performance ({\em CTA} score) of the backdoored model attacked by \naive and \doorping.
We use a tuple in the format of $\langle$Distillation Algorithm, Architecture, Dataset$\rangle$ for ease of presentation.
For instance, $\langle \DD$, AlexNet, CIFAR10$\rangle$ refers to an experiment that is carried out using the \DD distillation algorithm with Alexnet architecture to distill the CIFAR10 dataset.

\subsection{Attack Performance}

We first show the attack performance of both \naive and \doorping to answer {\em RQ1}.
To measure the attack performance of our attacks, we conduct a comparative evaluation of the {\em ASR} score between the backdoored model and the clean model trained by the normal dataset distillation procedure.
We expect the backdoored model misclassifies the input containing a specific trigger, while the clean model behaves normally.
\autoref{figure:basic_asr} reports the {\em ASR} score of \naive and \doorping on all datasets, model architectures, and dataset distillation methods.

\mypara{\naive}
As shown in~\autoref{figure:basic_asr}, we can clearly observe that the attack on the clean model achieves low {\em ASR} scores ranging between 0.042 and 0.120.
In contrast, in some cases, our \naive achieves higher {\em ASR} scores than the attack on the clean model.
For instance, the {\em ASR} score of $\langle \DD$, AlexNet, CIFAR10$\rangle$ is 0.692, and the {\em ASR} score of $\langle \DD$, ConvNet, SVHN$\rangle$ is 0.712.
These results show that our \naive generally performs well but fails in some cases, implying that fixed triggers simply added to the distilled data cannot be closely connected to the hidden behavior.

\mypara{\doorping}
As shown in~\autoref{figure:basic_asr}, almost all of the {\em ASR} scores are over 0.950 except for AlexNet trained by \DC distilling STL10 and SVHN.
For example, the {\em ASR} score of $\langle \DD$, AlexNet, CIFAR10$\rangle$ is 1.000.
Note that the lowest {\em ASR} score of $\langle \DC$, AlexNet, STL10$\rangle$ and $\langle \DC$, AlexNet, SVHN$\rangle$ are 0.811 and 0.693, respectively. 
These scores are also much higher than our \naive and the attack on the clean model.
On the other hand, the standard deviation of these two {\em ASR} scores is higher than the others.
These results indicate that the {\em ASR} scores are spread out.
More epochs may be required to optimize the triggers in these cases.
In general, the results demonstrate that iteratively optimizing triggers throughout the distillation process can establish a strong connection between the triggers and the hidden behavior injected into the backdoored model.

\mypara{Takeaways}
Our attack methods can successfully inject the predefined triggers into the model.
\naive generally performs well, though it fails in some settings.
In contrast, \doorping achieves remarkable performance among all the datasets, downstream model architectures, and dataset distillation methods.

\subsection{Distillation Model Utility}

Here, we focus on the utility performance of the backdoored model, i.e., measuring whether our attack leads to significant side effects on the primary task, to answer {\em RQ2}.
Ideally, a backdoored model should be as accurate as a clean model, given clean test data to ensure its stealthiness.
In this study, we evaluate the model utility from both quantitative and qualitative perspectives.

We first conduct a quantitative evaluation of the {\em CTA} score between the clean and backdoored models.
As shown in~\autoref{figure:basic_cta}, we find that the {\em CTA} scores of the backdoored models from \naive or \doorping are similar to that of the clean models.
For instance, the {\em CTA} score of clean model for $\langle \DD$, AlexNet, CIFAR10$\rangle$ is 0.649.
Meanwhile, the {\em CTA} scores for \naive and \doorping are 0.646 and 0.635, respectively, which drop only by 0.464\% and 2.1\% compared to the clean model.
Our results exemplify that the side effects caused by our backdoor attacks are within the acceptable performance variation of the model.
They have no significant impact on utility performance.
A similar observation can be drawn from other {\em CTA} scores.
Besides, for $\langle \DC$, AlexNet, SVHN$\rangle$, the clean model has the lowest {\em CTA} score, which is respectively 2.531\% and 6.751\% lower than \naive and \doorping.
As such, we carry out a Welsh $t$-test on the results (as we repeat the evaluation 10 times per our runtime configuration). 
Our null hypothesis is that the mean {\em CTA} score of \doorping and the clean model is the same. 
The $t$-test results show that Welch-Satterthwaite Degrees of Freedom is 17.986 and $\rho$-value is 0.894; hence we cannot reject our null hypothesis.
We conclude that such a difference is due to fluctuation.

\begin{figure}[!t]
\centering
\begin{subfigure}[t]{1\columnwidth}
\includegraphics[width=\linewidth]{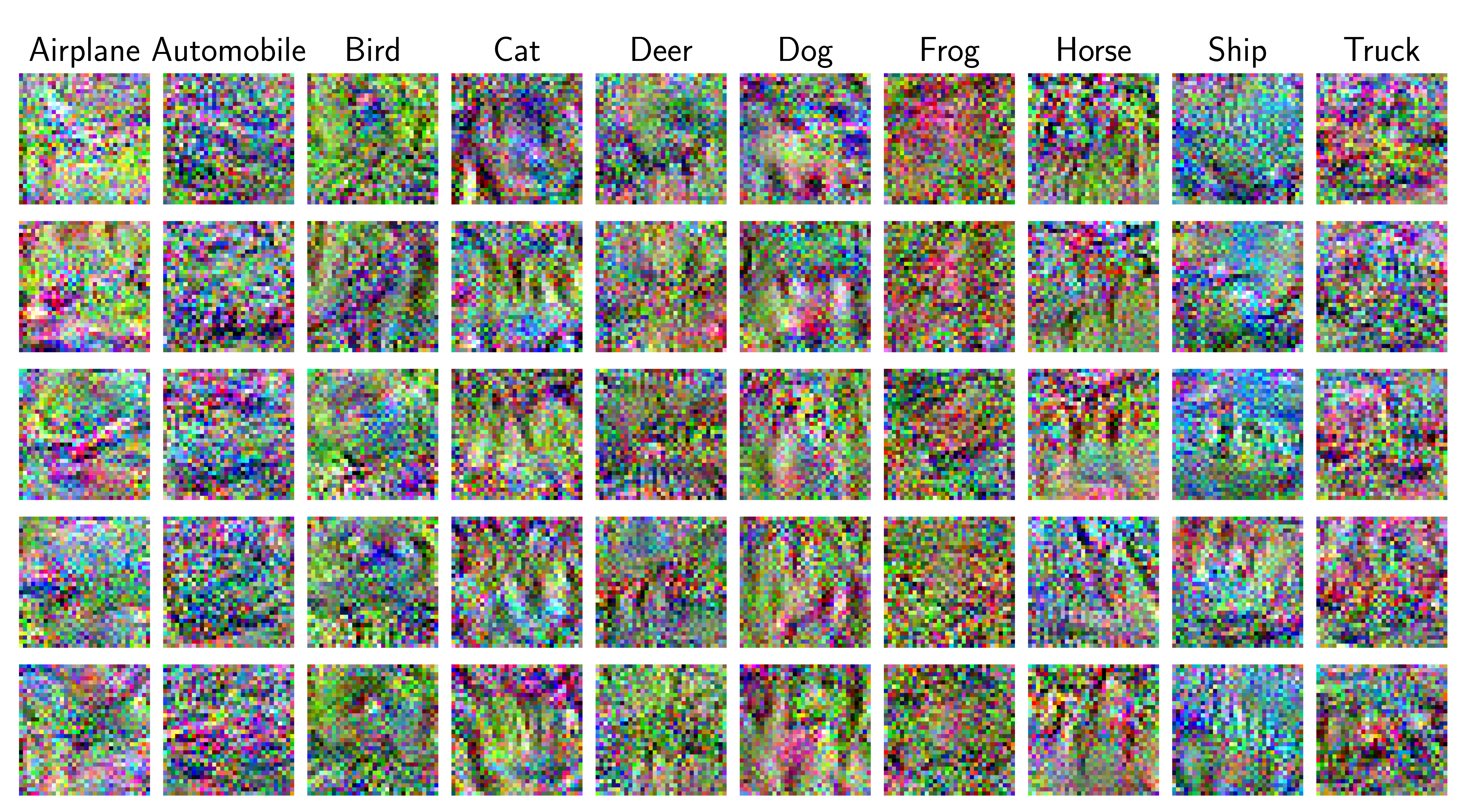}
\caption{\scriptsize Distilled clean images by \DD algorithm}
\label{figure:distilledimagesori}
\end{subfigure}\par\medskip
\begin{subfigure}[t]{1\columnwidth}
\includegraphics[width=\linewidth]{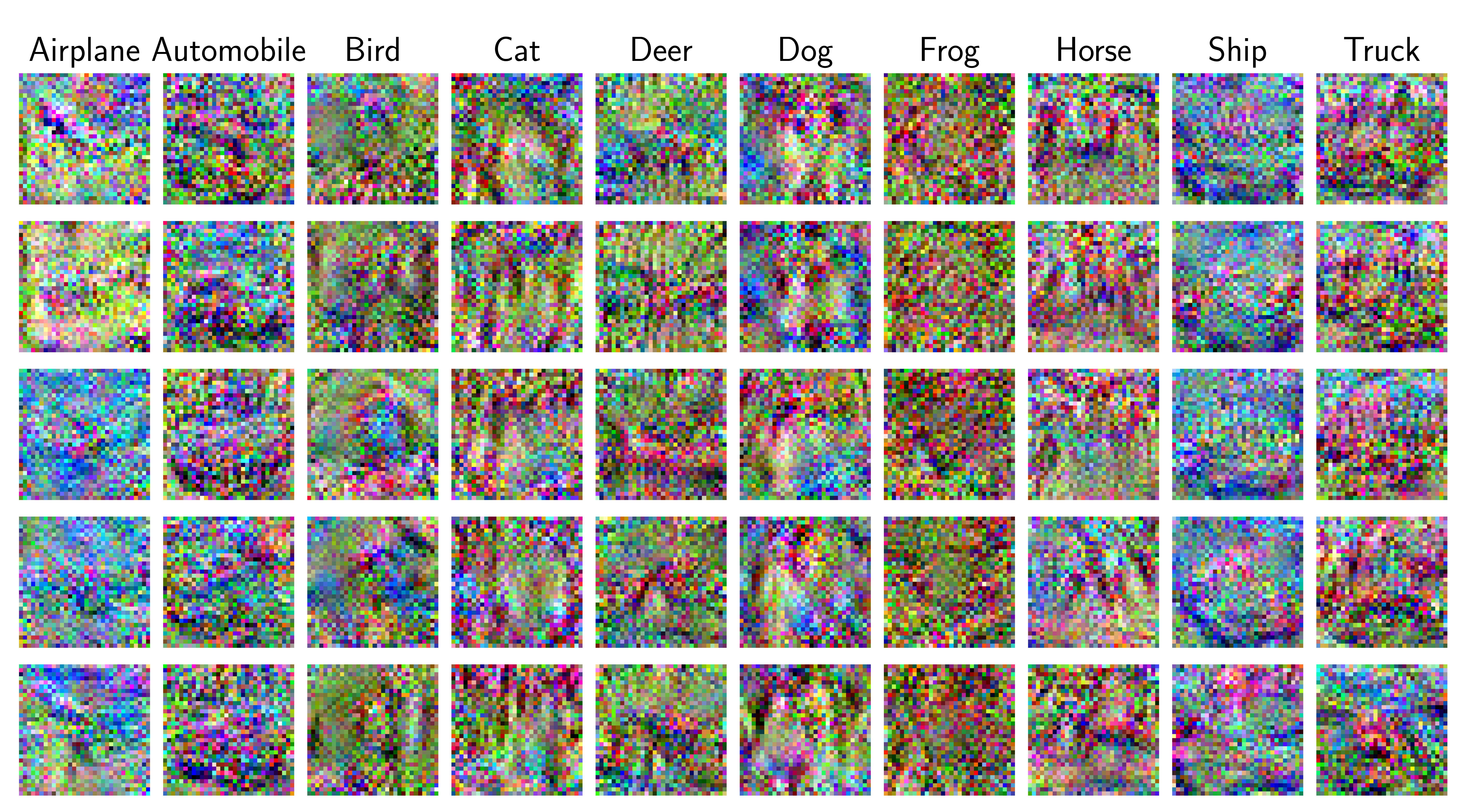}
\caption{\scriptsize Distilled images by \doorping and \DD algorithm}
\label{figure:distilledimagesback}
\end{subfigure}
\caption{Comparison of distilled images by \doorping given $\langle \DD$, AlexNet, CIFAR10$\rangle$.}
\label{figure:distilledimages}
\end{figure} 

\begin{figure}[!t]
\centering
\begin{subfigure}[t]{1\columnwidth}
\includegraphics[width=\linewidth]{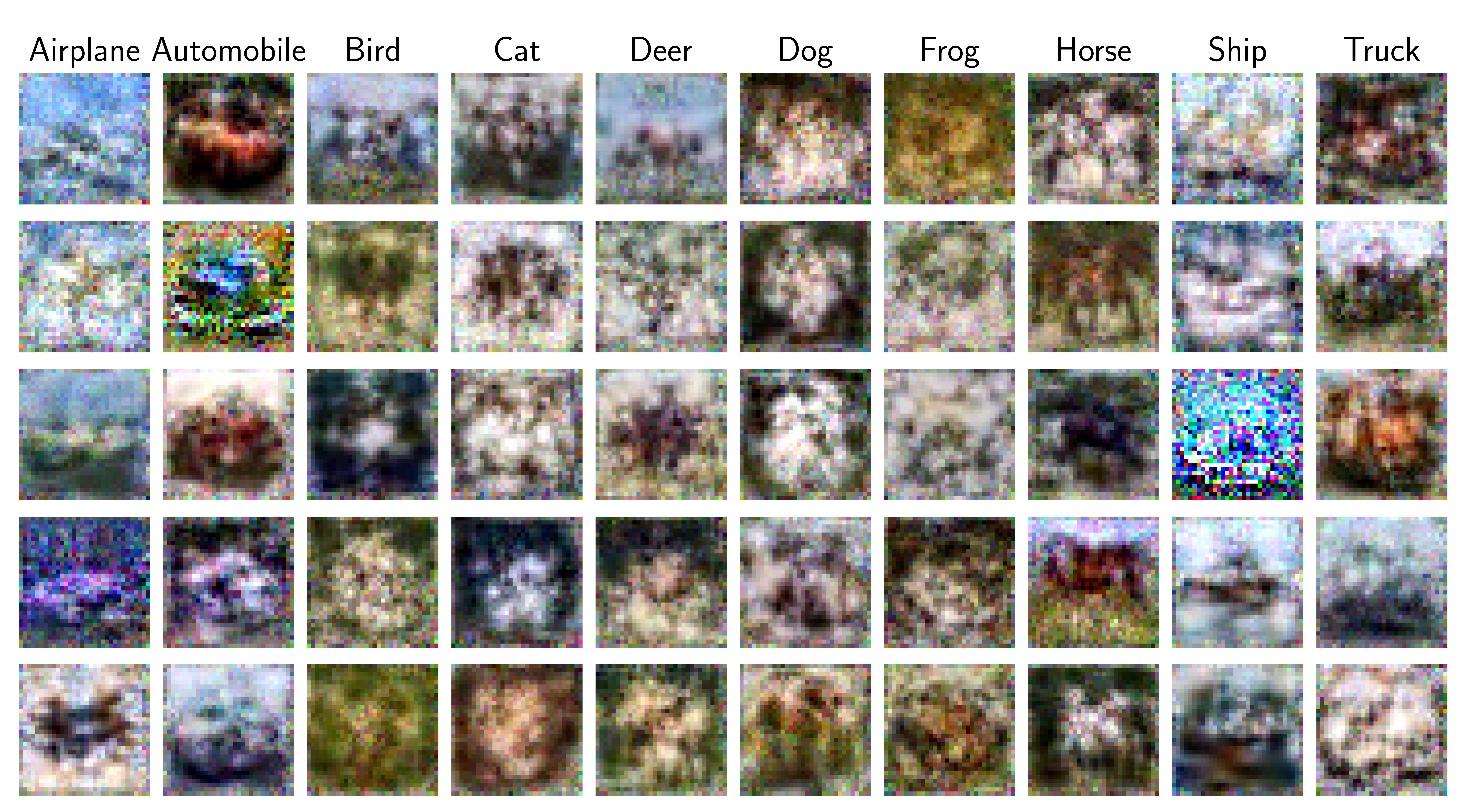}
\caption{\scriptsize Distilled clean images by \DC algorithm}
\label{figure:dcdistilledimagesori}
\end{subfigure}\par\medskip
\begin{subfigure}[t]{1\columnwidth}
\includegraphics[width=\linewidth]{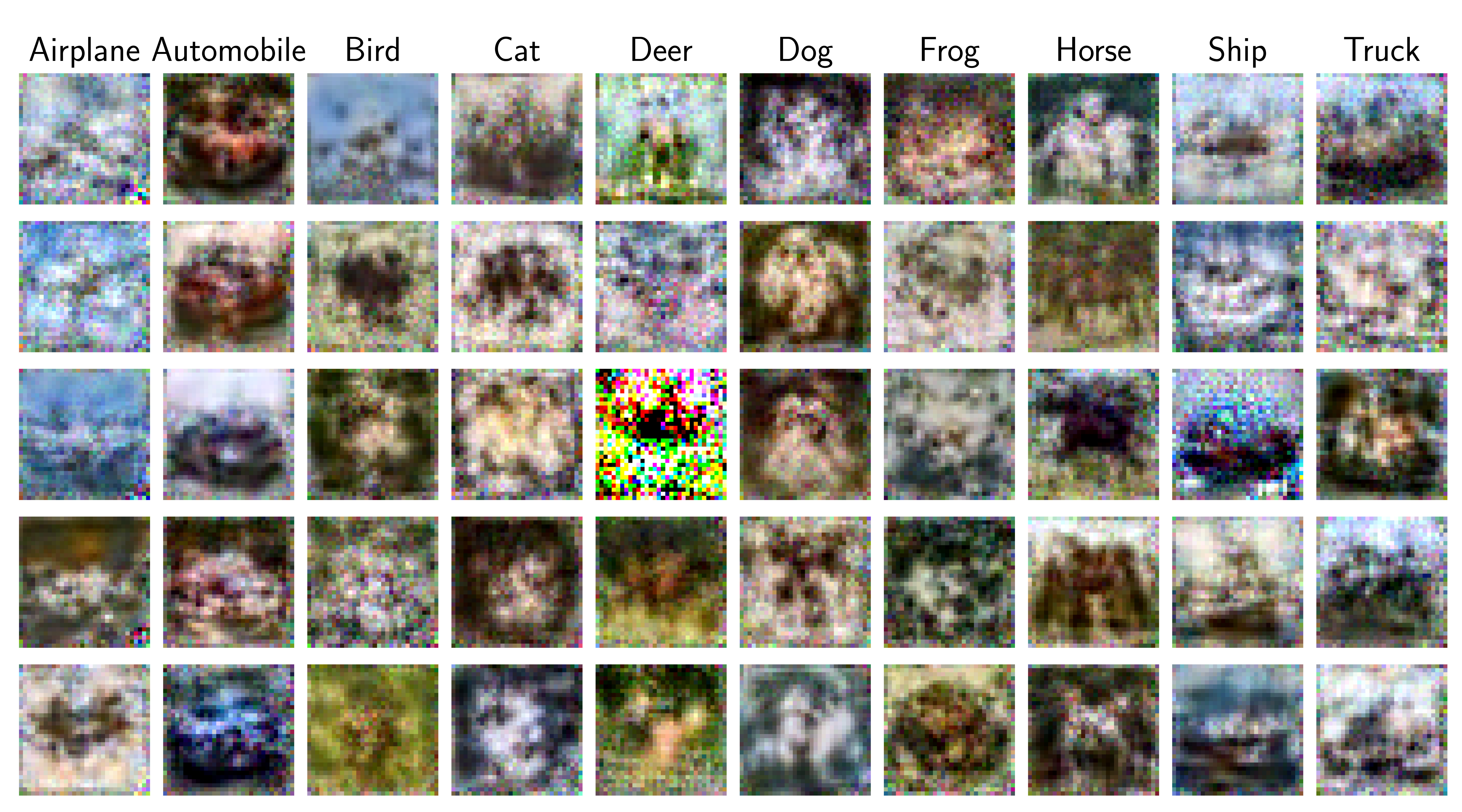}
\caption{\scriptsize Distilled images by \doorping and \DC algorithm}
\label{figure:dcdistilledimagestro}
\end{subfigure}
\caption{Comparison of distilled images by \doorping given $\langle \DC$, AlexNet, CIFAR10$\rangle$.}
\label{figure:dcdistilledimages}
\end{figure}

We then conduct a qualitative evaluation by visualizing some examples of distilled images by normal dataset distillation and our attack in~\autoref{figure:distilledimages} and~\autoref{figure:dcdistilledimages}.
The images distilled by \doorping are much similar to the ones in~\autoref{figure:distilledimagesori}
More propitiously, the backdoor trigger is totally unrecognizable to human inspection, meaning that the trigger is completely hidden in the synthetic image.

\mypara{Takeaways}
Our experimental results demonstrate that all the backdoored models still have the same level of utility performance as the clean model, i.e., our proposed backdoor attacks preserve the model’s utility.

\section{Ablation Study}
\label{section:ablationstudy}

\subsection{Additional Experimental Settings}
\label{section:hypara}

In~\autoref{table:par}, we list additional experimental settings which commonly used in our experiments.
There are some other hyper-parameters, such as distillation mode in \DD, which are not listed because we only use the default settings and never change them during the experiment.

\subsection{Effectiveness on Complex Datasets}

We also add the ablation study on the effectiveness of complex datasets.
We test our attack using CIFAR100~\cite{CIFAR}, which has 100 classes containing 600 images each.
For \DD, due to the GPU memory limitation, we distill five images per class and 500 synthetic images in total.
For \DC, the hyper-parameters and other settings remain the same as our main experiments.
\autoref{table:cifar100} illustrates the results of CIFAR100.
All {\em ASR} scores are larger than 0.900 without significant {\em CTA} degradation compared with those of the clean model and \naive. 
Our results show that \doorping can be easily extended to more complex datasets.

\mypara{Takeaway}
\doorping can be extended to more complex datasets with more classes and data samples.

\begin{table}[!t]
\centering
\customTableFont
\setlength{\tabcolsep}{2 pt}
\caption{{\em ASR} and {\em CTA} of \doorping with CIFAR100.}
\scalebox{0.95}{
\begin{tabular}{c | c | c c | c c}
\toprule
& & \multicolumn{2}{c |}{\DD} & \multicolumn{2}{c }{\DC}\\
& & {\em ASR} & {\em CTA}& {\em ASR} & {\em CTA}\\
\midrule
\multirow{2}{*}{Clean Model} & AlexNet & 0.014 $\!\! \pm\!\!$ 0.010 & 0.385 $\!\! \pm\!\!$ 0.009 & 0.012 $\!\! \pm\!\!$ 0.002 & 0.217 $\!\! \pm\!\!$ 0.007\\
& ConvNet & 0.029 $\!\! \pm\!\!$ 0.011 & 0.215 $\!\! \pm\!\!$ 0.014 & 0.012 $\!\! \pm\!\!$ 0.002 & 0.197 $\!\! \pm\!\!$ 0.007\\
\midrule
\multirow{2}{*}{\naive} & AlexNet & 0.128 $\!\! \pm\!\!$ 0.013 & 0.375 $\!\! \pm\!\!$ 0.010 & 0.007  $\!\! \pm\!\!$ 0.003 & 0.209  $\!\! \pm\!\!$ 0.005\\
& ConvNet & 0.011 $\!\! \pm\!\!$ 0.010 & 0.214 $\!\! \pm\!\!$ 0.011 & 0.006 $\!\! \pm\!\!$ 0.021 & 0.190 $\!\! \pm\!\!$ 0.004\\
\midrule
\multirow{2}{*}{\doorping} & AlexNet & 0.919 $\!\! \pm\!\!$ 0.014 & 0.373 $\!\! \pm\!\!$ 0.011 & 0.961 $\!\! \pm\!\!$ 0.024 & 0.209 $\!\! \pm\!\!$ 0.006\\
& ConvNet & 1.000 $\!\! \pm\!\!$ 0.000 & 0.205 $\!\! \pm\!\!$ 0.012 & 1.000 $\!\! \pm\!\!$ 0.000 & 0.196 $\!\! \pm\!\!$ 0.002\\
\bottomrule
\end{tabular}
}
\label{table:cifar100}
\end{table}

\subsection{Effectiveness on Cross Architectures}

Several dataset distillation methods~\cite{ZMB21,CWTEZ22} explore cross-architecture (CA) data distillation (i.e., the data distillation model is different from the downstream model).
To understand the effectiveness of \doorping on such cross-architecture scenarios, we choose three model architectures - AlexNet~\cite{KSH12}, ConvNet, and VGG11~\cite{SZ15} in our study.
We use AlexNet (ConvNet) as the distillation model and the other two architectures for evaluation.
As we can see in~\autoref{table:cross}, given \DC algorithm, \doorping achieves good {\em ASR} and {\em CTA} scores on VGG11 as the downstream model, which is trained on the synthetic data distilled by ConvNet.
In general, \doorping performs well on all cross-architecture models using the synthetic data distilled by ConvNet architecture.
However, \doorping does not perform well in most cross-architecture models using the synthetic data distilled by \DD algorithm.
We speculate that the root cause is the difference in the distillation algorithms.
For \DD, it compresses the image information (gradient calculated by the specific model) into the distilled dataset, i.e., model-specific. 
In contrast, \DC forces the synthetic dataset to learn the distribution of the original dataset, i.e., model-independent.
Therefore, \DC can better preserve the information of the original training images hence better preserving the trigger in the distilled dataset.
Consequently, \DC leaves the backdoor in a different model trained on this distilled dataset.

\begin{table}[!t]
\centering
\customTableFont
\setlength{\tabcolsep}{3pt}
\caption{Additional hyper-parameters used in backdoor attacks against \DD and \DC.}
\scalebox{0.95}{
\renewcommand{\arraystretch}{1.5}
\begin{tabular}{l | c | c }
\toprule
& \DD & \DC\\
\midrule
Batch size & 1024 & 256\\
Epochs & 400 & 1000\\
Distilled optimizer & SGD & SGD \\
Distilled loss function & Cross Entropy & Cross Entropy \\
Distilled images per class & 10 & 10\\
Distilled learning rate & 0.001 & 0.1 \\
Downstream training epochs & 30  & 300\\
Downstream model learning rate & 0.01 & 0.01\\
Downstream model optimizer & SGD & SGD\\
Downstream model loss function & Cross Entropy & Cross Entropy\\
Trigger learning rate & 0.08 & 0.08\\
Trigger optimizer & Adam & Adam\\
Trigger loss function & MSE Loss & MSE Loss\\
Top-$k$ & 1 & 1 \\
Poisoning ratio & 0.01 & 0.01\\
Alpha & 10 & 10\\
Threshold of trigger updating & 0.5 & 0.5\\
\bottomrule
\end{tabular}
}
\label{table:par}
\end{table}

\mypara{Takeaway}
\doorping can be used to attack cross-architecture models.
However, its effectiveness may be affected by the distillation models. 

\begin{table*}[!t]
\normalsize
\centering
\setlength{\tabcolsep}{3.2pt}
\caption{{\em ASR} and {\em CTA} of cross model architectures.}
\scalebox{0.49}{
\renewcommand{\arraystretch}{1.5}
\begin{tabular}{c c !{\vrule width 1pt} c c | c c !{\vrule width 1pt} c c | c c !{\vrule width 1pt} c c | c c !{\vrule width 1pt} c c | c c}
\toprule
&  & \multicolumn{4}{c!{\vrule width 1pt}}{\bf FMNIST}  & \multicolumn{4}{c!{\vrule width 1pt}}{\bf CIFAR10}  & \multicolumn{4}{c!{\vrule width 1pt}}{\bf STL10}  & \multicolumn{4}{c}{\bf SVHN}  \\
&  & \multicolumn{2}{c|}{AlexNet} & \multicolumn{2}{c!{\vrule width 1pt}}{ConvNet} & \multicolumn{2}{c|}{AlexNet} & \multicolumn{2}{c!{\vrule width 1pt}}{ConvNet} & \multicolumn{2}{c|}{AlexNet} & \multicolumn{2}{c!{\vrule width 1pt}}{ConvNet} & \multicolumn{2}{c|}{AlexNet} & \multicolumn{2}{c}{ConvNet} \\
& CA Model & {\em ASR} & {\em CTA} & {\em ASR} & {\em CTA} & {\em ASR} & {\em CTA} & {\em ASR} & {\em CTA} & {\em ASR} & {\em CTA} & {\em ASR} & {\em CTA} & {\em ASR} & {\em CTA} & {\em ASR} & {\em CTA} \\
\midrule
\multirow{3}{*}{\DD} & AlexNet & \cellcolor{gray!30}1.000 $\!\! \pm\!\!$ 0.000 & \cellcolor{gray!30}0.868 $\!\! \pm\!\!$ 0.013 & \cellcolor{gray!30}0.500 $\!\! \pm\!\!$ 0.500 & \cellcolor{gray!30}0.300 $\!\! \pm\!\!$ 0.118 & \cellcolor{gray!30}0.999 $\!\! \pm\!\!$ 0.000 & \cellcolor{gray!30}0.635 $\!\! \pm\!\!$ 0.009 & \cellcolor{gray!30}0.000 $\!\! \pm\!\!$ 0.000 & \cellcolor{gray!30}0.102 $\!\! \pm\!\!$ 0.005 & \cellcolor{gray!30}0.999 $\!\! \pm\!\!$ 0.000 & \cellcolor{gray!30}0.438 $\!\! \pm\!\!$ 0.010 & \cellcolor{gray!30}1.000 $\!\! \pm\!\!$ 0.000 & \cellcolor{gray!30}0.120 $\!\! \pm\!\!$ 0.011 & \cellcolor{gray!30}0.970 $\!\! \pm\!\!$ 0.009 & \cellcolor{gray!30}0.759 $\!\! \pm\!\!$ 0.010 & \cellcolor{gray!30}0.000 $\!\! \pm\!\!$ 0.000 & \cellcolor{gray!30}0.091 $\!\! \pm\!\!$ 0.008 \\
& ConvNet & 0.411 $\!\! \pm\!\!$ 0.069 & 0.342 $\!\! \pm\!\!$ 0.031 & 1.000 $\!\! \pm\!\!$ 0.000 & 0.804 $\!\! \pm\!\!$ 0.014 & 0.471 $\!\! \pm\!\!$ 0.014 & 0.214 $\!\! \pm\!\!$ 0.013 & 1.000 $\!\! \pm\!\!$ 0.000 & 0.476 $\!\! \pm\!\!$ 0.011 & 0.994 $\!\! \pm\!\!$ 0.002 & 0.216 $\!\! \pm\!\!$ 0.009 & 1.000 $\!\! \pm\!\!$ 0.000 & 0.371 $\!\! \pm\!\!$ 0.011 & 0.000 $\!\! \pm\!\!$ 0.000 & 0.195 $\!\! \pm\!\!$ 0.013 & 1.000 $\!\! \pm\!\!$ 0.000 & 0.375 $\!\! \pm\!\!$ 0.015 \\
& VGG11 & \cellcolor{gray!30}0.000 $\!\! \pm\!\!$ 0.000 & \cellcolor{gray!30}0.115 $\!\! \pm\!\!$ 0.012 & \cellcolor{gray!30}0.000 $\!\! \pm\!\!$ 0.000 & \cellcolor{gray!30}0.151 $\!\! \pm\!\!$ 0.077 & \cellcolor{gray!30}0.385 $\!\! \pm\!\!$ 0.472 & \cellcolor{gray!30}0.097 $\!\! \pm\!\!$ 0.010 & \cellcolor{gray!30}0.200 $\!\! \pm\!\!$ 0.400 & \cellcolor{gray!30}0.104 $\!\! \pm\!\!$ 0.013 & \cellcolor{gray!30}0.100 $\!\! \pm\!\!$ 0.300 & \cellcolor{gray!30}0.208 $\!\! \pm\!\!$ 0.018 & \cellcolor{gray!30}0.100 $\!\! \pm\!\!$ 0.300 & \cellcolor{gray!30}0.127 $\!\! \pm\!\!$ 0.027 & \cellcolor{gray!30}0.000 $\!\! \pm\!\!$ 0.000 & \cellcolor{gray!30}0.133 $\!\! \pm\!\!$ 0.040 & \cellcolor{gray!30}0.000 $\!\! \pm\!\!$ 0.000 & \cellcolor{gray!30}0.085 $\!\! \pm\!\!$ 0.038 \\
\midrule
\multirow{3}{*}{\DC} & AlexNet & 1.000 $\!\! \pm\!\!$ 0.000 & 0.751 $\!\! \pm\!\!$ 0.012 & 1.000 $\!\! \pm\!\!$ 0.000 & 0.716 $\!\! \pm\!\!$ 0.013 & 1.000 $\!\! \pm\!\!$ 0.000 & 0.364 $\!\! \pm\!\!$ 0.029 & 1.000 $\!\! \pm\!\!$ 0.000 & 0.301 $\!\! \pm\!\!$ 0.019 & 0.811 $\!\! \pm\!\!$ 0.146 & 0.317 $\!\! \pm\!\!$ 0.034 & 1.000 $\!\! \pm\!\!$ 0.000 & 0.312 $\!\! \pm\!\!$ 0.012 & 0.693 $\!\! \pm\!\!$ 0.177 & 0.419 $\!\! \pm\!\!$ 0.097 & 1.000 $\!\! \pm\!\!$ 0.000 & 0.446 $\!\! \pm\!\!$ 0.039 \\
& ConvNet & \cellcolor{gray!30}0.235 $\!\! \pm\!\!$ 0.200 & \cellcolor{gray!30}0.681 $\!\! \pm\!\!$ 0.027 & \cellcolor{gray!30}1.000 $\!\! \pm\!\!$ 0.000 & \cellcolor{gray!30}0.734 $\!\! \pm\!\!$ 0.008 & \cellcolor{gray!30}0.108 $\!\! \pm\!\!$ 0.016 & \cellcolor{gray!30}0.287 $\!\! \pm\!\!$ 0.011 & \cellcolor{gray!30}1.000 $\!\! \pm\!\!$ 0.000 & \cellcolor{gray!30}0.308 $\!\! \pm\!\!$ 0.008 & \cellcolor{gray!30}0.017 $\!\! \pm\!\!$ 0.010 & \cellcolor{gray!30}0.308 $\!\! \pm\!\!$ 0.021 & \cellcolor{gray!30}0.981 $\!\! \pm\!\!$ 0.000 & \cellcolor{gray!30}0.331 $\!\! \pm\!\!$ 0.013 & \cellcolor{gray!30}0.321 $\!\! \pm\!\!$ 0.131 & \cellcolor{gray!30}0.180 $\!\! \pm\!\!$ 0.009 & \cellcolor{gray!30}1.000 $\!\! \pm\!\!$ 0.000 & \cellcolor{gray!30}0.203 $\!\! \pm\!\!$ 0.021 \\
& VGG11 & 0.349 $\!\! \pm\!\!$ 0.434 & 0.744 $\!\! \pm\!\!$ 0.009 & 1.000 $\!\! \pm\!\!$ 0.000 & 0.759 $\!\! \pm\!\!$ 0.005 & 1.000 $\!\! \pm\!\!$ 0.000 & 0.312 $\!\! \pm\!\!$ 0.009 & 1.000 $\!\! \pm\!\!$ 0.000 & 0.287 $\!\! \pm\!\!$ 0.005 & 1.000 $\!\! \pm\!\!$ 0.000 & 0.306 $\!\! \pm\!\!$ 0.006 & 1.000 $\!\! \pm\!\!$ 0.000 & 0.281 $\!\! \pm\!\!$ 0.008 & 1.000 $\!\! \pm\!\!$ 0.000 & 0.477 $\!\! \pm\!\!$ 0.011 & 1.000 $\!\! \pm\!\!$ 0.000 & 0.380 $\!\! \pm\!\!$ 0.014 \\
\midrule
\end{tabular}
}
\label{table:cross}
\end{table*}

\subsection{Invisible Backdoor Attack}

\begin{figure}[!t]
\centering
\includegraphics[width=0.6\columnwidth]{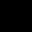}
\caption{Invisible trigger for \DD and AlexNet for CIFAR10 dataset.}
\label{figure:inv_trigger}
\end{figure}

\begin{figure*}[!t]
\centering
\includegraphics[width=1.9\columnwidth]{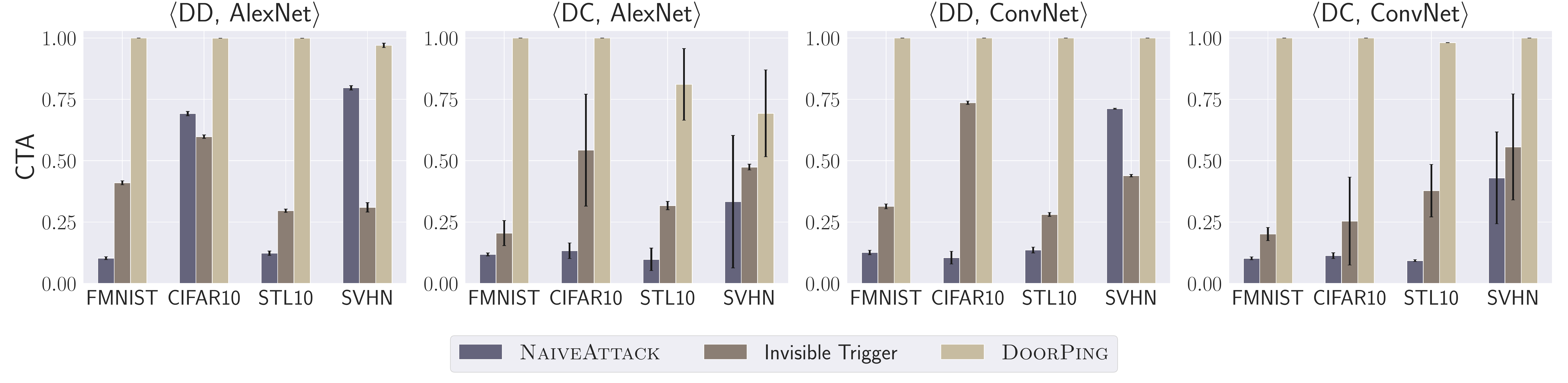}
\caption{{\em ASR} of Invisible backdoor attack against dataset distillation.}
\label{figure:inv_asr}
\end{figure*}

\begin{figure*}[!t]
\centering
\includegraphics[width=1.9\columnwidth]{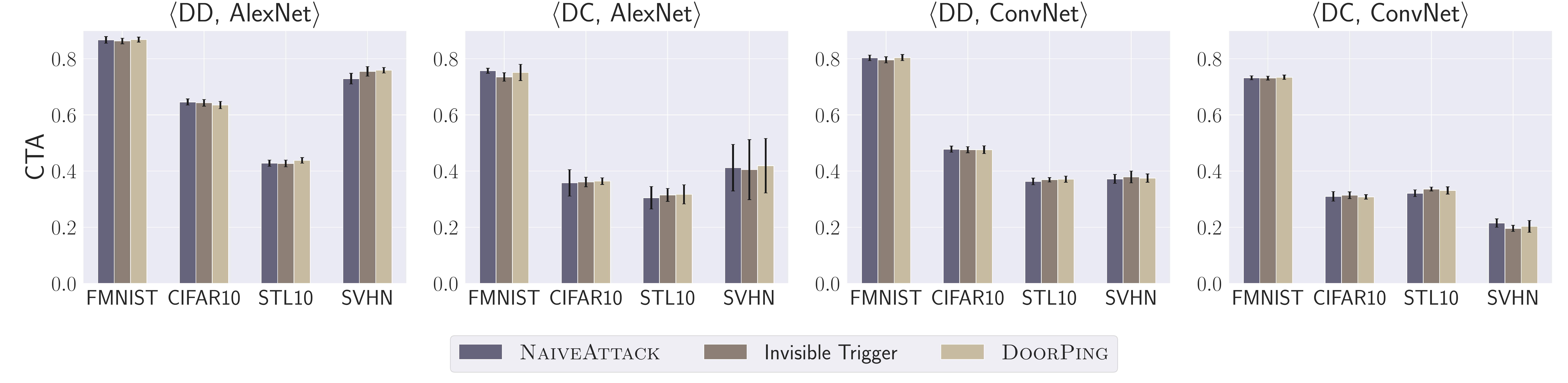}
\caption{{\em ASR} of Invisible backdoor attack against dataset distillation.}
\label{figure:inv_cta}
\end{figure*}

We also test another trigger pattern technique, invisible trigger~\cite{LXZZZ20}.
For \doorping with an invisible trigger, we choose a random image from the target class to which the adversary aims to map the backdoored images.
Specifically, in our experiments, we choose label 0 as the target class and randomly select an image from class 0 of the test dataset as the trigger.
In the original work, the trigger is optimized for about 2,000 epochs before the model re-training.
To simplify the workflow and save time, the trigger is optimized for 500 steps of each distillation epoch.
All other settings are the same as for the original \doorping.
~\autoref{figure:inv_asr} demonstrates the results of invisible backdoor attacks against dataset distillation.
As we can see, we cannot identify this trigger generated by an airplane with the naked eye, i.e., the invisible trigger.
For most cases, the invisible trigger will perform better than \naive without utility degradation from~\autoref{figure:inv_cta}.
We can see from~\autoref{figure:inv_asr} the invisible trigger cannot exceed our \doorping attacks for all the cases.
Furthermore,~\autoref{figure:inv_trigger} shows the trigger we optimized for $\langle \DD$, AlexNet, CIFAR10$\rangle$.

\mypara{Takeaways}
The invisible trigger cannot outperform the trigger patterns used by \doorping.
However, it still performs better than \naive in general.

\begin{algorithm}[t]
\caption{Invsible Algorithm}
\label{algorithm: invisible}
    \begin{algorithmic}[1]
    \REQUIRE The original training dataset $\mathbf{X}$, model/trigger learning rate $\eta$/$\eta_\mathbf{t}$, trigger position mask $\mathbf{m}$, pre-defined $\mathsf{threshold}$
    \ENSURE The distilled dataset $\tilde{\mathbf{X}}$
    \STATE Randomly initialize the distilled dataset $\tilde{\mathbf{X}}$, and randomly choose a backdoor trigger $\mathbf{t}$ from test dataset
    \WHILE {update distilled images}
        \STATE Initialize the model $\theta_0$
        \WHILE{update model}
            \STATE $\theta_{i+1} = \theta_{i}-\eta \nabla_{\theta_{i}} \ell(\tilde{\mathbf{X}}, \theta_{i})$
        \ENDWHILE
        \WHILE{update trigger}
            \STATE $\mathsf{out} = \mathsf{f}(\mathbf{t})$
            \STATE $\mathcal{L}_\mathbf{t} = \mathsf{MSE}(\mathsf{out}, 100) + |\mathbf{t} - $ black image$|$

            \STATE $\mathcal{L}_\mathbf{t}$ back-propagation
            \STATE $\mathbf{t} \leftarrow \update(\mathbf{t}, \eta_\mathbf{t}, \mathcal{L}_\mathbf{t}, \mathbf{m})$
        \ENDWHILE
        \STATE Inject updated trigger $\mathbf{t}$ into $\epsilon |\mathbf{X}|$ samples in $\mathbf{X}$ to build the backdoored dataset $\hat{\mathbf{X}}$.
        \STATE $\mathcal{L}=\ell(\hat{\mathbf{X}}, \theta)$, $\tilde{\mathcal{L}}=\ell(\tilde{\mathbf{X}}, \theta)$
        \STATE $\tilde{\mathbf{X}}$ $\leftarrow$ $\update(\tilde{\mathbf{X}}, \mathcal{L}, \tilde{\mathcal{L}})$
    \ENDWHILE
    \end{algorithmic}
\end{algorithm}

\subsection{Number of Distilled Samples per Class}

Previous distillation work~\cite{ZMB21,ZB21,NNXL21,NCL21,ZB212,CWTEZ22} has proven that better {\em CTA} can be achieved by increasing the number of distilled samples.
It motivated us to investigate the effect of the number of distilled samples per class on our attacks.
Concretely, we select 1, 10, and 50 samples per class to assess the effect on both \naive and \doorping.
We show the backdoor attack performance in~\autoref{table:ipc}.
In general, we can see that the {\em ASR} score increases with the number of distilled samples in each class.
We can also find that the attack performances of \naive and \doorping are suboptimal when the number of distilled samples is 1, especially on the \DC algorithm.
So, if the gradient distribution of distilled image has a significant standard deviation compared to that of the original training samples, this distilled image cannot fully represent these training samples.
However, when the number of distilled images varies from 10 to 50, the {\em ASR} scores become more stable, with only one below 0.970 (approximately 0.861).
As for the {\em CTA}, we can observe a similar trend. 
More distilled samples lead to higher model utility performance.
Yet, the model utility does not improve much in most cases when attacking the \DD algorithm, except for $\langle \DD$, AlexNet, SVHN$\rangle$.

\mypara{Takeaways}
The increasing number of distilled samples leads to better {\em ASR} and {\em CTA} scores.
This is expected since the downstream model is trained on more distilled training samples (hence more backdoored samples).

\begin{table*}[!t]
\normalsize
\centering
\setlength{\tabcolsep}{3.2pt}
\caption{Attack performance under different target models and distilled samples per class.}
\scalebox{0.49}{
\renewcommand{\arraystretch}{1.5}
\begin{tabular}{c | c c !{\vrule width 1pt} c c | c c !{\vrule width 1pt} c c | c c !{\vrule width 1pt} c c | c c !{\vrule width 1pt} c c | c c}
\toprule
\multicolumn{2}{c}{}&  & \multicolumn{4}{c!{\vrule width 1pt}}{\bf FMNIST}  & \multicolumn{4}{c!{\vrule width 1pt}}{\bf CIFAR10}  & \multicolumn{4}{c!{\vrule width 1pt}}{\bf STL10}  & \multicolumn{4}{c}{\bf SVHN}  \\
\multicolumn{2}{c}{} &  & \multicolumn{2}{c|}{\naive} & \multicolumn{2}{c!{\vrule width 1pt}}{\doorping} & \multicolumn{2}{c|}{\naive} & \multicolumn{2}{c!{\vrule width 1pt}}{\doorping} & \multicolumn{2}{c|}{\naive} & \multicolumn{2}{c!{\vrule width 1pt}}{\doorping} & \multicolumn{2}{c|}{\naive} & \multicolumn{2}{c}{\doorping} \\
\multicolumn{2}{c}{} & \# & {\em ASR} & {\em CTA} & {\em ASR} & {\em CTA} & {\em ASR} & {\em CTA} & {\em ASR} & {\em CTA} & {\em ASR} & {\em CTA} & {\em ASR} & {\em CTA} & {\em ASR} & {\em CTA} & {\em ASR} & {\em CTA} \\
\midrule
\multirow{6}{*}{AlexNet} & \multirow{3}{*}{\DD} & 1  & 0.102 $\!\! \pm\!\!$ 0.025  & 0.828 $\!\! \pm\!\!$ 0.010  & 1.000 $\!\! \pm\!\!$ 0.000  & 0.821 $\!\! \pm\!\!$ 0.009  & 0.692 $\!\! \pm\!\!$ 0.016  & 0.646 $\!\! \pm\!\!$ 0.014  & 1.000 $\!\! \pm\!\!$ 0.000  & 0.633 $\!\! \pm\!\!$ 0.015  & 0.113 $\!\! \pm\!\!$ 0.011  & 0.407 $\!\! \pm\!\!$ 0.010  & 1.000 $\!\! \pm\!\!$ 0.000  & 0.424 $\!\! \pm\!\!$ 0.011  & 0.577 $\!\! \pm\!\!$ 0.015  & 0.374 $\!\! \pm\!\!$ 0.012  & 0.997 $\!\! \pm\!\!$ 0.002  & 0.406 $\!\! \pm\!\!$ 0.014 \\
& &  10 & \cellcolor{gray!30}0.103 $\!\! \pm\!\!$ 0.006 & \cellcolor{gray!30}0.867 $\!\! \pm\!\!$ 0.012 & \cellcolor{gray!30}1.000 $\!\! \pm\!\!$ 0.000 & \cellcolor{gray!30}0.868 $\!\! \pm\!\!$ 0.009 & \cellcolor{gray!30}0.692 $\!\! \pm\!\!$ 0.009 & \cellcolor{gray!30}0.646 $\!\! \pm\!\!$ 0.011 & \cellcolor{gray!30}0.999 $\!\! \pm\!\!$ 0.000 & \cellcolor{gray!30}0.635 $\!\! \pm\!\!$ 0.013 & \cellcolor{gray!30}0.123 $\!\! \pm\!\!$ 0.009 & \cellcolor{gray!30}0.428 $\!\! \pm\!\!$ 0.011 & \cellcolor{gray!30}0.999 $\!\! \pm\!\!$ 0.000 & \cellcolor{gray!30}0.438 $\!\! \pm\!\!$ 0.010 & \cellcolor{gray!30}0.797 $\!\! \pm\!\!$ 0.009 & \cellcolor{gray!30}0.729 $\!\! \pm\!\!$ 0.019 & \cellcolor{gray!30}0.970 $\!\! \pm\!\!$ 0.009 & \cellcolor{gray!30}0.759 $\!\! \pm\!\!$ 0.010 \\
& &  50  & 0.186 $\!\! \pm\!\!$ 0.010  & 0.871 $\!\! \pm\!\!$ 0.011  & 1.000 $\!\! \pm\!\!$ 0.000  & 0.882 $\!\! \pm\!\!$ 0.011  & 0.809 $\!\! \pm\!\!$ 0.014  & 0.652 $\!\! \pm\!\!$ 0.013  & 0.972 $\!\! \pm\!\!$ 0.028  & 0.650 $\!\! \pm\!\!$ 0.015  & 0.110 $\!\! \pm\!\!$ 0.007  & 0.454 $\!\! \pm\!\!$ 0.010  & 1.000 $\!\! \pm\!\!$ 0.000  & 0.436 $\!\! \pm\!\!$ 0.012  & 0.782 $\!\! \pm\!\!$ 0.020  & 0.765 $\!\! \pm\!\!$ 0.015  & 1.000 $\!\! \pm\!\!$ 0.000  & 0.755 $\!\! \pm\!\!$ 0.017 \\
& \multirow{3}{*}{\DC} & 1 & \cellcolor{gray!30}0.085 $\!\! \pm\!\!$ 0.008 & \cellcolor{gray!30}0.536 $\!\! \pm\!\!$ 0.032 & \cellcolor{gray!30}0.098 $\!\! \pm\!\!$ 0.006 & \cellcolor{gray!30}0.541 $\!\! \pm\!\!$ 0.023 & \cellcolor{gray!30}0.214 $\!\! \pm\!\!$ 0.059 & \cellcolor{gray!30}0.229 $\!\! \pm\!\!$ 0.016 & \cellcolor{gray!30}0.268 $\!\! \pm\!\!$ 0.054 & \cellcolor{gray!30}0.234 $\!\! \pm\!\!$ 0.014 & \cellcolor{gray!30}0.126 $\!\! \pm\!\!$ 0.294 & \cellcolor{gray!30}0.196 $\!\! \pm\!\!$ 0.044 & \cellcolor{gray!30}0.179 $\!\! \pm\!\!$ 0.120 & \cellcolor{gray!30}0.180 $\!\! \pm\!\!$ 0.033 & \cellcolor{gray!30}0.332 $\!\! \pm\!\!$ 0.118 & \cellcolor{gray!30}0.111 $\!\! \pm\!\!$ 0.013 & \cellcolor{gray!30}0.284 $\!\! \pm\!\!$ 0.131 & \cellcolor{gray!30}0.111 $\!\! \pm\!\!$ 0.012 \\
& &  10  & 0.118 $\!\! \pm\!\!$ 0.006  & 0.757 $\!\! \pm\!\!$ 0.009  & 1.000 $\!\! \pm\!\!$ 0.000  & 0.751 $\!\! \pm\!\!$ 0.029  & 0.133 $\!\! \pm\!\!$ 0.032  & 0.358 $\!\! \pm\!\!$ 0.047  & 1.000 $\!\! \pm\!\!$ 0.000  & 0.364 $\!\! \pm\!\!$ 0.012  & 0.098 $\!\! \pm\!\!$ 0.046  & 0.305 $\!\! \pm\!\!$ 0.040  & 0.811 $\!\! \pm\!\!$ 0.146  & 0.317 $\!\! \pm\!\!$ 0.034  & 0.333 $\!\! \pm\!\!$ 0.270  & 0.412 $\!\! \pm\!\!$ 0.083  & 0.693 $\!\! \pm\!\!$ 0.177  & 0.419 $\!\! \pm\!\!$ 0.097 \\
& &  50 & \cellcolor{gray!30}0.126 $\!\! \pm\!\!$ 0.013 & \cellcolor{gray!30}0.828 $\!\! \pm\!\!$ 0.004 & \cellcolor{gray!30}1.000 $\!\! \pm\!\!$ 0.000 & \cellcolor{gray!30}0.813 $\!\! \pm\!\!$ 0.003 & \cellcolor{gray!30}0.151 $\!\! \pm\!\!$ 0.021 & \cellcolor{gray!30}0.467 $\!\! \pm\!\!$ 0.006 & \cellcolor{gray!30}0.990 $\!\! \pm\!\!$ 0.009 & \cellcolor{gray!30}0.470 $\!\! \pm\!\!$ 0.006 & \cellcolor{gray!30}0.153 $\!\! \pm\!\!$ 0.016 & \cellcolor{gray!30}0.462 $\!\! \pm\!\!$ 0.005 & \cellcolor{gray!30}0.861 $\!\! \pm\!\!$ 0.062 & \cellcolor{gray!30}0.471 $\!\! \pm\!\!$ 0.007 & \cellcolor{gray!30}0.763 $\!\! \pm\!\!$ 0.030 & \cellcolor{gray!30}0.741 $\!\! \pm\!\!$ 0.010 & \cellcolor{gray!30}0.979 $\!\! \pm\!\!$ 0.010 & \cellcolor{gray!30}0.735 $\!\! \pm\!\!$ 0.007 \\
\midrule
\multirow{6}{*}{ConvNet} & \multirow{3}{*}{\DD} & 1  & 0.124 $\!\! \pm\!\!$ 0.007  & 0.784 $\!\! \pm\!\!$ 0.014  & 1.000 $\!\! \pm\!\!$ 0.000  & 0.800 $\!\! \pm\!\!$ 0.010  & 0.137 $\!\! \pm\!\!$ 0.014  & 0.450 $\!\! \pm\!\!$ 0.012  & 1.000 $\!\! \pm\!\!$ 0.000  & 0.453 $\!\! \pm\!\!$ 0.014  & 0.151 $\!\! \pm\!\!$ 0.008  & 0.357 $\!\! \pm\!\!$ 0.013  & 1.000 $\!\! \pm\!\!$ 0.000  & 0.367 $\!\! \pm\!\!$ 0.010  & 0.612 $\!\! \pm\!\!$ 0.016  & 0.332 $\!\! \pm\!\!$ 0.021  & 1.000 $\!\! \pm\!\!$ 0.000  & 0.340 $\!\! \pm\!\!$ 0.013 \\
& &  10 & \cellcolor{gray!30}0.126 $\!\! \pm\!\!$ 0.009 & \cellcolor{gray!30}0.803 $\!\! \pm\!\!$ 0.010 & \cellcolor{gray!30}1.000 $\!\! \pm\!\!$ 0.000 & \cellcolor{gray!30}0.804 $\!\! \pm\!\!$ 0.011 & \cellcolor{gray!30}0.105 $\!\! \pm\!\!$ 0.026 & \cellcolor{gray!30}0.478 $\!\! \pm\!\!$ 0.011 & \cellcolor{gray!30}1.000 $\!\! \pm\!\!$ 0.000 & \cellcolor{gray!30}0.476 $\!\! \pm\!\!$ 0.014 & \cellcolor{gray!30}0.136 $\!\! \pm\!\!$ 0.012 & \cellcolor{gray!30}0.363 $\!\! \pm\!\!$ 0.012 & \cellcolor{gray!30}1.000 $\!\! \pm\!\!$ 0.000 & \cellcolor{gray!30}0.371 $\!\! \pm\!\!$ 0.011 & \cellcolor{gray!30}0.712 $\!\! \pm\!\!$ 0.002 & \cellcolor{gray!30}0.372 $\!\! \pm\!\!$ 0.016 & \cellcolor{gray!30}1.000 $\!\! \pm\!\!$ 0.000 & \cellcolor{gray!30}0.375 $\!\! \pm\!\!$ 0.015 \\
& &  50  & 0.242 $\!\! \pm\!\!$ 0.010  & 0.828 $\!\! \pm\!\!$ 0.013  & 1.000 $\!\! \pm\!\!$ 0.000  & 0.828 $\!\! \pm\!\!$ 0.009  & 0.121 $\!\! \pm\!\!$ 0.008  & 0.482 $\!\! \pm\!\!$ 0.013  & 1.000 $\!\! \pm\!\!$ 0.000  & 0.489 $\!\! \pm\!\!$ 0.014  & 0.134 $\!\! \pm\!\!$ 0.005  & 0.378 $\!\! \pm\!\!$ 0.013  & 1.000 $\!\! \pm\!\!$ 0.000  & 0.370 $\!\! \pm\!\!$ 0.010  & 0.912 $\!\! \pm\!\!$ 0.025  & 0.477 $\!\! \pm\!\!$ 0.017  & 1.000 $\!\! \pm\!\!$ 0.000  & 0.482 $\!\! \pm\!\!$ 0.016 \\
& \multirow{3}{*}{\DC} & 1 & \cellcolor{gray!30}0.081 $\!\! \pm\!\!$ 0.007 & \cellcolor{gray!30}0.535 $\!\! \pm\!\!$ 0.023 & \cellcolor{gray!30}0.091 $\!\! \pm\!\!$ 0.004 & \cellcolor{gray!30}0.545 $\!\! \pm\!\!$ 0.013 & \cellcolor{gray!30}0.222 $\!\! \pm\!\!$ 0.043 & \cellcolor{gray!30}0.230 $\!\! \pm\!\!$ 0.006 & \cellcolor{gray!30}0.225 $\!\! \pm\!\!$ 0.051 & \cellcolor{gray!30}0.229 $\!\! \pm\!\!$ 0.007 & \cellcolor{gray!30}0.166 $\!\! \pm\!\!$ 0.024 & \cellcolor{gray!30}0.223 $\!\! \pm\!\!$ 0.007 & \cellcolor{gray!30}0.181 $\!\! \pm\!\!$ 0.037 & \cellcolor{gray!30}0.217 $\!\! \pm\!\!$ 0.009 & \cellcolor{gray!30}0.090 $\!\! \pm\!\!$ 0.056 & \cellcolor{gray!30}0.113 $\!\! \pm\!\!$ 0.007 & \cellcolor{gray!30}0.114 $\!\! \pm\!\!$ 0.029 & \cellcolor{gray!30}0.111 $\!\! \pm\!\!$ 0.006 \\
& &  10  & 0.102 $\!\! \pm\!\!$ 0.006  & 0.732 $\!\! \pm\!\!$ 0.007  & 1.000 $\!\! \pm\!\!$ 0.000  & 0.734 $\!\! \pm\!\!$ 0.008  & 0.113 $\!\! \pm\!\!$ 0.012  & 0.310 $\!\! \pm\!\!$ 0.017  & 1.000 $\!\! \pm\!\!$ 0.000  & 0.308 $\!\! \pm\!\!$ 0.008  & 0.093 $\!\! \pm\!\!$ 0.004  & 0.321 $\!\! \pm\!\!$ 0.012  & 0.981 $\!\! \pm\!\!$ 0.000  & 0.331 $\!\! \pm\!\!$ 0.013  & 0.430 $\!\! \pm\!\!$ 0.187  & 0.215 $\!\! \pm\!\!$ 0.015  & 1.000 $\!\! \pm\!\!$ 0.000  & 0.203 $\!\! \pm\!\!$ 0.021 \\
& &  50 & \cellcolor{gray!30}0.107 $\!\! \pm\!\!$ 0.005 & \cellcolor{gray!30}0.776 $\!\! \pm\!\!$ 0.005 & \cellcolor{gray!30}1.000 $\!\! \pm\!\!$ 0.000 & \cellcolor{gray!30}0.774 $\!\! \pm\!\!$ 0.004 & \cellcolor{gray!30}0.117 $\!\! \pm\!\!$ 0.008 & \cellcolor{gray!30}0.380 $\!\! \pm\!\!$ 0.007 & \cellcolor{gray!30}1.000 $\!\! \pm\!\!$ 0.000 & \cellcolor{gray!30}0.361 $\!\! \pm\!\!$ 0.007 & \cellcolor{gray!30}0.106 $\!\! \pm\!\!$ 0.007 & \cellcolor{gray!30}0.413 $\!\! \pm\!\!$ 0.011 & \cellcolor{gray!30}0.998 $\!\! \pm\!\!$ 0.002 & \cellcolor{gray!30}0.421 $\!\! \pm\!\!$ 0.007 & \cellcolor{gray!30}0.851 $\!\! \pm\!\!$ 0.059 & \cellcolor{gray!30}0.480 $\!\! \pm\!\!$ 0.027 & \cellcolor{gray!30}1.000 $\!\! \pm\!\!$ 0.000 & \cellcolor{gray!30}0.490 $\!\! \pm\!\!$ 0.021 \\
\bottomrule
\end{tabular}
}
\label{table:ipc}
\end{table*}

\subsection{Number of Distillation Epochs}

\begin{figure*}[!t]
\centering
\includegraphics[width=1.9\columnwidth]{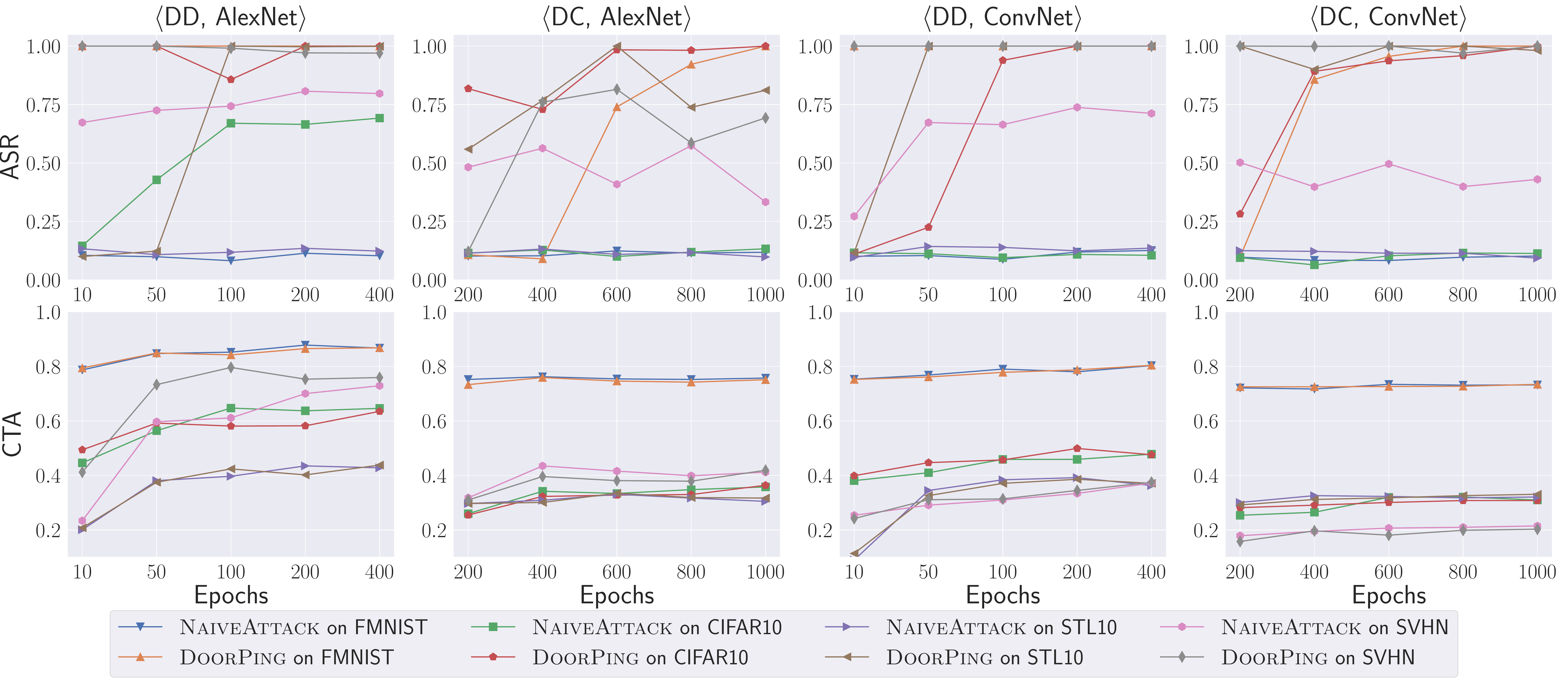}
\caption{{\em ASR} and {\em CTA} of \naive and \doorping under different training epochs and different model architectures.}
\label{figure:epoch}
\end{figure*}

We further investigate the impact of the number of distillation epochs on attack and utility performance.
The rationale is that the number of distillation epochs has a significant impact on the final distillation image.
Therefore, we report the attack and utility performance by varying the number of distilled epochs from 10 to 400 for \DD and from 200 to 1000 for \DC, respectively.
As depicted in~\autoref{figure:epoch}, we can observe that in general, both {\em ASR} and {\em CTA} scores first increase and stabilize after a certain number of distillation epochs.
We can also find that the {\em ASR} scores of some cases stabilize throughout the whole distilled procedure.
For example, the {\em ASR} score of $\langle \DC$, ConvNet, CIFAR10$\rangle$ is around 0.100 from beginning to end, but the {\em CTA} score soars from 0.254 to 0.320 after 400 epochs.

\mypara{Takeaways}
Our results suggest that performing the dataset distillation process is worthwhile through a larger number of distillation epochs since, in most cases, both attack and utility performance increase with the number of distillation epochs.

\subsection{Number of Original Samples}

\begin{figure*}[!t]
\centering
\includegraphics[width=1.9\columnwidth]{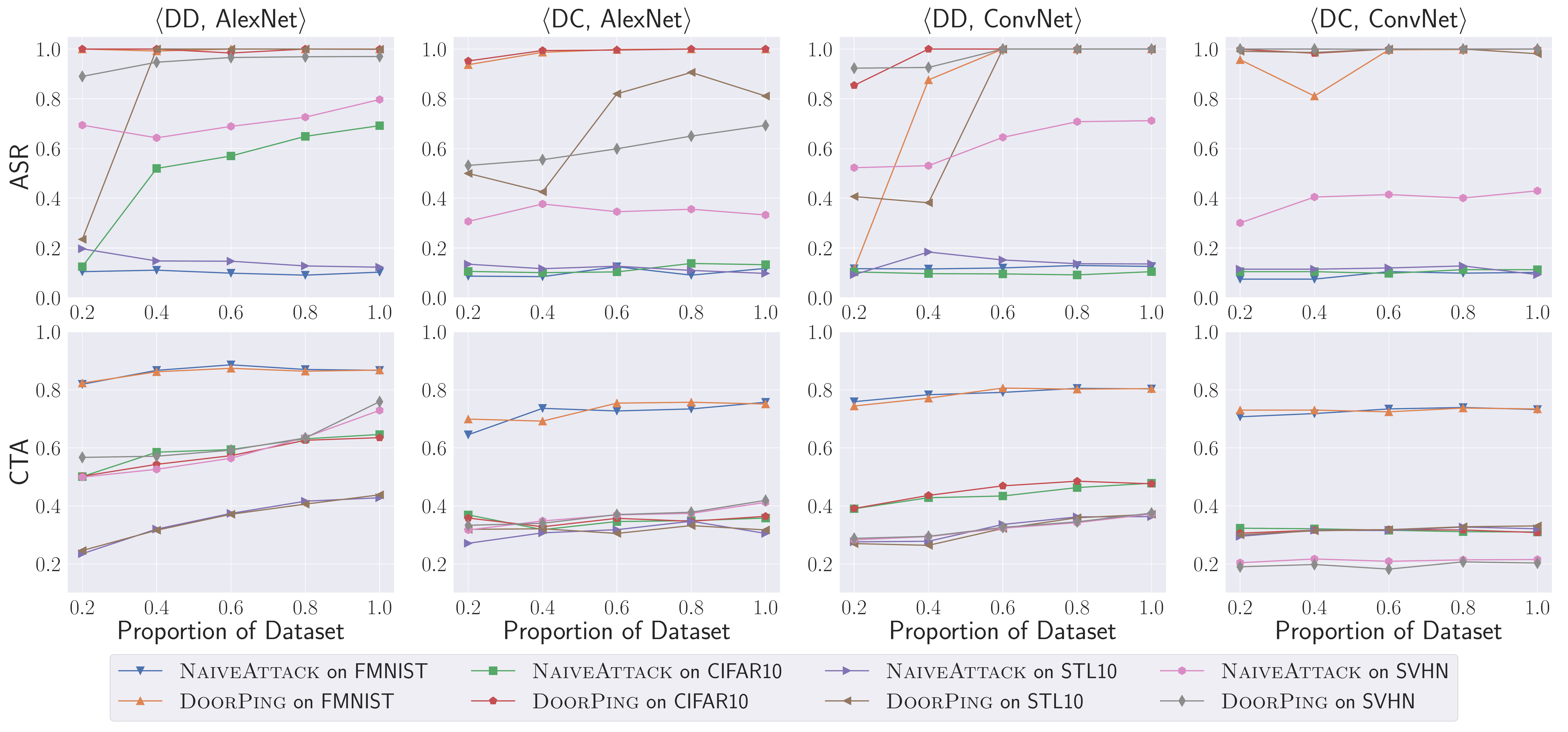}
\caption{{\em ASR} and {\em CTA} of \naive and \doorping under the different proportions of original training datasets and different model architectures. 
The X-axis represents the proportion of the original samples used in the dataset distillation process.}
\label{figure:sample}
\end{figure*}

Dataset distillation aims to reduce the redundancy in the training datasets.
One more possible redundancy is the number of original training samples.
Here, we study the impact of the number of original training samples on the performance of our attacks.
Concretely, we vary the proportion of the entire dataset from 0.2 to 1 to report the {\em ASR} and {\em CTA} scores.
We show the trends with the increase of the number of the original training samples in~\autoref{figure:sample}.
As we can see, the {\em ASR} score generally grows with the upswing of the sample numbers.
In particular, the {\em ASR} score of some cases remains stationary throughout.
However, almost all cases start with a much lower {\em ASR} score, i.e., when only 20\% of the samples from the original training dataset are used to distill images, both of our attacks only achieve relatively poor attack performance.
For instance, the {\em ASR} score of $\langle \DD$, AlexNet, STL10$\rangle$ is only 0.235.
These results demonstrate that the number of original training dataset affect the attack performance significantly.
As expected for the {\em CTA} score, the majority of the model utility increase with the increasing number of original training samples. 
Only some {\em CTA} scores oscillate in an ultra-fine range.
For example, the {\em CTA} score of $\langle \DC$, ConvNet, SVHN$\rangle$ fluctuates from 0.190 to 0.217, and the {\em ASR} score is 1.000 when using \doorping.
Besides, we can also find that for $\langle \DC$, AlexNet, STL10$\rangle$, the {\em CTA} score only increases from 0.305 to 0.319, but the {\em ASR} score surges to over 0.811 after the proportion gets larger than 0.4.
This means it is acceptable to distill less training data, as the model utility does not change much but still achieves an acceptable attack performance.

\mypara{Takeaways}
The increasing size of the original training dataset leads to better {\em ASR} and {\em CTA} scores.
This is also expected since the distillation models have learned sufficient patterns from additional samples.

\subsection{Number of Selected Neurons}
\label{section:top_k_selection}

\begin{figure*}[!t]
\centering
\includegraphics[width=1.9\columnwidth]{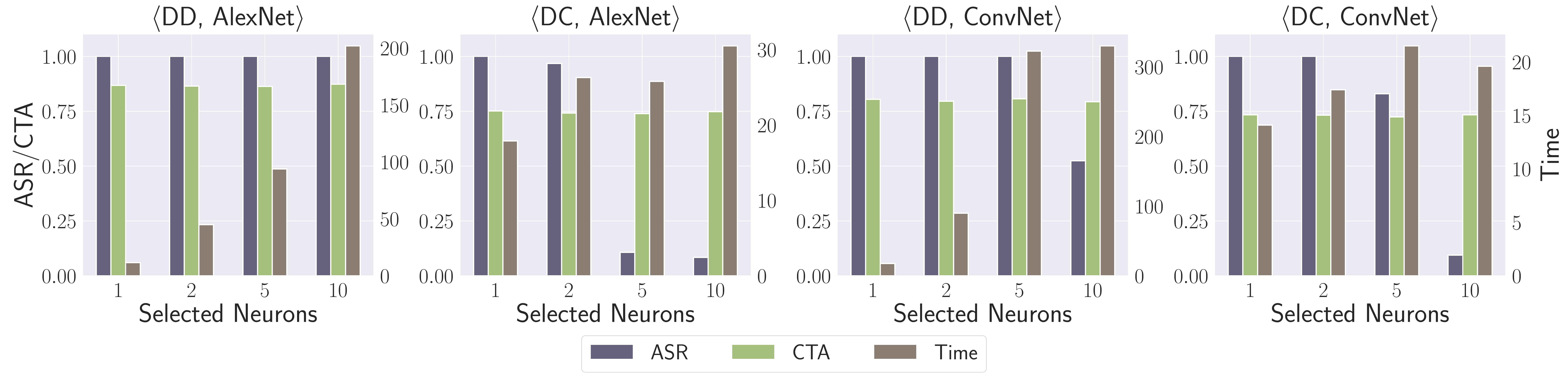}
\caption{{\em ASR}, {\em CTA}, and the running time of \doorping with different selected neurons for FMNIST dataset.}
\label{figure:fmnist_neurons}
\end{figure*}

\begin{figure*}[!t]
\centering
\includegraphics[width=1.9\columnwidth]{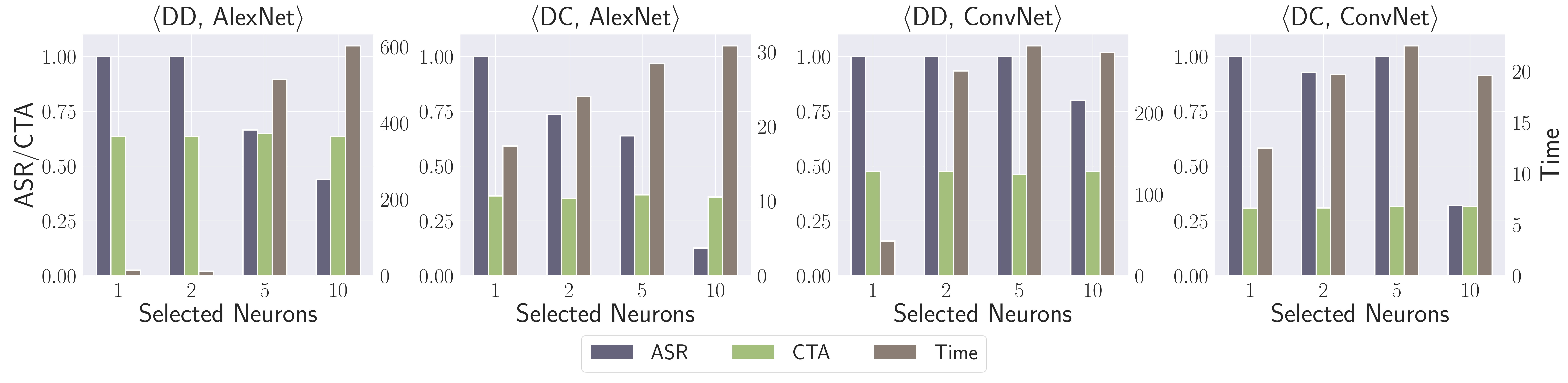}
\caption{{\em ASR}, {\em CTA}, and runtime of \doorping with different selected neurons for CIFAR10 dataset.}
\label{figure:cifar_neurons}
\end{figure*}

\begin{figure*}[!t]
\centering
\includegraphics[width=1.9\columnwidth]{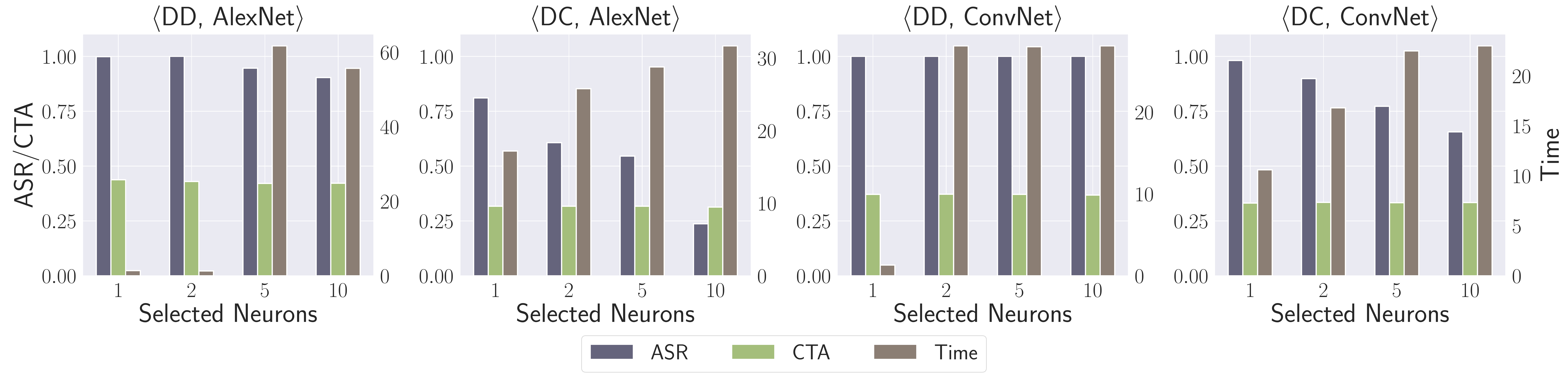}
\caption{{\em ASR}, {\em CTA}, and the running time of \doorping with different selected neurons for STL10 dataset.}
\label{figure:stl10_neurons}
\end{figure*}

\begin{figure*}[!t]
\centering
\includegraphics[width=1.9\columnwidth]{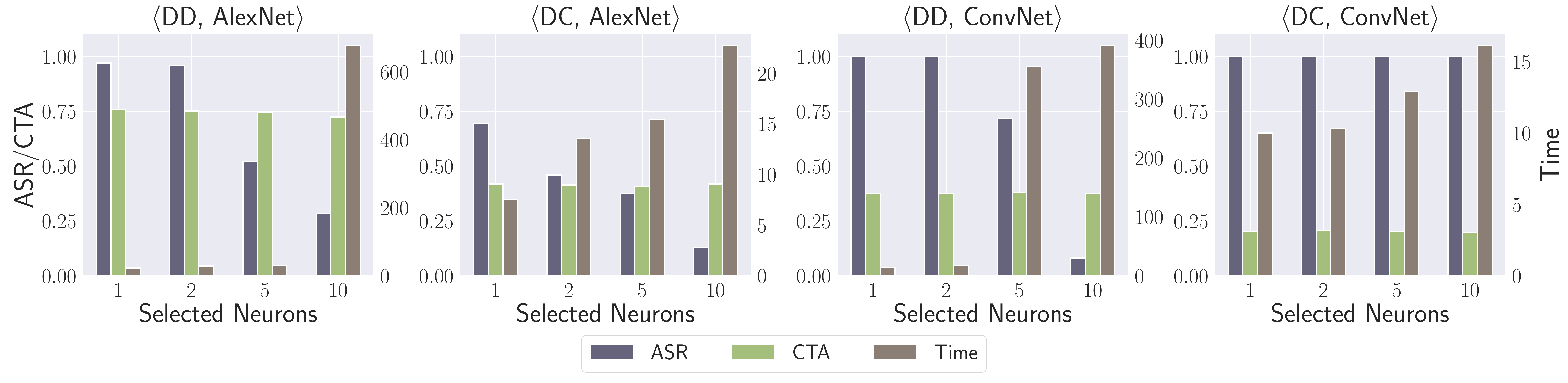}
\caption{{\em ASR}, {\em CTA}, and the running time of \doorping with different selected neurons for SVHN dataset.}
\label{figure:svhn_neurons}
\end{figure*}

We aim to understand the impact of the number of selected neurons to optimize the trigger in \doorping (i.e., the impact of top-$k$).
Especially in the penultimate layer of the models, we conduct the evaluation by setting the number of selected neurons to 1, 2, 5, and 10, respectively.
We report the {\em ASR} score, the {\em CTA} score, and the average running time per epoch in~\autoref{figure:fmnist_neurons},~\autoref{figure:cifar_neurons},~\autoref{figure:stl10_neurons}, and~\autoref{figure:svhn_neurons}.
As we can see from these figures, with the number of selected neurons increasing, the {\em ASR} scores decrease while the runtime increases.
The experiments also show that the {\em CTA} scores are almost unchanged.
For example, the {\em ASR} scores of $\langle \DD$, AlexNet, CIFAR10$\rangle$ are 0.999, 1.000, 0.665, and 0.440.
Their respective runtime increases from 15s to 602s, while the {\em CTA} scores remain stable (0.635, 0.636, 0.648, 0.635. respectively).
The reason behind this is that the weights of some selected neurons connecting these neurons to the preceding and following layers are smaller than others.
In other words, we need to refrain from exploiting these neurons to enhance the triggers.
We, therefore, set the number of the selected neurons to 1 throughout all experiments in the paper.

\mypara{Takeaways}
The increasing number of selected neurons harms the attack performance while increasing the runtime.
This observation is in line with previous work~\cite{LMALZWZ18}.

\subsection{Poisoning Ratio}

\begin{figure*}[!t]
\centering
\includegraphics[width=1.9\columnwidth]{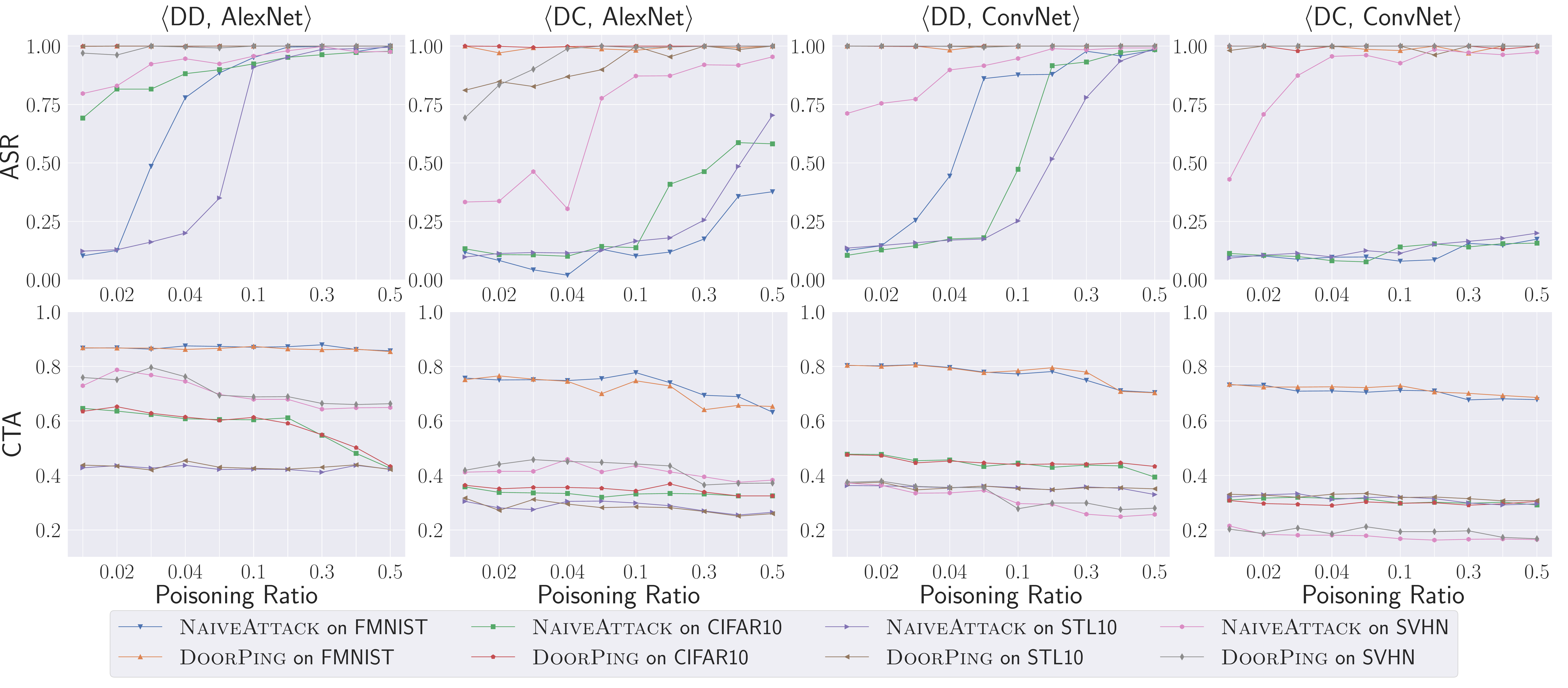}
\caption{{\em ASR} and {\em CTA} of \naive and \doorping under the different poisoning ratios and different model architectures. 
The X-axis represents the poisoning ratio in the whole training dataset.}
\label{figure:proportion}
\end{figure*}

Here, we investigate the impact of the poisoning ratio (i.e., $\epsilon$ in~\autoref{section:doorpingattack}) in the entire training dataset.
We vary the poisoning ratio from 0.01 to 0.5 and report both attack and utility performance in~\autoref{figure:proportion}.
For the attack performance, we can find that the {\em ASR} scores vary significantly in general.
For instance, the {\em ASR} scores increase from 0.811 to 1.000 and from 0.693 to 1.000 for $\langle \DC$, AlexNet, STL10$\rangle$, and $\langle \DC$, AlexNet, SVHN$\rangle$ in \doorping, respectively.
For the majority of cases in \naive, with the poisoning ratio increasing from 0.05 to 0.5, the {\em ASR} scores are also increasing.
Taking $\langle \DD$, AlexNet, STL10$\rangle$ as an example, for the poisoning ratio of \naive increases from 0.01 to 0.05, the {\em ASR} score fluctuates between 0.136 and 0.175.
When the poisoning ratio increases to 0.1 and expends to 0.5, the {\em ASR} score rises and eventually reaches 0.990.
However, unlike \doorping, we find it challenging to achieve a 1.000 {\em ASR} score in \naive, which exemplifies that \doorping is more effective than \naive.
For the model utility performance, we can observe a general downward trend.
For example, the {\em CTA} score of \doorping decreases from 0.364 to 0.325 on $\langle \DC$, AlexNet, CIFAR10$\rangle$.
We also observe similar trends in \naive.

\mypara{Takeaways}
The poisoning ratio impacts the {\em ASR} scores, especially for \naive. 
However, the {\em CTA} scores may vary given different backdoor sample ratios incurred by the respective properties of distillation models (e.g., hyperparameters, architectures, etc.).

\subsection{Trigger Size}

Previous work~\cite{LMALZWZ18} has shown that the larger the trigger size is, the higher the attack performance is.
Thus, we first investigate the impact of trigger size on attack performance.
We set the trigger size to $2 \times 2$, $3 \times 3$, and $4 \times 4$ to investigate their impacts on the attack and utility performance.
\autoref{table:triggersize} shows the {\em ASR} and the {\em CTA} scores with respect to different trigger sizes.
For \naive, as the trigger size increases, the {\em ASR} score also increases, especially from $3 \times 3$ to $4 \times 4$.
For instance, the {\em ASR} score of \naive for $\langle \DC$, ConvNet, SVHN$\rangle$ increases from 0.430 to 0.769.
For \doorping, we can see that the {\em ASR} score is close to 1.0 in most cases, regardless of the trigger size setting.
For example, given $\langle \DD$, AlexNet, STL10$\rangle$, when the trigger size is increased from $2 \times 2$ to $4 \times 4$, the {\em ASR} score increases from 0.811 to 0.984. 
Similarly, the {\em ASR} score increases from 0.693 to 0.805 for SVHN.
These results show that larger triggers generally lead to higher attack performance in our attacks.
In terms of the impact of trigger size on utility performance, we find that the majority of {\em CTA} scores slightly decrease with the increase of the trigger size.
To this end, we calculate the Pearson correlation coefficient between the trigger size and the {\em CTA} score.
In total, we have 32 correlation values. 
Among those values, 9 are positive, and 23 are negative.
The average of the correlations is -0.370.
Therefore, the {\em CTA} negatively correlates with the trigger size.
Despite the side effects caused by larger trigger sizes, the {\em CTA} scores are still within the acceptable performance variation of the model.
They do not significantly impact the model utility performance.

\mypara{Takeaways}
When the trigger size becomes more prominent and larger, the final synthetic image contains more trigger information but less information of the original images.
It may lead to the inevitable trade-off between attack performance and model utility.

\begin{table*}[!t]
\normalsize
\centering
\setlength{\tabcolsep}{3.2pt}
\caption{Attack performance of different target models and trigger sizes.}
\scalebox{0.49}{
\renewcommand{\arraystretch}{1.5}
\begin{tabular}{c | c c !{\vrule width 1pt} c c | c c !{\vrule width 1pt} c c | c c !{\vrule width 1pt} c c | c c !{\vrule width 1pt} c c | c c}
\toprule
\multicolumn{2}{c}{}&  & \multicolumn{4}{c!{\vrule width 1pt}}{\bf FMNIST}  & \multicolumn{4}{c!{\vrule width 1pt}}{\bf CIFAR10}  & \multicolumn{4}{c!{\vrule width 1pt}}{\bf STL10}  & \multicolumn{4}{c}{\bf SVHN}  \\
\multicolumn{2}{c}{} &  & \multicolumn{2}{c|}{\naive} & \multicolumn{2}{c!{\vrule width 1pt}}{\doorping} & \multicolumn{2}{c|}{\naive} & \multicolumn{2}{c!{\vrule width 1pt}}{\doorping} & \multicolumn{2}{c|}{\naive} & \multicolumn{2}{c!{\vrule width 1pt}}{\doorping} & \multicolumn{2}{c|}{\naive} & \multicolumn{2}{c}{\doorping} \\
\multicolumn{2}{c}{} & \# & {\em ASR} & {\em CTA} & {\em ASR} & {\em CTA} & {\em ASR} & {\em CTA} & {\em ASR} & {\em CTA} & {\em ASR} & {\em CTA} & {\em ASR} & {\em CTA} & {\em ASR} & {\em CTA} & {\em ASR} & {\em CTA} \\
\midrule
\multirow{6}{*}{AlexNet} & \multirow{3}{*}{\DD} & 2 & \cellcolor{gray!30}0.103 $\!\! \pm\!\!$ 0.006 & \cellcolor{gray!30}0.867 $\!\! \pm\!\!$ 0.012 & \cellcolor{gray!30}1.000 $\!\! \pm\!\!$ 0.000 & \cellcolor{gray!30}0.868 $\!\! \pm\!\!$ 0.009 & \cellcolor{gray!30}0.692 $\!\! \pm\!\!$ 0.009 & \cellcolor{gray!30}0.646 $\!\! \pm\!\!$ 0.011 & \cellcolor{gray!30}0.999 $\!\! \pm\!\!$ 0.000 & \cellcolor{gray!30}0.635 $\!\! \pm\!\!$ 0.013 & \cellcolor{gray!30}0.123 $\!\! \pm\!\!$ 0.009 & \cellcolor{gray!30}0.428 $\!\! \pm\!\!$ 0.011 & \cellcolor{gray!30}0.999 $\!\! \pm\!\!$ 0.000 & \cellcolor{gray!30}0.438 $\!\! \pm\!\!$ 0.010 & \cellcolor{gray!30}0.797 $\!\! \pm\!\!$ 0.009 & \cellcolor{gray!30}0.729 $\!\! \pm\!\!$ 0.019 & \cellcolor{gray!30}0.970 $\!\! \pm\!\!$ 0.009 & \cellcolor{gray!30}0.759 $\!\! \pm\!\!$ 0.010 \\
& &  3  & 0.729 $\!\! \pm\!\!$ 0.007  & 0.833 $\!\! \pm\!\!$ 0.013  & 1.000 $\!\! \pm\!\!$ 0.000  & 0.814 $\!\! \pm\!\!$ 0.013  & 0.789 $\!\! \pm\!\!$ 0.013  & 0.646 $\!\! \pm\!\!$ 0.014  & 1.000 $\!\! \pm\!\!$ 0.000  & 0.614 $\!\! \pm\!\!$ 0.012  & 0.126 $\!\! \pm\!\!$ 0.008  & 0.413 $\!\! \pm\!\!$ 0.011  & 1.000 $\!\! \pm\!\!$ 0.000  & 0.418 $\!\! \pm\!\!$ 0.010  & 0.805 $\!\! \pm\!\!$ 0.005  & 0.699 $\!\! \pm\!\!$ 0.015  & 0.982 $\!\! \pm\!\!$ 0.011  & 0.702 $\!\! \pm\!\!$ 0.020 \\
& &  4 & \cellcolor{gray!30}0.929 $\!\! \pm\!\!$ 0.010 & \cellcolor{gray!30}0.809 $\!\! \pm\!\!$ 0.012 & \cellcolor{gray!30}1.000 $\!\! \pm\!\!$ 0.000 & \cellcolor{gray!30}0.809 $\!\! \pm\!\!$ 0.011 & \cellcolor{gray!30}0.809 $\!\! \pm\!\!$ 0.010 & \cellcolor{gray!30}0.582 $\!\! \pm\!\!$ 0.016 & \cellcolor{gray!30}1.000 $\!\! \pm\!\!$ 0.000 & \cellcolor{gray!30}0.606 $\!\! \pm\!\!$ 0.014 & \cellcolor{gray!30}0.154 $\!\! \pm\!\!$ 0.010 & \cellcolor{gray!30}0.423 $\!\! \pm\!\!$ 0.009 & \cellcolor{gray!30}1.000 $\!\! \pm\!\!$ 0.000 & \cellcolor{gray!30}0.442 $\!\! \pm\!\!$ 0.012 & \cellcolor{gray!30}0.857 $\!\! \pm\!\!$ 0.020 & \cellcolor{gray!30}0.677 $\!\! \pm\!\!$ 0.026 & \cellcolor{gray!30}0.993 $\!\! \pm\!\!$ 0.005 & \cellcolor{gray!30}0.639 $\!\! \pm\!\!$ 0.017 \\
& \multirow{3}{*}{\DC} & 2  & 0.118 $\!\! \pm\!\!$ 0.006  & 0.757 $\!\! \pm\!\!$ 0.009  & 1.000 $\!\! \pm\!\!$ 0.000  & 0.751 $\!\! \pm\!\!$ 0.029  & 0.133 $\!\! \pm\!\!$ 0.032  & 0.358 $\!\! \pm\!\!$ 0.047  & 1.000 $\!\! \pm\!\!$ 0.000  & 0.364 $\!\! \pm\!\!$ 0.012  & 0.098 $\!\! \pm\!\!$ 0.046  & 0.305 $\!\! \pm\!\!$ 0.040  & 0.811 $\!\! \pm\!\!$ 0.146  & 0.317 $\!\! \pm\!\!$ 0.034  & 0.333 $\!\! \pm\!\!$ 0.270  & 0.412 $\!\! \pm\!\!$ 0.083  & 0.693 $\!\! \pm\!\!$ 0.177  & 0.419 $\!\! \pm\!\!$ 0.097 \\
& &  3 & \cellcolor{gray!30}0.062 $\!\! \pm\!\!$ 0.028 & \cellcolor{gray!30}0.749 $\!\! \pm\!\!$ 0.036 & \cellcolor{gray!30}1.000 $\!\! \pm\!\!$ 0.000 & \cellcolor{gray!30}0.734 $\!\! \pm\!\!$ 0.011 & \cellcolor{gray!30}0.105 $\!\! \pm\!\!$ 0.024 & \cellcolor{gray!30}0.347 $\!\! \pm\!\!$ 0.019 & \cellcolor{gray!30}1.000 $\!\! \pm\!\!$ 0.000 & \cellcolor{gray!30}0.347 $\!\! \pm\!\!$ 0.029 & \cellcolor{gray!30}0.123 $\!\! \pm\!\!$ 0.203 & \cellcolor{gray!30}0.296 $\!\! \pm\!\!$ 0.066 & \cellcolor{gray!30}0.965 $\!\! \pm\!\!$ 0.000 & \cellcolor{gray!30}0.318 $\!\! \pm\!\!$ 0.054 & \cellcolor{gray!30}0.323 $\!\! \pm\!\!$ 0.016 & \cellcolor{gray!30}0.437 $\!\! \pm\!\!$ 0.057 & \cellcolor{gray!30}0.633 $\!\! \pm\!\!$ 0.303 & \cellcolor{gray!30}0.448 $\!\! \pm\!\!$ 0.118 \\
& &  4  & 0.124 $\!\! \pm\!\!$ 0.019  & 0.750 $\!\! \pm\!\!$ 0.010  & 1.000 $\!\! \pm\!\!$ 0.011  & 0.753 $\!\! \pm\!\!$ 0.008  & 0.160 $\!\! \pm\!\!$ 0.087  & 0.333 $\!\! \pm\!\!$ 0.065  & 0.988 $\!\! \pm\!\!$ 0.011  & 0.342 $\!\! \pm\!\!$ 0.010  & 0.154 $\!\! \pm\!\!$ 0.054  & 0.293 $\!\! \pm\!\!$ 0.030  & 0.984 $\!\! \pm\!\!$ 0.000  & 0.308 $\!\! \pm\!\!$ 0.039  & 0.475 $\!\! \pm\!\!$ 0.254  & 0.434 $\!\! \pm\!\!$ 0.098  & 0.805 $\!\! \pm\!\!$ 0.185  & 0.424 $\!\! \pm\!\!$ 0.122 \\
\midrule
\multirow{6}{*}{ConvNet} & \multirow{3}{*}{\DD} & 2 & \cellcolor{gray!30}0.126 $\!\! \pm\!\!$ 0.009 & \cellcolor{gray!30}0.803 $\!\! \pm\!\!$ 0.010 & \cellcolor{gray!30}1.000 $\!\! \pm\!\!$ 0.000 & \cellcolor{gray!30}0.804 $\!\! \pm\!\!$ 0.011 & \cellcolor{gray!30}0.105 $\!\! \pm\!\!$ 0.026 & \cellcolor{gray!30}0.478 $\!\! \pm\!\!$ 0.011 & \cellcolor{gray!30}1.000 $\!\! \pm\!\!$ 0.000 & \cellcolor{gray!30}0.476 $\!\! \pm\!\!$ 0.014 & \cellcolor{gray!30}0.136 $\!\! \pm\!\!$ 0.012 & \cellcolor{gray!30}0.363 $\!\! \pm\!\!$ 0.012 & \cellcolor{gray!30}1.000 $\!\! \pm\!\!$ 0.000 & \cellcolor{gray!30}0.371 $\!\! \pm\!\!$ 0.011 & \cellcolor{gray!30}0.712 $\!\! \pm\!\!$ 0.002 & \cellcolor{gray!30}0.372 $\!\! \pm\!\!$ 0.016 & \cellcolor{gray!30}1.000 $\!\! \pm\!\!$ 0.000 & \cellcolor{gray!30}0.375 $\!\! \pm\!\!$ 0.015 \\
& &  3  & 0.216 $\!\! \pm\!\!$ 0.003  & 0.791 $\!\! \pm\!\!$ 0.006  & 1.000 $\!\! \pm\!\!$ 0.000  & 0.784 $\!\! \pm\!\!$ 0.012  & 0.125 $\!\! \pm\!\!$ 0.014  & 0.467 $\!\! \pm\!\!$ 0.013  & 0.999 $\!\! \pm\!\!$ 0.000  & 0.477 $\!\! \pm\!\!$ 0.014  & 0.133 $\!\! \pm\!\!$ 0.006  & 0.353 $\!\! \pm\!\!$ 0.013  & 1.000 $\!\! \pm\!\!$ 0.000  & 0.346 $\!\! \pm\!\!$ 0.012  & 0.770 $\!\! \pm\!\!$ 0.006  & 0.365 $\!\! \pm\!\!$ 0.019  & 0.999 $\!\! \pm\!\!$ 0.000  & 0.356 $\!\! \pm\!\!$ 0.017 \\
& &  4 & \cellcolor{gray!30}0.560 $\!\! \pm\!\!$ 0.009 & \cellcolor{gray!30}0.800 $\!\! \pm\!\!$ 0.013 & \cellcolor{gray!30}1.000 $\!\! \pm\!\!$ 0.000 & \cellcolor{gray!30}0.798 $\!\! \pm\!\!$ 0.012 & \cellcolor{gray!30}0.167 $\!\! \pm\!\!$ 0.013 & \cellcolor{gray!30}0.465 $\!\! \pm\!\!$ 0.013 & \cellcolor{gray!30}1.000 $\!\! \pm\!\!$ 0.000 & \cellcolor{gray!30}0.456 $\!\! \pm\!\!$ 0.013 & \cellcolor{gray!30}0.139 $\!\! \pm\!\!$ 0.009 & \cellcolor{gray!30}0.366 $\!\! \pm\!\!$ 0.011 & \cellcolor{gray!30}1.000 $\!\! \pm\!\!$ 0.000 & \cellcolor{gray!30}0.374 $\!\! \pm\!\!$ 0.010 & \cellcolor{gray!30}0.771 $\!\! \pm\!\!$ 0.036 & \cellcolor{gray!30}0.368 $\!\! \pm\!\!$ 0.017 & \cellcolor{gray!30}0.975 $\!\! \pm\!\!$ 0.021 & \cellcolor{gray!30}0.378 $\!\! \pm\!\!$ 0.012 \\
& \multirow{3}{*}{\DC} & 2  & 0.102 $\!\! \pm\!\!$ 0.006  & 0.732 $\!\! \pm\!\!$ 0.007  & 1.000 $\!\! \pm\!\!$ 0.000  & 0.734 $\!\! \pm\!\!$ 0.008  & 0.113 $\!\! \pm\!\!$ 0.012  & 0.310 $\!\! \pm\!\!$ 0.017  & 1.000 $\!\! \pm\!\!$ 0.000  & 0.308 $\!\! \pm\!\!$ 0.008  & 0.093 $\!\! \pm\!\!$ 0.004  & 0.321 $\!\! \pm\!\!$ 0.012  & 0.981 $\!\! \pm\!\!$ 0.000  & 0.331 $\!\! \pm\!\!$ 0.013  & 0.430 $\!\! \pm\!\!$ 0.187  & 0.215 $\!\! \pm\!\!$ 0.015  & 1.000 $\!\! \pm\!\!$ 0.000  & 0.203 $\!\! \pm\!\!$ 0.021 \\
& &  3 & \cellcolor{gray!30}0.079 $\!\! \pm\!\!$ 0.009 & \cellcolor{gray!30}0.737 $\!\! \pm\!\!$ 0.010 & \cellcolor{gray!30}1.000 $\!\! \pm\!\!$ 0.000 & \cellcolor{gray!30}0.737 $\!\! \pm\!\!$ 0.006 & \cellcolor{gray!30}0.107 $\!\! \pm\!\!$ 0.023 & \cellcolor{gray!30}0.315 $\!\! \pm\!\!$ 0.015 & \cellcolor{gray!30}0.998 $\!\! \pm\!\!$ 0.001 & \cellcolor{gray!30}0.317 $\!\! \pm\!\!$ 0.008 & \cellcolor{gray!30}0.128 $\!\! \pm\!\!$ 0.011 & \cellcolor{gray!30}0.320 $\!\! \pm\!\!$ 0.011 & \cellcolor{gray!30}0.991 $\!\! \pm\!\!$ 0.000 & \cellcolor{gray!30}0.333 $\!\! \pm\!\!$ 0.008 & \cellcolor{gray!30}0.588 $\!\! \pm\!\!$ 0.043 & \cellcolor{gray!30}0.228 $\!\! \pm\!\!$ 0.016 & \cellcolor{gray!30}1.000 $\!\! \pm\!\!$ 0.000 & \cellcolor{gray!30}0.210 $\!\! \pm\!\!$ 0.040 \\
& &  4  & 0.110 $\!\! \pm\!\!$ 0.011  & 0.732 $\!\! \pm\!\!$ 0.006  & 0.998 $\!\! \pm\!\!$ 0.001  & 0.707 $\!\! \pm\!\!$ 0.006  & 0.120 $\!\! \pm\!\!$ 0.016  & 0.321 $\!\! \pm\!\!$ 0.012  & 1.000 $\!\! \pm\!\!$ 0.000  & 0.305 $\!\! \pm\!\!$ 0.011  & 0.136 $\!\! \pm\!\!$ 0.015  & 0.334 $\!\! \pm\!\!$ 0.015  & 1.000 $\!\! \pm\!\!$ 0.000  & 0.327 $\!\! \pm\!\!$ 0.012  & 0.769 $\!\! \pm\!\!$ 0.153  & 0.168 $\!\! \pm\!\!$ 0.022  & 1.000 $\!\! \pm\!\!$ 0.000  & 0.211 $\!\! \pm\!\!$ 0.033 \\
\bottomrule
\end{tabular}
}
\label{table:triggersize}
\end{table*}

\subsection{Trigger Trajectory}
\label{section:trajectory}

During the distillation process, as we update the backdoor trigger based on the model parameters during the distillation process, these triggers should be different theoretically at different distilled epochs.
We collect all the generated backdoor triggers from different distilled epochs.
We use the distilled images to train the downstream model after the distillation and test all the trigger images we collect.
We find that the {\em ASR} scores from these triggers can also achieve similar results as the results after the distillation.
For example, the triggers generated in the 10 distilled epochs lead to a {\em ASR} score of 1.000 on the backdoored model trained on 400 distilled-epoch images of $\langle \DD$, AlexNet, CIFAR10$\rangle$.
This reality is actually an advantage of our attack, especially facing a defense like De-trigger that needs to know the exact triggers.
In our situation, we have numerous triggers from different distilled epochs, which makes the defense much harder.
This also means our distilled images contain information about the triggers in the distillation process, even though these triggers are different in each distilled epoch.
This trajectory during the training procedure can result in a more challenging trigger detection.
We name this phenomenon \emph{trigger trajectory}.

\mypara{Takeaways}
The trigger trajectory is a unique feature offered by \doorping.
It enables the attackers to record a set of triggers that can later be used to attack the downstream models.

\subsection{Value of Magnifying Factor \texorpdfstring{$\alpha$}{}}
\label{section:alpha}

\begin{figure*}[!t]
\centering
\includegraphics[width=1.9\columnwidth]{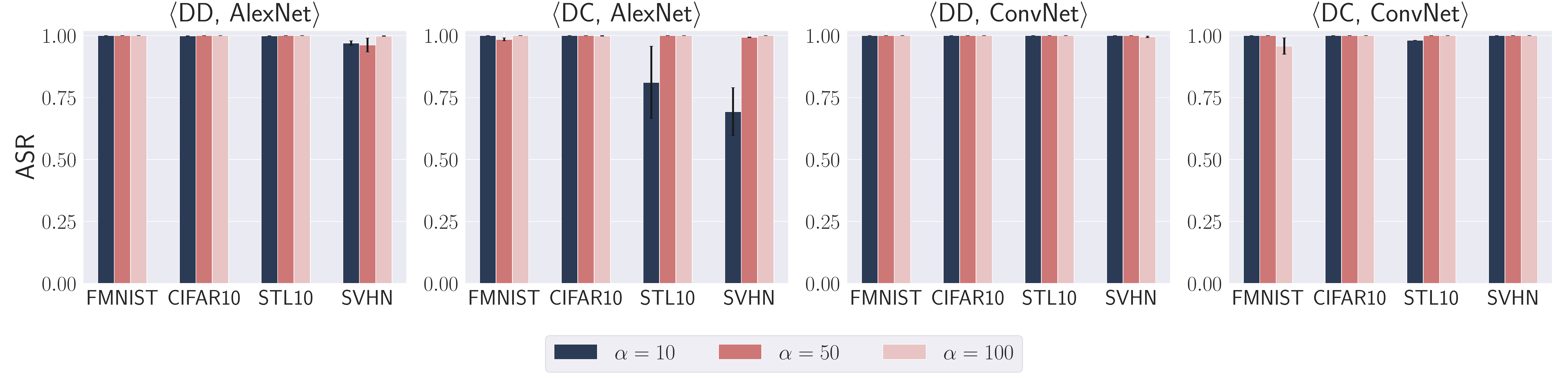}
\caption{{\em ASR} of \doorping with different $\alpha$.}
\label{figure:alphaasr}
\end{figure*}

\begin{figure*}[!t]
\centering
\includegraphics[width=1.9\columnwidth]{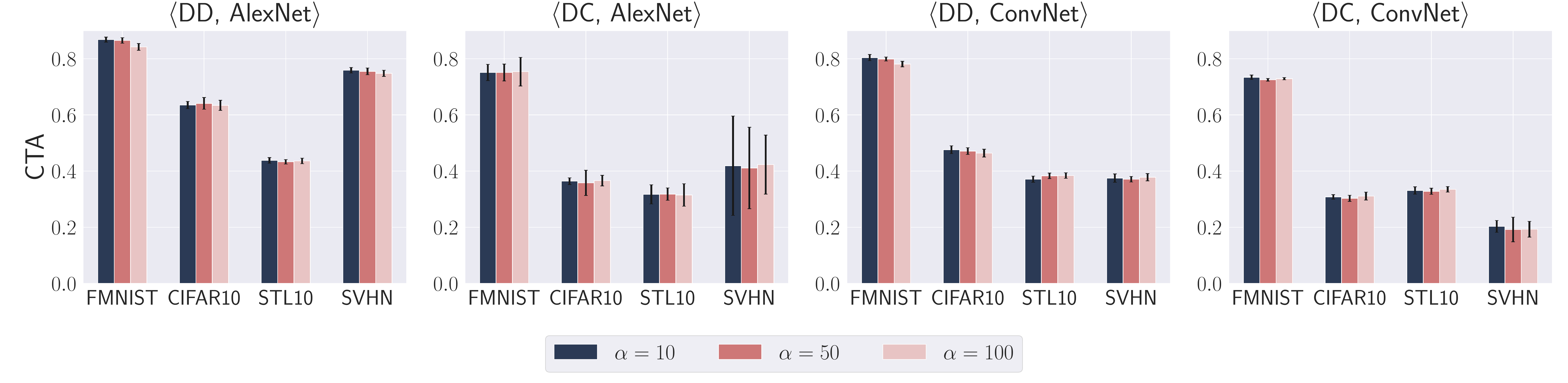}
\caption{{\em CTA} of \doorping with different $\alpha$.}
\label{figure:alphacta}
\end{figure*}

We calculate the MSE Loss between the output and the output magnified by $\alpha$.
Thus, we should know how $\alpha$ acts on the optimized trigger.
Theoretically, the larger $\alpha$ is, the better the triggers are during the distillation process.
Previous works~\cite{LMALZWZ18,LXZZZ20} set the magnifying value to a specific constant, 100, whereas the optimizer will allow the outputs from the selected neurons in the penultimate larger closer to this number.
However, the weakness of setting constants is that the optimizer will not work when the output itself is close to 100.
To solve this problem, our method chooses to magnify the output to $\alpha$ times so that the triggers will always substantially affect these selected neurons.
In particular, we choose $\alpha$ from 10, 50, and 100.
\autoref{figure:alphaasr} and~\autoref{figure:alphacta} reports the results among different $\alpha$.
We can see from~\autoref{figure:alphaasr}, the {\em ASR} scores are almost close to 1.000 with the alpha increasing, and the utilities are stable in~\autoref{figure:alphacta}.
In our experiments, we simply set it to 10.

\mypara{Takeways}
The {\em ASR} increases with the increasing magnifying factor $\alpha$. 
However, the attack performance plateaus after $\alpha$ is greater than 50.

\section{Defenses}
\label{section:defenses}

To mitigate the threat of backdoor attacks, many defense mechanisms have been proposed in the literature. 
These defenses can be broadly categorized into three detection levels~\cite{XWLBGL21}, i.e., \textit{model-level} (if a model is backdoored)~\cite{LLTMAZ19,LLKLLM212,WYSLVZZ19}, \textit{input-level} (if the test time input contains triggers)~\cite{CJOK20,K21,GXWCRN19}, and \textit{dataset-level} (if a training dataset is backdoored)~\cite{TWTZ21,TLM18,HKSO21}.
In this section, we evaluate if our attacks can be defended by the existing mechanisms at all three levels. 
For each detection level, we select three representative approaches.
Note that we only evaluate \doorping here due to its good attack performance (see~\autoref{section:experiments}).

\subsection{Model-Level Defense}

\mypara{ABS~\cite{LLTMAZ19}}
ABS analyzes the inner neuron behavior by determining how the output activation changes when introducing different levels of stimulation to a neuron.
The neurons that substantially elevate the activation of a particular output label, regardless of the input, are considered potentially compromised.
We apply ABS to identify these neurons in the backdoored models. 
For all experiments, ABS does not identify any backdoor neurons or layers for all the models.
All the compromised neuron candidate lists are empty. 
We conclude that ABS cannot defend \doorping.

\mypara{Neural Attention Distillation (NAD)~\cite{LLKLLM212}}
NAD is an architecture to erase backdoors from backdoored models.
It utilizes a teacher model to fine-tune the backdoored student model using a small subset of clean data. 
In this way, the intermediate-layer attention of the student model aligns with that of the teacher model.
The backdoor is then effectively removed.
In our experiments, we choose a subset of the clean dataset with a proportion of 0.050 and a clean model trained by the clean dataset as our teacher model.
We report the {\em ASR} and {\em CTA} scores after the fine-tuning process in~\autoref{table:nad}.
We can clearly see that all the fine-tuned models classify the input into one specific class.
This behavior leads to low {\em CTA} scores ($\sim$0.100) and {\em ASR} scores of 1.000 or 0.000.
Our results show that NAD is not an effective defense against \doorping either.

\begin{table}[!t]
\centering
\customTableFont
\setlength{\tabcolsep}{3 pt}
\caption{{\em ASR} and {\em CTA} of \doorping backdoored models after NAD process.}
\scalebox{0.95}{
\begin{tabular}{c | c c | c c | c c | c c}
\toprule
& \multicolumn{4}{c| }{AlexNet} & \multicolumn{4}{c}{ConvNet}\\
& \multicolumn{2}{c| }\DD & \multicolumn{2}{c| }\DC & \multicolumn{2}{c| }\DD & \multicolumn{2}{c }\DC\\
& {\em ASR} & {\em CTA} & {\em ASR} & {\em CTA} & {\em ASR} & {\em CTA} & {\em ASR} & {\em CTA}\\
\midrule
{\bf FMNIST} & 0.000 & 0.100 & 1.000 & 0.100 & 1.000 & 0.100 & 1.000 & 0.100\\
{\bf CIFAR10} & 0.000 & 0.100 & 1.000 & 0.100 & 1.000 & 0.100 & 1.000 & 0.100\\
{\bf STL10} & 1.000 & 0.100 & 1.000 & 0.100 & 1.000 & 0.100 & 1.000 & 0.100\\
{\bf SVHN} &  1.000 & 0.067 & 1.000 & 0.067 & 1.000 & 0.067 & 1.000 & 0.067\\
\bottomrule
\end{tabular}
}
\label{table:nad}
\end{table}

\mypara{Neural Cleanse~\cite{WYSLVZZ19}}
Neural Cleanse generates \emph{Anomaly Index} of neuron units for a given classifier.
If the Anomaly Index is greater than 2, the classifier is considered a backdoored model.
We adopt the default parameter settings of Neural Cleanse and use the original testing dataset as a clean dataset in the evaluation.
\autoref{table:neuralcleanse} reports the Anomaly Index produced by Neural Cleanse for \doorping.
We can see that all the Anomaly Indices are consistently smaller than 2. 
It indicates that Neural Cleanse cannot detect our backdoor attacks in the distilled classifiers.

\begin{table}[!t]
\centering
\customTableFont
\setlength{\tabcolsep}{3 pt}
\caption{Anomaly Indices produced by Neural Cleanse for \doorping. 
A classifier is predicted to be backdoored if the Anomaly Index is larger than 2.}
\scalebox{0.95}{
\begin{tabular}{c | c | c | c | c}
\toprule
& \multicolumn{2}{c |}{AlexNet} & \multicolumn{2}{c}{ConvNet}\\
& \DD & \DC & \DD & \DC\\
\midrule
{\bf FMNIST} & 1.466 & 1.670 & 1.304 & 0.995\\
{\bf CIFAR10}& 1.338 & 0.919 & 0.745 & 1.895\\
{\bf STL10}& 0.879 & 0.676 & 0.676 & 1.218\\
{\bf SVHN}& 1.835 & 1.908 & 1.266 & 0.739\\
\bottomrule
\end{tabular}
}
\label{table:neuralcleanse}
\end{table}

\subsection{Input-Level Defense}
\label{section:input_level_defense}

\mypara{De-noising Autoencoder~\cite{CJOK20}}
De-noising Autoencoder builds a deep autoencoder model by learning from paired clean images and their counterparts with added Gaussian noise.
Upon the model being trained, De-noising Autoencoder removes the noise (i.e., the trigger) of the model input by feeding them to the autoencoder model, as we believe some triggers are recognized as noise in \naive and \doorping.
We follow the same procedure outlined in~\cite{CJOK20} to train the autoencoder in our experiments and then use the autoencoder to remove the noise of our backdoored distilled images.
We evaluate the {\em ASR} and {\em CTA} of the backdoored model at the test time (i.e., the test time samples are first filtered by De-noising Autoencoder).
As we can see in~\autoref{table:denoising_naive} and~\autoref{table:denoising}, most of {\em ASR} scores decrease.
However, the {\em CTA} score also drops significantly in most cases.
For example, the {\em CTA} score is only 0.191, which is lower than the original model utility of 0.654 of $\langle \DD$, AlexNet, CIFAR10$\rangle$.
The reason is that De-noising Autoencoder may also remove helpful information from the input images besides the trigger.
There is a clear utility-defense trade-off when applying De-noising Autoencoder.

\begin{table}[!t]
\centering
\customTableFont
\setlength{\tabcolsep}{3 pt}
\caption{{\em ASR} and {\em CTA} of \naive backdoored models after De-noising Autoencoder process.}
\scalebox{0.95}{
\begin{tabular}{c| c c| c c| c c| c c}
\toprule
& \multicolumn{4}{c| }{AlexNet} & \multicolumn{4}{c}{ConvNet}\\
& \multicolumn{2}{c| }\DD & \multicolumn{2}{c| }\DC & \multicolumn{2}{c| }\DD & \multicolumn{2}{c }\DC\\
& {\em ASR} & {\em CTA} & {\em ASR} & {\em CTA} & {\em ASR} & {\em CTA} & {\em ASR} & {\em CTA}\\
\midrule
{\bf FMNIST} & 0.050 & 0.414 & 0.000 & 0.669 & 0.207 & 0.504 & 0.059 & 0.660\\
{\bf CIFAR10} & 0.098 & 0.279 & 0.000 & 0.337 & 0.320 & 0.261 & 0.008 & 0.284\\
{\bf STL10} & 0.011 & 0.305 & 0.000 & 0.289 & 0.263 & 0.275 & 0.004 & 0.119\\
{\bf SVHN} &  0.000 & 0.347 & 0.000 & 0.484 & 0.000 & 0.143 & 0.000 & 0.133\\
\bottomrule
\end{tabular}
}
\label{table:denoising_naive}
\end{table}

\begin{table}[!t]
\centering
\customTableFont
\setlength{\tabcolsep}{3 pt}
\caption{{\em ASR} and {\em CTA} of \doorping backdoored models after De-noising Autoencoder process.}
\scalebox{0.95}{
\begin{tabular}{c| c c| c c| c c| c c}
\toprule
& \multicolumn{4}{c| }{AlexNet} & \multicolumn{4}{c}{ConvNet}\\
& \multicolumn{2}{c| }\DD & \multicolumn{2}{c| }\DC & \multicolumn{2}{c| }\DD & \multicolumn{2}{c }\DC\\
& {\em ASR} & {\em CTA} & {\em ASR} & {\em CTA} & {\em ASR} & {\em CTA} & {\em ASR} & {\em CTA}\\
\midrule
{\bf FMNIST} & 0.000 & 0.615 & 0.000 & 0.679 & 0.000 & 0.486 & 0.001 & 0.653\\
{\bf CIFAR10} & 0.000 & 0.264 & 0.013 & 0.344 & 0.081 & 0.285 & 0.000 & 0.287\\
{\bf STL10} & 0.000 & 0.334 & 0.000 & 0.297 & 1.000 & 0.283 & 0.000 & 0.262\\
{\bf SVHN} &  0.000 & 0.413 & 0.028 & 0.261 & 0.385 & 0.073 & 0.857 & 0.153\\
\bottomrule
\end{tabular}
}
\label{table:denoising}
\end{table}

\mypara{De-trigger Autoencoder~\cite{K21}}
Similar to De-noising Autoencoder, De-trigger Autoencoder learns from both clean images and clean images with the trigger to reconstruct the clean images.
The defenders must know the trigger information (i.e., pattern and location) to train a De-trigger Autoencoder.
All the testing procedures are the same as we outline in De-noising Autoencoder.
We report the results in~\autoref{table:detrigger}.
All {\em ASR} scores and {\em CTA} scores decrease sharply. 
The majority are even worse than the result of De-noising Autoencoder.
In conclusion, De-trigger Autoencoder cannot defend the \doorping as it suffers from the same utility-defense trade-off.

\begin{table}[!t]
\centering
\customTableFont
\setlength{\tabcolsep}{3 pt}
\caption{{\em ASR} and {\em CTA} of \doorping backdoored models after De-trigger Autoencoder process.}
\scalebox{0.95}{
\begin{tabular}{c| c c| c c| c c| c c}
\toprule
& \multicolumn{4}{c| }{AlexNet} & \multicolumn{4}{c}{ConvNet}\\
& \multicolumn{2}{c| }\DD & \multicolumn{2}{c| }\DC & \multicolumn{2}{c| }\DD & \multicolumn{2}{c }\DC\\
& {\em ASR} & {\em CTA} & {\em ASR} & {\em CTA} & {\em ASR} & {\em CTA} & {\em ASR} & {\em CTA}\\
\midrule
{\bf FMNIST} & 0.039 & 0.508 & 0.049 & 0.574 & 0.003 & 0.287 & 0.002 & 0.390\\
{\bf CIFAR10} & 0.145 & 0.191 & 0.282 & 0.202 & 0.066 & 0.154 & 0.088 & 0.165\\
{\bf STL10} & 0.169 & 0.144 & 0.075 & 0.203 & 0.000 & 0.100 & 0.264 & 0.161\\
{\bf SVHN} & 0.122 & 0.143 & 0.260 & 0.197 & 0.065 & 0.195 & 0.126 & 0.133\\
\bottomrule
\end{tabular}
}
\label{table:detrigger}
\end{table}

\mypara{STRIP~\cite{GXWCRN19}}
STRIP filters triggered samples at the test time based on the predicted randomness of perturbated samples (i.e., by applying different image patterns to suspicious images).
Its detection capability is assessed by two metrics: false rejection rate (FRR) and false acceptance rate (FAR).
The FRR is the probability when the benign input is regarded as a backdoored input by the STRIP detection system.
The FAR is the probability that the backdoored input is recognized as the benign input by the STRIP detection system.
A detection system usually attempts to minimize the FAR while using a slightly higher FRR as the trade-off.
STRIP algorithm chooses the detection threshold by using the percent point function (PPF) on the distribution of the entropy of benign samples. 

\begin{figure}[!t]
\centering
\includegraphics[width=0.6\columnwidth]{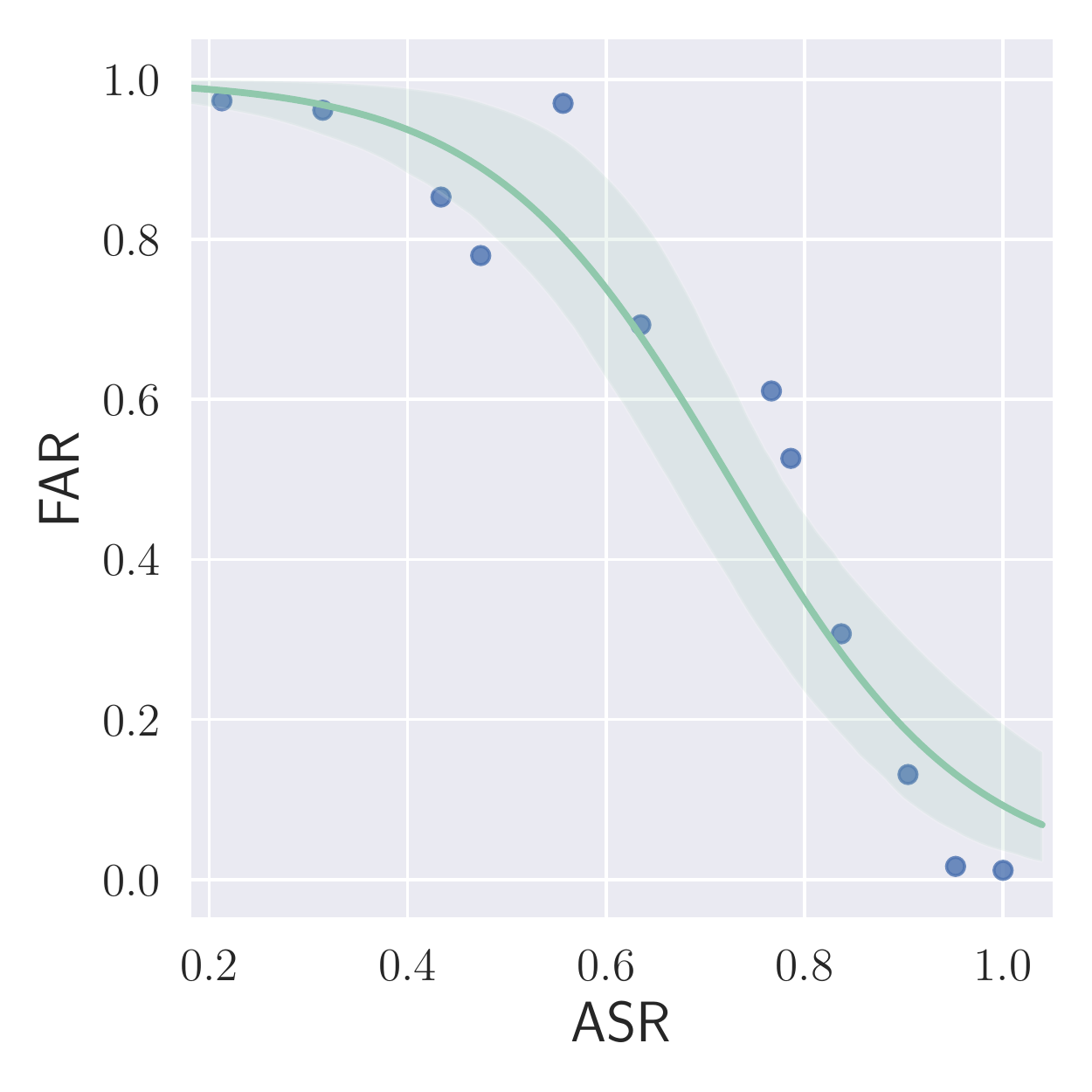}
\caption{Relationship between the {\em ASR} and FAR.}
\label{figure:relationshipasrfar}
\end{figure}

We use STRIP to check if the defender can use it to identify triggered samples in the test data.
\autoref{table:strip_test} reports the FRR and FAR scores of STRIP detecting the testing dataset (clean and backdoor).
Here, we add the trigger to 2,000 images in the testing dataset and employ another 2,000 as benign ones.
10 images are employed as the overlay samples, which are used for replicating with the inputs to measure the randomness (entropy) of predicted labels.
As we can observe in~\autoref{table:strip_test}, STRIP can achieve good detection performance for the testing images with triggers, i.e., both FRR and FAR is close to 0.
In light of this finding, we further investigate why STRIP performs so well and how to reduce its detection performance from the perspective of an attacker.
Recall that the critical insight of STRIP is that the predictions of all perturbed inputs of triggered images tend to be always consistent (i.e., the target class).
In other words, the high detection performance, as shown in~\autoref{table:strip_test}, indicates that our optimized triggers can be stably preserved in the perturbed images.
Crucially, our \doorping attack enables the attacker to keep a trigger trajectory (see~\autoref{section:trajectory}) whereby different triggers are preserved. 
Instead of using the final optimized trigger, we test if other triggers along the trajectory can be employed to find a balance point between attack and detection performance.
We show the relationship between the {\em ASR} and FAR in~\autoref{figure:relationshipasrfar}.
As we can see, the FAR score of STRIP can be significantly increased (i.e., poor detection performance) when we apply triggers that lead to a suboptimal {\em ASR}.
For example, when the FAR score is around 0.595, the {\em ASR} score is about 0.767, attaining a decent attack performance.
Our finding indicates that the \doorping attack can practically evade STRIP detection by trading off some attack performance. 

\begin{table}[!t]
\centering
\customTableFont
\setlength{\tabcolsep}{3 pt}
\caption{FRR and FAR of STRIP detecting test samples (clean and backdoor).
We add the trigger into 40\% of the original testing dataset and use another 40\% as the benign samples.
10 images are treated as the overlay indices to evaluate FRR and FAR.}
\scalebox{0.95}{
\begin{tabular}{c | c c | c c | c c | c c}
\toprule
& \multicolumn{4}{c| }{AlexNet} & \multicolumn{4}{c}{ConvNet}\\
& \multicolumn{2}{c| }{\DD} & \multicolumn{2}{c|}{\DC} & \multicolumn{2}{c| }{\DD} & \multicolumn{2}{c}{\DC}\\
& FRR & FAR & FRR & FAR & FRR & FAR & FRR & FAR\\
\midrule
{\bf FMNIST} & 0.000 & 0.000 & 0.000 & 1.000 & 0.000 & 0.000 & 0.000 & 1.000\\
{\bf CIFAR10}& 0.000 & 0.015 & 0.013 & 0.004 & 0.000 & 0.005 & 0.012 & 0.000\\
{\bf STL10}& 0.015 & 0.000 & 0.023 & 0.000 & 0.000 & 0.000 & 0.006 & 0.000\\
{\bf SVHN}& 0.015 & 0.000 & 0.016 & 0.024 & 0.020 & 0.110 & 0.016 & 0.000\\
\bottomrule
\end{tabular}
}
\label{table:strip_test}
\end{table}

\subsection{Dataset-Level Defense}

\mypara{Statistical Analysis of DNNs (SCAn)~\cite{TWTZ21}}
SCAn leverages an Expectation-Maximization (EM) algorithm to decompose an image into its identity part (e.g., person) and variation part (e.g., poses).
Based on the global information of all categories, the distribution of variations is exploited by a likelihood ratio test to analyze the representations in each category, identifying those that are more likely to be described by a mixture model by adding attack samples into legitimate images of the current category.
When the test statistic of the class $T$ (denoted as $J^{*}_{T}$) is larger than 7.389, the class $T$ is reported as being contaminated.
Here, 7.389 is actually $e^{2}$ determined by SCAn.
\autoref{table:scan} reports the test statistic for our backdoor target class (class 0).
As we can see in~\autoref{table:scan}, none of the $J^{*}_{T}$ scores are larger than 7.389. 
The results show that SCAn cannot detect our backdoor target class effectively by \doorping.

\begin{table}[!t]
\centering
\customTableFont
\setlength{\tabcolsep}{3 pt}
\caption{$J^{*}_{T}$ of the target class from different model architectures and datasets by SCAn.}
\scalebox{0.95}{
\begin{tabular}{c | c | c | c | c}
\toprule
& \multicolumn{2}{c | }{AlexNet} & \multicolumn{2}{c}{ConvNet}\\
& \DD & \DC & \DD & \DC\\
\midrule
{\bf FMNIST} & 1.868 & 2.224 & 6.817 & 7.075\\
{\bf CIFAR10} & 0.573 & 1.003 & 0.074 & 2.879\\
{\bf STL10} & 0.283 & 0.270 & 0.736 & 3.424\\
{\bf SVHN} & 0.671 & 0.015 & 0.954 & 0.226\\
\bottomrule
\end{tabular}
}
\label{table:scan}
\end{table}

\mypara{Spectral Signature~\cite{TLM18}}
Spectral signature builds on top of the idea that a classifier amplifies signals that are critical to classification.
It finds that backdoored training datasets used in backdoor attacks can leave detectable traces in the covariance spectrum of the feature representation, i.e., the clean sample leads to a small covariance value.
In contrast, the backdoor sample leads to an immense covariance value.
Spectral signature calculates the outlier score of each sample and the mean value for the backdoor and clean samples.
The results are shown in~\autoref{table:spectralsignature}.
As we can see in~\autoref{table:spectralsignature}, most of the average outlier scores of backdoor samples are smaller than the clean ones.
As such, Spectral Signature cannot detect the backdoored distilled datasets generated by \doorping attack.

\begin{table}[!t]
\tiny
\centering
\customTableFont
\setlength{\tabcolsep}{2 pt}
\caption{Average outlier score of samples generated by Spectral Signature on \doorping.
The smaller the score is, the more likely the sample is clean.}
\scalebox{0.95}{
\begin{tabular}{c | c c | c c | c c | c c}
\toprule
& \multicolumn{4}{c| }{AlexNet} & \multicolumn{4}{c}{ConvNet}\\
& \multicolumn{2}{c| }{\DD} & \multicolumn{2}{c|}{\DC} & \multicolumn{2}{c| }{\DD} & \multicolumn{2}{c}{\DC}\\
& Backdoor & Clean & Backdoor & Clean & Backdoor & Clean & Backdoor & Clean\\
\midrule
{\bf FMNIST} & 7.623 & 9.476 & 4.584 & 2.203 & 6.829 & 9.077 & 5.411 & 2.883\\
{\bf CIFAR10}& 7.022 & 10.247 & 1.088 & 2.447 & 4.753 & 5.713 & 11.046 & 11.680\\
{\bf STL10}& 5.153 & 5.620 & 9.951 & 8.720 & 4.466 & 4.055 & 2.572 & 4.092\\
{\bf SVHN}& 6.416 & 10.972 & 4.198 & 15.856 & 5.131 & 9.406 & 23.041 & 10.793\\
\bottomrule
\end{tabular}
}
\label{table:spectralsignature}
\end{table}

\begin{figure*}[!t]
\centering
\includegraphics[width=1.9\columnwidth]{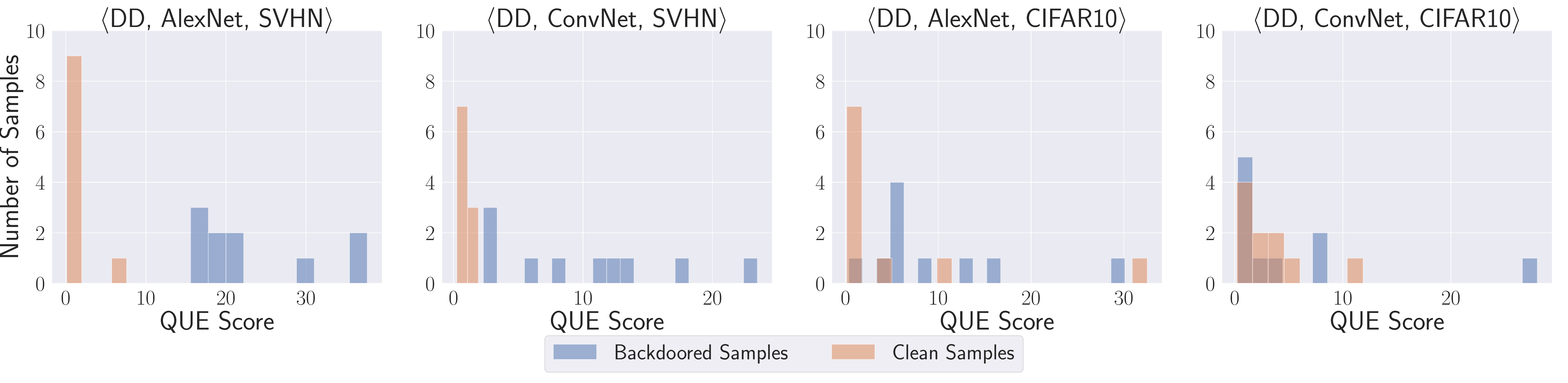}
\caption{QUE scores of \doorping for SVHN and CIFAR10 dataset by using \DD algorithm.}
\label{figure:que}
\end{figure*}

\mypara{SPECTRE~\cite{HKSO21}}
SPECTRE is a defense algorithm using robust covariance estimation to amplify the spectral signature of backdoored data in the training dataset. 
The mean QUantum Entropy (QUE) score of a backdoor sample is usually higher than the clean sample. 
SPECTRE then marks such backdoor samples with a robust spectral signature.
In the original settings, SPECTRE detects the backdoor sample in each class.
For the \doorping attack, all backdoor images are included in the target class we pre-defined. 
It means that there are no backdoor images in the other classes.
To this end, we modify the settings of SPECTRE and only detect the backdoor samples in the target class.
We include an equal number of backdoored and clean distilled images in the target class.
\autoref{table:spectre} reports the accuracy of the SPECTRE detection.
We can observe that SPECTRE performs well in some cases. 
For example, given $\langle \DD$, SVHN$\rangle$, SPECTRE can detect the triggers by \doorping.
However, SPECTRE can not successfully identify the triggers in the other datasets.
To better understand the root cause, we plot the QUE Score of SVHN compared with CIFAR10 in~\autoref{figure:que}.
As shown in~\autoref{figure:que}, SPECTRE can separate the clean and backdoored samples given the SVHN dataset.
However, the robust statistics used by SPECTRE can not separate the backdoored samples from the clean ones given the CIFAR10 dataset. 
A similar trend can be found in STL10 and FMNIST.
The signature of the backdoored samples is amplified effectively by SPECTRE.
However, in some other cases, it works poorly. 
For example, given STL10, the accuracy scores of SPECTRE are no greater than 50\% in all cases.
We conclude that SPECTRE is not robust for all of the datasets we test.
This inconsistency indicates that SPECTRE is not a reliable defense mechanism against our \doorping attack.

\begin{table}[!t]
\centering
\customTableFont
\setlength{\tabcolsep}{3 pt}
\caption{Accuracy of detecting backdoor samples by using SPECTRE.}
\scalebox{0.95}{
\begin{tabular}{c | c | c | c | c}
\toprule
& \multicolumn{2}{c| }{AlexNet} & \multicolumn{2}{c}{ConvNet}\\
& \DD & \DC & \DD & \DC\\
\midrule
{\bf FMNIST} & 60\% & 60\% & 90\% & 50\% \\
{\bf CIFAR10}& 80\% & 70\% & 50\% & 40\% \\
{\bf STL10}& 20\% & 50\% & 40\% & 50\% \\
{\bf SVHN}& 100\% & 50\% & 100\% & 70\% \\
\bottomrule
\end{tabular}
}
\label{table:spectre}
\end{table}

\section{Related Work}
\label{section:related_work}

\mypara{Backdoor Attack}
Backdoor attack~\cite{CLLLS17,GDG17,LMALZWZ18,GTXCSSMLG22,KS21} is a training time attack and has emerged as a major security threat to deep neural networks (DNNs) in many application areas (e.g., natural language processing~\cite{CSCBMSWZ21,SSTS21}, image classification~\cite{DLZL21,DLL21}, face recognition~\cite{CLLLS17}, point clouds~\cite{XMCLK21,LCZTZLZ21}, etc.).
It implants a hidden backdoor (also called neural trojan~\cite{KS21,LMALZWZ18}) into the target model via poisoning training samples (i.e., attacker modified input-label pairs). 
The injected backdoor can be activated during inference time if an attacker-specific trigger (either pre-defined or optimization-based) is presented. 
Previous works mainly focus on the effectiveness of backdoor attacks on DNN-based classifiers~\cite{CLLLS17,GDG17}, graph neural networks~\cite{ZJWG21,XPJW21}, pre-trained encoders~\cite{JLG22,SJZLCSFYW21}, contrastive learning-based models~\cite{CT22}, transfer learning~\cite{YLZZ19}, etc.
In recent years, many efforts also adopt the concepts and techniques in adversarial examples~\cite{ISS14,FGCGCG21} to improve the stealthiness of the triggers and make them imperceptible to human moderators~\cite{DLZL21,DLL21,LLWLHL21}.
Furthermore, previous works mainly inject triggers to the original training dataset during the model training procedure, which cannot be applied to the distilled datasets as aforementioned.
Thus, we take the first step to inject triggers into the synthetic data during the dataset distillation process.

\mypara{Defense Against Backdoor Attacks}
Defense mechanisms against backdoor attacks~\cite{KS21,GTXCSSMLG22,LWJLX20} can be broadly grouped into two categories.
The first category of defense mechanisms is identifying backdoored data samples and filtering them out before training a model.
Their central intuition is that the backdoored data samples, due to the manipulation from attackers, are statistically different from non-backdoored counterparts either in the input space~\cite{SKL17,LXS17,DK19,DAR20} or in the feature space~\cite{TLM18,PGHFZFGD20,KL17}.
The second category of defense mechanisms orbits around the models.
Given the assumption that the model holders cannot pre-filter the training data, these mechanisms secure the models by eliminating the triggers at the training/test time~\cite{LXS17,DJS19,GXWCRN19,SAKNT19}, certifying their robustness to input perturbations~\cite{WCG20,ZJWG21,RWRK20,XCCL21}, identifying backdoored models~\cite{WYSLVZZ19,WZLCXW20,ZZWGC21}, removing the backdoors from the backdoored models~\cite{WYSLVZZ19,LLKLLM21,LKLLLM21,LLWL21}, etc.
We refer the audience to~\cite{KS21,GTXCSSMLG22,LWJLX20} for comprehensive surveys on backdoor attacks and defenses.
Our experimental results indicate that existing defense mechanisms provide insufficient robustness guarantees under \doorping.

\mypara{Dataset Distillation}
Dataset distillation~\cite{WZTE18,ZMB21,ZB21,NNXL21,NCL21,ZB212,CWTEZ22} is a technique for data-efficient learning, which does not rely on large datasets.
The first work of dataset distillation~\cite{WZTE18} calculates the loss gradient from a model trained by the distilled dataset.
Some other works related to Dataset Condensation~\cite{ZMB21,ZB21} are proposed to improve the quality of the distilled dataset.
These works match the gradient of the original training dataset with distilled datasets to achieve similar performance.
They also use differentiable siamese augmentation~\cite{ZB21} to improve the result but not much.
Zhao and Bilen~\cite{ZB212} then provide a method for minimizing the distribution discrepancy between real and synthetic data in these sampled embedding spaces.
KIP~\cite{NNXL21,NCL21} is another method using large-scale Neural Tangent Kernel computation.
Another work~\cite{CWTEZ22} uses the trajectory of pre-trained models and matches the parameters from a select model and the model trained by distilled dataset.
Nevertheless, this work has such a tremendous learning rate (as large as 1000) for updating distilled images that the matching loss will become NaN for many situations.
Note that the model architecture used in the dataset distillation processing must be the same as the downstream model architecture, which is required by most current dataset distillation techniques.

\section{Limitation}
\label{section:discussion}

In this section, we discuss our attack limitations in two aspects.
The first is the limitations of \DD and \DC themselves.
For \DD and \DC, neither work can utilize the model with the BatchNorm (BN) layer as an upstream model.
In fact, the BN layer is one of the most widely used layers in neural networks to accelerate convergence and avoid loss into NaN.
This drawback vastly limits the choice of upstream models.
Besides, for \DD, it is hard for the users to distill an extensive dataset.
For example, the loss becomes NaN when distilling large datasets such as SVHN (which contains over 70,000 samples).
The second aspect is how they limit our attacks.
We present that the attack cannot be deployed in a federated learning environment.
The root cause is that both \DD and \DC cannot be trivially deployed in collaborative systems since they re-initialize the model parameters in every epoch.
For different samples in different clients, the results differ significantly.
Simply combining the distilled datasets or model parameters from the clients is impracticable.
Note that there are preliminary efforts in federated dataset distillation~\cite{HGMZLL22,XWCYH22,SLCFTSK22}.
We consider backdoor attacks against these federated systems as our future research direction.

\section{Conclusion}

In this paper, we propose the first backdoor attack against the machine learning models via a malicious dataset distillation service provider. 
We inject triggers into the synthetic data during the distillation process rather than during the model training phase, where all previous attacks are performed.
Immense evaluations are conducted on multiple datasets, architectures, and dataset distillation techniques. 
Our results demonstrate that our proposed attacks achieve remarkable attack and utility performance.
We hope this study highlights the need to understand the security and privacy issues of dataset distillation, especially the consequences of using distilled datasets from third parties.

\mypara{Acknowledgments}
The authors wish to thank Shaofeng Li and Tian Dong for their valuable discussions and feedback.
This work is partially funded by the Helmholtz Association within the project ``Trustworthy Federated Data Analytics'' (TFDA) (funding number ZT-I-OO1 4) and by the European Health and Digital Executive Agency (HADEA) within the project ``Understanding the individual host response against Hepatitis D Virus to develop a personalized approach for the management of hepatitis D'' (D-Solve).

\bibliographystyle{plain}
\begin{small}
\bibliography{simple_generated_py3}    

\begin{thebibliography}{10}

\bibitem{CIFAR}
\url{https://www.cs.toronto.edu/~kriz/cifar.html}.

\bibitem{BNL12}
Battista Biggio, Blaine Nelson, and Pavel Laskov.
\newblock {Poisoning Attacks against Support Vector Machines}.
\newblock In {\em {ICML}}, 2012.

\bibitem{c21}
Nicholas Carlini.
\newblock {Poisoning the Unlabeled Dataset of Semi-Supervised Learning}.
\newblock In {\em {USENIX Security}}, 2021.

\bibitem{CT22}
Nicholas Carlini and Andreas Terzis.
\newblock Poisoning and backdooring contrastive learning.
\newblock In {\em International Conference on Learning Representations (ICLR)},
  2022.

\bibitem{CWTEZ22}
George Cazenavette, Tongzhou Wang, Antonio Torralba, Alexei~A. Efros, and
  Jun-Yan Zhu.
\newblock {Dataset Distillation by Matching Training Trajectories}.
\newblock In {\em {CVPR}}, 2022.

\bibitem{CSBMSWZ21}
Xiaoyi Chen, Ahmed Salem, Michael Backes, Shiqing Ma, Qingni Shen, Zhonghai Wu,
  and Yang Zhang.
\newblock {BadNL: Backdoor Attacks Against NLP Models with Semantic-preserving
  Improvements}.
\newblock In {\em {ACSAC}}, pages 554--569, 2021.

\bibitem{CSCBMSWZ21}
Xiaoyi Chen, Ahmed Salem, Dingfan Chen, Michael Backes, Shiqing Ma, Qingni
  Shen, Zhonghai Wu, and Yang Zhang.
\newblock Badnl: Backdoor attacks against nlp models with semantic-preserving
  improvements.
\newblock In {\em Annual Computer Security Applications Conference}, pages
  554--569, 2021.

\bibitem{CLLLS17}
Xinyun Chen, Chang Liu, Bo~Li, Kimberly Lu, and Dawn Song.
\newblock Targeted backdoor attacks on deep learning systems using data
  poisoning.
\newblock {\em arXiv preprint arXiv:1712.05526}, 2017.

\bibitem{CJOK20}
Seungju Cho, Tae~Joon Jun, Byungsoo Oh, and Daeyoung Kim.
\newblock {DAPAS: Denoising autoencoder to prevent adversarial attack in
  semantic segmentation}.
\newblock {\em {International Joint Conference on Neural Networks}}, 2020.

\bibitem{CNL11}
Adam Coates, Andrew~Y. Ng, and Honglak Lee.
\newblock {An Analysis of Single-Layer Networks in Unsupervised Feature
  Learning}.
\newblock In {\em {AISTATS}}, pages 215--223, 2011.

\bibitem{DCLT19}
Jacob Devlin, Ming{-}Wei Chang, Kenton Lee, and Kristina Toutanova.
\newblock {{BERT:} Pre-training of Deep Bidirectional Transformers for Language
  Understanding}.
\newblock In {\em {NAACL-HLT}}, pages 4171--4186, 2019.

\bibitem{DK19}
Ilias Diakonikolas and Daniel~M Kane.
\newblock Recent advances in algorithmic high-dimensional robust statistics.
\newblock {\em arXiv preprint arXiv:1911.05911}, 2019.

\bibitem{DAR20}
Bao~Gia Doan, Ehsan Abbasnejad, and Damith~C. Ranasinghe.
\newblock {Februus: Input Purification Defense Against Trojan Attacks on Deep
  Neural Network Systems}.
\newblock In {\em {ACSAC}}, pages 897--912, 2020.

\bibitem{DLL21}
Khoa Doan, Yingjie Lao, and Ping Li.
\newblock Backdoor attack with imperceptible input and latent modification.
\newblock In {\em Advances in Neural Information Processing Systems (NeurIPS)},
  volume~34, 2021.

\bibitem{DLZL21}
Khoa Doan, Yingjie Lao, Weijie Zhao, and Ping Li.
\newblock Lira: Learnable, imperceptible and robust backdoor attacks.
\newblock In {\em IEEE/CVF International Conference on Computer Vision (ICCV)},
  pages 11966--11976, 2021.

\bibitem{DJS19}
Min Du, Ruoxi Jia, and Dawn Song.
\newblock Robust anomaly detection and backdoor attack detection via
  differential privacy.
\newblock {\em arXiv preprint arXiv:1911.07116}, 2019.

\bibitem{FGCGCG21}
Liam Fowl, Micah Goldblum, Ping{-}Yeh Chiang, Jonas Geiping, Wojtek Czaja, and
  Tom Goldstein.
\newblock {Adversarial Examples Make Strong Poisons}.
\newblock In {\em {NeurIPS}}, 2021.

\bibitem{FJR15}
Matt Fredrikson, Somesh Jha, and Thomas Ristenpart.
\newblock {Model Inversion Attacks that Exploit Confidence Information and
  Basic Countermeasures}.
\newblock In {\em {CCS}}, pages 1322--1333, 2015.

\bibitem{GXWCRN19}
Yansong Gao, Change Xu, Derui Wang, Shiping Chen, Damith~C Ranasinghe, and
  Surya Nepal.
\newblock {STRIP: A Defence Against Trojan Attacks on Deep Neural Networks}.
\newblock In {\em {ACSAC}}, pages 113--125, 2019.

\bibitem{GTXCSSMLG22}
Micah Goldblum, Dimitris Tsipras, Chulin Xie, Xinyun Chen, Avi Schwarzschild,
  Dawn Song, Aleksander Madry, Bo~Li, and Tom Goldstein.
\newblock Dataset security for machine learning: Data poisoning, backdoor
  attacks, and defenses.
\newblock {\em IEEE Transactions on Pattern Analysis and Machine Intelligence},
  2022.

\bibitem{GSS15}
Ian Goodfellow, Jonathon Shlens, and Christian Szegedy.
\newblock {Explaining and Harnessing Adversarial Examples}.
\newblock In {\em {ICLR}}, 2015.

\bibitem{ISS14}
Ian~J Goodfellow, Jonathon Shlens, and Christian Szegedy.
\newblock Explaining and harnessing adversarial examples.
\newblock {\em arXiv preprint arXiv:1412.6572}, 2014.

\bibitem{GYMT21}
Jianping Gou, Baosheng Yu, Stephen~J Maybank, and Dacheng Tao.
\newblock Knowledge distillation: A survey.
\newblock {\em International Journal of Computer Vision}, 129(6):1789--1819,
  2021.

\bibitem{GDG17}
Tianyu Gu, Brendan Dolan-Gavitt, and Siddharth Grag.
\newblock {Badnets: Identifying Vulnerabilities in the Machine Learning Model
  Supply Chain}.
\newblock {\em {CoRR abs/1708.06733}}, 2017.

\bibitem{HKSO21}
Jonathan Hayase, Weihao Kong, Raghav Somani, and Sewoong Oh.
\newblock {SPECTRE: Defending against backdoor attacks using robust covariance
  estimation}.
\newblock In {\em {ICML}}, 2021.

\bibitem{HJBGZ21}
Xinlei He, Jinyuan Jia, Michael Backes, Neil~Zhenqiang Gong, and Yang Zhang.
\newblock {Stealing Links from Graph Neural Networks}.
\newblock In {\em {USENIX Security}}, pages 2669--2686, 2021.

\bibitem{HVD15}
Geoffrey~E. Hinton, Oriol Vinyals, and Jeffrey Dean.
\newblock {Distilling the Knowledge in a Neural Network}.
\newblock {\em {CoRR abs/1503.02531}}, 2015.

\bibitem{HYXRGWTM20}
Qingyong Hu, Bo~Yang, Linhai Xie, Stefano Rosa, Yulan Guo, Zhihua Wang, Niki
  Trigoni, and Andrew Markham.
\newblock {RandLA-Net: Efficient Semantic Segmentation of Large-Scale Point
  Clouds}.
\newblock In {\em {CVPR}}, pages 11105--11114, 2020.

\bibitem{HGMZLL22}
Shengyuan Hu, Jack Goetz, Kshitiz Malik, Hongyuan Zhan, Zhe Liu, and Yue Liu.
\newblock {FedSynth: Gradient Compression via Synthetic Data in Federated
  Learning}.
\newblock {\em {CoRR abs/2204.01273}}, 2022.

\bibitem{JLG22}
Jinyuan Jia, Yupei Liu, and Neil~Zhenqiang Gong.
\newblock {BadEncoder: Backdoor Attacks to Pre-trained Encoders in
  Self-Supervised Learning}.
\newblock In {\em {S\&P}}, 2022.

\bibitem{KS21}
Sara Kaviani and Insoo Sohn.
\newblock Defense against neural trojan attacks: A survey.
\newblock {\em Neurocomputing}, 423:651--667, 2021.

\bibitem{KL17}
Pang~Wei Koh and Percy Liang.
\newblock Understanding black-box predictions via influence functions.
\newblock In {\em International conference on machine learning}, pages
  1885--1894, 2017.

\bibitem{KSH12}
Alex Krizhevsky, Ilya Sutskever, and Geoffrey~E. Hinton.
\newblock {ImageNet Classification with Deep Convolutional Neural Networks}.
\newblock In {\em {NIPS}}, pages 1106--1114, 2012.

\bibitem{K21}
Hyun Kwon.
\newblock {Defending Deep Neural Networks against Backdoor Attack by Using
  De-trigger Autoencoder}.
\newblock {\em {IEEE Access}}, 2021.

\bibitem{LV15}
Bo~Li and Yevgeniy Vorobeychik.
\newblock {Scalable Optimization of Randomized Operational Decisions in
  Adversarial Classification Settings}.
\newblock In {\em {AISTATS}}, pages 599--607, 2015.

\bibitem{LXZZZ20}
Shaofeng Li, Minhui Xue, Benjamin Zi~Hao Zhao, Haojin Zhu, and Xinpeng Zhang.
\newblock {Invisible Backdoor Attacks on Deep Neural Networks via Steganography
  and Regularization}.
\newblock {\em {IEEE Transactions on Dependable and Secure Computing}}, 2020.

\bibitem{LCZTZLZ21}
Xinke Li, Zhirui Chen, Yue Zhao, Zekun Tong, Yabang Zhao, Andrew Lim, and
  Joey~Tianyi Zhou.
\newblock Pointba: Towards backdoor attacks in 3d point cloud.
\newblock In {\em IEEE/CVF International Conference on Computer Vision (ICCV)},
  pages 16492--16501, 2021.

\bibitem{LKLLLM21}
Yige Li, Nodens Koren, Lingjuan Lyu, Xixiang Lyu, Bo~Li, and Xingjun Ma.
\newblock Neural attention distillation: Erasing backdoor triggers from deep
  neural networks.
\newblock In {\em International Conference on Learning Representations}, 2021.

\bibitem{LLKLLM21}
Yige Li, Xixiang Lyu, Nodens Koren, Lingjuan Lyu, Bo~Li, and Xingjun Ma.
\newblock Anti-backdoor learning: Training clean models on poisoned data.
\newblock In {\em Advances in Neural Information Processing Systems}, 2021.

\bibitem{LLKLLM212}
Yige Li, Xixiang Lyu, Nodens Koren, Lingjuan Lyu, Bo~Li, and Xingjun Ma.
\newblock {Neural Attention Distillation: Erasing Backdoor Triggers from Deep
  Neural Networks}.
\newblock In {\em {ICLR}}, 2021.

\bibitem{LWJLX20}
Yiming Li, Baoyuan Wu, Yong Jiang, Zhifeng Li, and Shu-Tao Xia.
\newblock Backdoor learning: A survey.
\newblock {\em arXiv preprint arXiv:2007.08745}, 2020.

\bibitem{LLWLHL21}
Yuezun Li, Yiming Li, Baoyuan Wu, Longkang Li, Ran He, and Siwei Lyu.
\newblock Invisible backdoor attack with sample-specific triggers.
\newblock In {\em Proceedings of the IEEE/CVF International Conference on
  Computer Vision}, pages 16463--16472, 2021.

\bibitem{LJZWWLW19}
Xiang Ling, Shouling Ji, Jiaxu Zou, Jiannan Wang, Chunming Wu, Bo~Li, and Ting
  Wang.
\newblock {DEEPSEC: A Uniform Platform for Security Analysis of Deep Learning
  Model}.
\newblock In {\em {S\&P}}, pages 673--690, 2019.

\bibitem{LLWL21}
Xuankai Liu, Fengting Li, Bihan Wen, and Qi~Li.
\newblock Removing backdoor-based watermarks in neural networks with limited
  data.
\newblock In {\em 2020 25th International Conference on Pattern Recognition
  (ICPR)}, pages 10149--10156. IEEE, 2021.

\bibitem{LLTMAZ19}
Yingqi Liu, Wen-Chuan Lee, Guanhong Tao, Shiqing Ma, Yousra Aafer, and Xiangyu
  Zhang.
\newblock {ABS: Scanning Neural Networks for Back-Doors by Artificial Brain
  Stimulation}.
\newblock In {\em {CCS}}, pages 1265--1282, 2019.

\bibitem{LMALZWZ18}
Yingqi Liu, Shiqing Ma, Yousra Aafer, Wen-Chuan Lee, Juan Zhai, Weihang Wang,
  and Xiangyu Zhang.
\newblock {Trojaning Attack on Neural Networks}.
\newblock In {\em {NDSS}}, 2018.

\bibitem{LWHSZBCFZ22}
Yugeng Liu, Rui Wen, Xinlei He, Ahmed Salem, Zhikun Zhang, Michael Backes,
  Emiliano~De Cristofaro, Mario Fritz, and Yang Zhang.
\newblock {ML-Doctor: Holistic Risk Assessment of Inference Attacks Against
  Machine Learning Models}.
\newblock In {\em {USENIX Security}}, 2022.

\bibitem{LXS17}
Yuntao Liu, Yang Xie, and Ankur Srivastava.
\newblock Neural trojans.
\newblock In {\em 2017 IEEE International Conference on Computer Design
  (ICCD)}, pages 45--48. IEEE, 2017.

\bibitem{MBG21}
Mohammad Malekzadeh, Anastasia Borovykh, and Deniz G{\"u}nd{\"u}z.
\newblock Honest-but-curious nets: Sensitive attributes of private inputs can
  be secretly coded into the entropy of classifiers' outputs.
\newblock In {\em ACM CCS}, 2021.

\bibitem{NBCJRSSTX08}
Blaine Nelson, Marco Barreno, Fuching~Jack Chi, Anthony~D. Joseph, enjamin
  I.~P.~Rubinstein, Udam Saini, Charles Sutton, J.~Doug Tygar, and Kai Xia.
\newblock {Exploiting Machine Learning to Subvert Your Spam Filter}.
\newblock In {\em {First USENIX Workshop on Large-Scale Exploits and Emergent
  Threats}}, 2008.

\bibitem{NWCBWN11}
Yuval Netzer, Tao Wang, Adam Coates, Alessandro Bissacco, Bo~Wu, and Andrew~Y.
  Ng.
\newblock {Reading Digits in Natural Images with Unsupervised Feature
  Learning}.
\newblock In {\em {NIPS}}, 2011.

\bibitem{NCL21}
Timothy Nguyen, Zhourong Chen, and Jaehoon Lee.
\newblock {Dataset Meta-Learning from Kernel Ridge-Regression}.
\newblock In {\em {ICLR}}, 2021.

\bibitem{NNXL21}
Timothy Nguyen, Roman Novak, Lechao Xiao, and Jaehoon Lee.
\newblock {Dataset Distillation with Infinitely Wide Convolutional Networks}.
\newblock In {\em {NeurIPS}}, 2021.

\bibitem{PMSW18}
Nicolas Papernot, Patrick McDaniel, Arunesh Sinha, and Michael Wellman.
\newblock {SoK: Towards the Science of Security and Privacy in Machine
  Learning}.
\newblock In {\em {Euro S\&P}}, pages 399--414, 2018.

\bibitem{PMJFCS16}
Nicolas Papernot, Patrick~D. McDaniel, Somesh Jha, Matt Fredrikson, Z.~Berkay
  Celik, and Ananthram Swami.
\newblock {The Limitations of Deep Learning in Adversarial Settings}.
\newblock In {\em {Euro S\&P}}, pages 372--387, 2016.

\bibitem{PGHFZFGD20}
Neehar Peri, Neal Gupta, W~Ronny Huang, Liam Fowl, Chen Zhu, Soheil Feizi, Tom
  Goldstein, and John~P Dickerson.
\newblock Deep k-nn defense against clean-label data poisoning attacks.
\newblock In {\em European Conference on Computer Vision}, pages 55--70.
  Springer, 2020.

\bibitem{RBXXWHH20}
Yu~Rong, Yatao Bian, Tingyang Xu, Weiyang Xie, Ying Wei, Wenbing Huang, and
  Junzhou Huang.
\newblock {Self-Supervised Graph Transformer on Large-Scale Molecular Data}.
\newblock In {\em {NeurIPS}}, 2020.

\bibitem{RWRK20}
Elan Rosenfeld, Ezra Winston, Pradeep Ravikumar, and Zico Kolter.
\newblock Certified robustness to label-flipping attacks via randomized
  smoothing.
\newblock In {\em International Conference on Machine Learning}, pages
  8230--8241. PMLR, 2020.

\bibitem{SBBFZ20}
Ahmed Salem, Apratim Bhattacharya, Michael Backes, Mario Fritz, and Yang Zhang.
\newblock {Updates-Leak: Data Set Inference and Reconstruction Attacks in
  Online Learning}.
\newblock In {\em {USENIX Security}}, pages 1291--1308, 2020.

\bibitem{SWBMZ22}
Ahmed Salem, Rui Wen, Michael Backes, Shiqing Ma, and Yang Zhang.
\newblock {Dynamic Backdoor Attacks Against Machine Learning Models}.
\newblock In {\em {Euro S\&P}}, 2022.

\bibitem{SZHBFB19}
Ahmed Salem, Yang Zhang, Mathias Humbert, Pascal Berrang, Mario Fritz, and
  Michael Backes.
\newblock {ML-Leaks: Model and Data Independent Membership Inference Attacks
  and Defenses on Machine Learning Models}.
\newblock In {\em {NDSS}}, 2019.

\bibitem{SSTS21}
Roei Schuster, Congzheng Song, Eran Tromer, and Vitaly Shmatikov.
\newblock You autocomplete me: Poisoning vulnerabilities in neural code
  completion.
\newblock In {\em 30th USENIX Security Symposium (USENIX Security)}, pages
  1559--1575, 2021.

\bibitem{SDSE20}
Roy Schwartz, Jesse Dodge, Noah~A. Smith, and Oren Etzioni.
\newblock {Green {AI}}.
\newblock {\em {Commun. of the ACM}}, 2020.

\bibitem{SHNSSDG18}
Ali Shafahi, W~Ronny Huang, Mahyar Najibi, Octavian Suciu, Christoph Studer,
  Tudor Dumitras, and Tom Goldstein.
\newblock {Poison Frogs! Targeted Clean-Label Poisoning Attacks on Neural
  Networks}.
\newblock In {\em {NeurIPS}}, pages 6103--6113, 2018.

\bibitem{SJZLCSFYW21}
Lujia Shen, Shouling Ji, Xuhong Zhang, Jinfeng Li, Jing Chen, Jie Shi,
  Chengfang Fang, Jianwei Yin, and Ting Wang.
\newblock Backdoor pre-trained models can transfer to all.
\newblock {\em arXiv preprint arXiv:2111.00197}, 2021.

\bibitem{SSSS17}
Reza Shokri, Marco Stronati, Congzheng Song, and Vitaly Shmatikov.
\newblock {Membership Inference Attacks Against Machine Learning Models}.
\newblock In {\em {S\&P}}, pages 3--18, 2017.

\bibitem{SZ15}
Karen Simonyan and Andrew Zisserman.
\newblock {Very Deep Convolutional Networks for Large-Scale Image Recognition}.
\newblock In {\em {ICLR}}, 2015.

\bibitem{SRS17}
Congzheng Song, Thomas Ristenpart, and Vitaly Shmatikov.
\newblock {Machine Learning Models that Remember Too Much}.
\newblock In {\em {CCS}}, pages 587--601, 2017.

\bibitem{SLCFTSK22}
Rui Song, Dai Liu, Dave~Zhenyu Chen, Andreas Festag, Carsten Trinitis, Martin
  Schulz, and Alois~C. Knoll.
\newblock {Federated Learning via Decentralized Dataset Distillation in
  Resource-Constrained Edge Environments}.
\newblock {\em {CoRR abs/2208.11311}}, 2022.

\bibitem{SKL17}
Jacob Steinhardt, Pang Wei~W Koh, and Percy~S Liang.
\newblock Certified defenses for data poisoning attacks.
\newblock {\em Advances in neural information processing systems}, 30, 2017.

\bibitem{SAKNT19}
Mahesh Subedar, Nilesh Ahuja, Ranganath Krishnan, Ibrahima~J Ndiour, and Omesh
  Tickoo.
\newblock Deep probabilistic models to detect data poisoning attacks.
\newblock {\em arXiv preprint arXiv:1912.01206}, 2019.

\bibitem{TWTZ21}
Di~Tang, XiaoFeng Wang, Haixu Tang, and Kehuan Zhang.
\newblock {Demon in the Variant: Statistical Analysis of DNNs for Robust
  Backdoor Contamination Detection}.
\newblock In {\em {USENIX Security}}, pages 1541--1558, 2021.

\bibitem{TLZM16}
Jian Tang, Jingzhou Liu, Ming Zhang, and Qiaozhu Mei.
\newblock {Visualizing Large-scale and High-dimensional Data}.
\newblock In {\em {WWW}}, pages 287--297, 2016.

\bibitem{TLM18}
Brandon Tran, Jerry Li, and Aleksander Madry.
\newblock Spectral signatures in backdoor attacks.
\newblock {\em Advances in neural information processing systems}, 31, 2018.

\bibitem{WCG20}
Binghui Wang, Xiaoyu Cao, Neil~Zhenqiang Gong, et~al.
\newblock On certifying robustness against backdoor attacks via randomized
  smoothing.
\newblock {\em arXiv preprint arXiv:2002.11750}, 2020.

\bibitem{WYSLVZZ19}
Bolun Wang, Yuanshun Yao, Shawn Shan, Huiying Li, Bimal Viswanath, Haitao
  Zheng, and Ben~Y. Zhao.
\newblock {Neural Cleanse: Identifying and Mitigating Backdoor Attacks in
  Neural Networks}.
\newblock In {\em {S\&P}}, pages 707--723, 2019.

\bibitem{WY21}
Lin Wang and Kuk-Jin Yoon.
\newblock Knowledge distillation and student-teacher learning for visual
  intelligence: A review and new outlooks.
\newblock {\em IEEE Transactions on Pattern Analysis and Machine Intelligence},
  2021.

\bibitem{WZLCXW20}
Ren Wang, Gaoyuan Zhang, Sijia Liu, Pin-Yu Chen, Jinjun Xiong, and Meng Wang.
\newblock Practical detection of trojan neural networks: Data-limited and
  data-free cases.
\newblock In {\em European Conference on Computer Vision}, pages 222--238.
  Springer, 2020.

\bibitem{WZTE18}
Tongzhou Wang, Jun-Yan Zhu, Antonio Torralba, and Alexei~A. Efros.
\newblock {Dataset Distillation}.
\newblock {\em {CoRR abs/1811.10959}}, 2018.

\bibitem{XPJW21}
Zhaohan Xi, Ren Pang, Shouling Ji, and Ting Wang.
\newblock {Graph Backdoor}.
\newblock In {\em {USENIX Security}}, 2021.

\bibitem{XMCLK21}
Zhen Xiang, David~J. Miller, Siheng Chen, Xi~Li, and George Kesidis.
\newblock A backdoor attack against 3d point cloud classifiers.
\newblock In {\em IEEE/CVF International Conference on Computer Vision (ICCV)},
  pages 7597--7607, 2021.

\bibitem{XRV17}
Han Xiao, Kashif Rasul, and Roland Vollgraf.
\newblock {Fashion-MNIST: a Novel Image Dataset for Benchmarking Machine
  Learning Algorithms}.
\newblock {\em {CoRR abs/1708.07747}}, 2017.

\bibitem{XCCL21}
Chulin Xie, Minghao Chen, Pin-Yu Chen, and Bo~Li.
\newblock Crfl: Certifiably robust federated learning against backdoor attacks.
\newblock In {\em International Conference on Machine Learning}, 2021.

\bibitem{XWCYH22}
Yuanhao Xiong, Ruochen Wang, Minhao Cheng, Felix Yu, and Cho-Jui Hsieh.
\newblock {FedDM: Iterative Distribution Matching for Communication-Efficient
  Federated Learning}.
\newblock {\em {CoRR abs/2207.09653}}, 2022.

\bibitem{XWLBGL21}
Xiaojun Xu, Qi~Wang, Huichen Li, Nikita Borisov, Carl~A. Gunter, and Bo~Li.
\newblock {Detecting AI Trojans Using Meta Neural Analysis}.
\newblock In {\em {S\&P}}, 2021.

\bibitem{YLZZ19}
Yuanshun Yao, Huiying Li, Haitao Zheng, and Ben~Y. Zhao.
\newblock {Latent Backdoor Attacks on Deep Neural Networks}.
\newblock In {\em {CCS}}, pages 2041--2055, 2019.

\bibitem{ZJWG21}
Zaixi Zhang, Jinyuan Jia, Binghui Wang, and Neil~Zhenqiang Gong.
\newblock {Backdoor Attacks to Graph Neural Networks}.
\newblock In {\em {SACMAT}}, pages 15--26, 2021.

\bibitem{ZB21}
Bo~Zhao and Hakan Bilen.
\newblock {Dataset Condensation with Differentiable Siamese Augmentatio}.
\newblock In {\em {ICML}}, 2021.

\bibitem{ZB212}
Bo~Zhao and Hakan Bilen.
\newblock {Dataset Condensation with Distribution Matching}.
\newblock {\em {CoRR abs/2110.04181}}, 2021.

\bibitem{ZMB21}
Bo~Zhao, Konda~Reddy Mopuri, and Hakan Bilen.
\newblock {Dataset Condensation With Gradient Matching}.
\newblock In {\em {ICLR}}, 2021.

\bibitem{ZZWGC21}
Songzhu Zheng, Yikai Zhang, Hubert Wagner, Mayank Goswami, and Chao Chen.
\newblock Topological detection of trojaned neural networks.
\newblock {\em Advances in Neural Information Processing Systems}, 34, 2021.

\bibitem{ZHLTSG192}
Chen Zhu, W.~Ronny Huang, Hengduo Li, Gavin Taylor, Christoph Studer, and Tom
  Goldstein.
\newblock {Transferable Clean-Label Poisoning Attacks on Deep Neural Nets}.
\newblock In {\em {International Conference on Machine Learning}}, pages
  7614--7623, 2019.

\end{thebibliography}
\end{small}

\end{document}